\input lanlmac
\newskip\defaultbaselineskip\defaultbaselineskip=12pt

\def\GeV{\mathord{\rm \;GeV}}
\def\MeV{\mathord{\rm \;MeV}}

\def\sinh{\mathop{\rm sinh}\nolimits}
\def\bar{\overline}

\def\gsim{\mathrel{\raise2pt\hbox to 8pt{\raise -5pt\hbox{$\sim$}\hss{$>$}}}}
\def\rsim{\mathrel{\raise2pt\hbox to 8pt{\raise -5pt\hbox{$\sim$}\hss{$>$}}}}
\def\lsim{\mathrel{\raise2pt\hbox to 8pt{\raise -5pt\hbox{$\sim$}\hss{$<$}}}}
\def\ssqr#1#2{{\vbox{\hrule height.#2pt
      \hbox{\vrule width.#2pt height#1pt \kern#1pt\vrule width.#2pt}
      \hrule height.#2pt}\kern-.#2pt}}

\def\listrefsnomod{\footatend\vfill\supereject\immediate\closeout\rfile\writestoppt
\centerline{{\bf References}}\bigskip{\frenchspacing%
\parindent=20pt\escapechar=` \input \jobname.refs\vfill\eject}}
\ifx\hyperref\undefined\def\hyperref#1#2#3#4{#4}\fi
\ifx\hyperdef\undefined\def\hyperdef#1#2#3#4{#4}\fi
\ifx\href\undefined\def\href#1#2{#2}\fi
\global\newcount\mtabno
\mtabno=1
\def\table#1#2#3{\DefWarn#1%
\xdef #1{\noexpand\hyperref{}{table}{\the\mtabno}%
{\the\mtabno}}\goodbreak\midinsert
$$#2$$\nobreak\centerline{Table~\hyperdef\hypernoname{table}%
{\the\mtabno}{\the\mtabno}. {\sl #3}}\bigskip\endinsert
\writedef{#1\leftbracket#1}\global\advance\mtabno by1}
\def\figure#1#2#3{\DefWarn#1\xdef#1{\noexpand\hyperref{}{figure}%
{\the\figno}{\the\figno}}\writedef{#1\leftbracket#1}%
\figinsert\figin{\centerline{#2}}\medskip\centerline{\vbox{\baselineskip
\defaultbaselineskip
\advance\hsize by -1truein\noindent\wrlabeL{#1=#1}\centerline{\sl
{\bf Fig.~\hyperdef\hypernoname{figure}{\the\figno}{\the\figno}:} #3}}}
\bigskip\endinsert\global\advance\figno by1}
\newread\myread
\def\ignore#1{}%
\def\readdefs{\ifx\writedef\ignore
 \immediate\openin\myread=\jobname.defs
 \ifeof\myread\message{No file \jobname.defs yet.}\else
   \closein\myread\relax\input\jobname.defs\fi\else
 \errmessage{can't \string\readdefs\space after \string\writedefs!!!}
\fi\relax}
\def\makelatexlike#1{\expandafter\let\csname old \string#1\endcsname=#1%
                 \def#1##1{%
                  \edef\tempname{\ifcat\relax\noexpand##1\noexpand##1\else
                                 \expandafter\noexpand\csname##1\endcsname\fi}%
                  \expandafter\expandafter\csname old \string#1\endcsname
                  \tempname}}
\def\nameuse#1{\edef\tempname{\ifcat\relax\noexpand#1\noexpand#1\else
                              \expandafter\noexpand\csname#1\endcsname\fi}%
               \expandafter\ifx\tempname\undefined
                  \message{YET UNDEFINED NAME \string`\string#1\string' USED.}%
                  ???%
               \else\expandafter\tempname\fi}
                                          

\newwrite\figsfile \newwrite\tabsfile
\newif\ifinplacealso\inplacealsofalse
\def\forjournal{
\defaultbaselineskip=24pt
\immediate\openout\figsfile\jobname.figs
\immediate\openout\tabsfile\jobname.tabs
\let\oldfigure=\figure
\let\oldtable=\table
\let\oldend=\end
\def\fourthoffive##1##2##3##4##5{##4}
\def\fourthoffour##1##2##3##4{##4}
\def\getno##1{\expandafter\fourthoffive##1}
\def\ignore##1{}
\ifinplacealso
\def\figure##1##2##3{{\toks0={{\figno=\getno##1\let##1=\undefined
  \let\wrlabeL\ignore\let\writedef\ignore
  \let\hyperref\fourthoffour\let\hyperdef\fourthoffour
  \oldfigure##1{##2}{##3}}\vfill\eject}\immediate\write\figsfile{\the\toks0}%
  \oldfigure##1{##2}{##3}}}
\def\table##1##2##3{{\toks0={{\mtabno=\getno##1\let##1=\undefined
  \let\wrlabeL\ignore\let\writedef\ignore
  \let\hyperref\fourthoffour\let\hyperdef\fourthoffour
  \oldtable##1{##2}{##3}}\vfill\eject}\immediate\write\tabsfile{\the\toks0}%
  \oldtable##1{##2}{##3}}}
\else
\def\figure##1##2##3{{\toks0={{%
  \oldfigure##1{##2}{##3}}\vfill\eject}\immediate\write\figsfile{\the\toks0}}}
\def\table##1##2##3{{\toks0={{%
  \oldtable##1{##2}{##3}}\vfill\eject}\immediate\write\tabsfile{\the\toks0}}}
\fi
\def\end{\vfill\eject\nopagenumbers
         \immediate\closeout\figsfile\input\jobname.figs
         \immediate\closeout\tabsfile\input\jobname.tabs
         \oldend}
}








\def\author{\bigskip\centerline{David Daniel} 
    \smallskip\centerline{\it T-8, MS-B285,
                     Los Alamos National Laboratory, Los Alamos, NM 87545}}



\def\NPB#1{{\it Nucl. Phys.} {\bf B#1}}
\def\NPBPS#1{{\it Nucl. Phys.} {\bf B} ({\it Proc. Suppl.}) {\bf #1}}
\def\PRL#1{{\it Phys. Rev. Lett.} {\bf #1}}

\def\PRD#1{{\it Phys. Rev.} {\bf D#1}}

\def\PLB#1{{\it Phys. Lett.} {\bf #1B}}

\def\etal{{\it et al.\ }}

\def\tsukuba{Int. Symp. {\it ``LATTICE 91''}, Proceedings of the 
             International Symposium on Lattice Field Theory, Tsukuba, Japan, 
             1991, Eds. Fukugita $et\ al.$, \NPBPS{26}, (1992) }

\def\dallas{{\it ``LATTICE 93''}, Proceedings of
             the International Symposium on Lattice Field Theory, Dallas, 
             U.S.A., 1993, Eds. T.~Draper  $et\ al.$, 
             \NPBPS{34}, (1994) }
\def\bielefeld{{\it ``LATTICE 94''}, Proceedings of
             the International Symposium on Lattice Field Theory, Bielefeld, 
             Germany, 1994, Eds. F.~Karsch $et\ al.$, 
             \NPBPS{42}, (1995) }



\input epsf
\readdefs\writedefs


\ifx\href\undefined\def\href#1#2{{#2}}\fi
\def\spireshome%
{http://www.slac.stanford.edu/cgi-bin/spiface/find/hep/www?FORMAT=WWW&}
{\catcode`\%=12
\xdef\spiresjournal#1#2#3{\noexpand\href{\spireshome
                          rawcmd=find+journal+#1%2C+#2%2C+#3}}
\xdef\spireseprint#1#2{\noexpand\href{\spireshome rawcmd=find+eprint+#1%2F#2}}
\xdef\spiresreport#1{\noexpand\href{\spireshome rawcmd=find+rept+#1}}
}
\def\eprint#1#2{\spireseprint{#1}{#2}{#1/#2}}


\def\GMO{{\rm GMO}}
\def\eff{{\rm eff}}
\def\MSbar{{\ifmmode\let\finish=\relax\else$\let\finish=$\fi
            \mathpalette{\hbox\bgroup$}{\overline{MS}\egroup$}%
            \finish}}
\def\mbar{\hbox{$\overline{m}$}}

\def\lilj{\hbox{$\{ U_i U_j \}$}}

\def\sslj{\hbox{$ \{ SS,\ S U_j   \}$}}
\def\clj{\hbox{$ \{ C U_j   \}$}}
\def\decu#1#2#3{_{\{#1#2#3\}}}


\baselineskip=12pt 

\Title{
  LA UR-95-2354}{
Hadron Spectrum with Wilson fermions}

\centerline{
  Tanmoy Bhattacharya and Rajan Gupta}
\smallskip
\centerline{\it
  T-8, MS-B285, Los Alamos National Laboratory, Los Alamos, NM 87545}

\bigskip

\centerline{
  Gregory Kilcup}
\smallskip
\centerline{\it
  Physics Department, The Ohio State University, Columbus, OH 43210}

\bigskip

\centerline{
  Stephen Sharpe}
\smallskip
\centerline{\it
  Physics Department, University of Washington, Seattle, WA 98195}

\bigskip
\bigskip
\bigskip

We present results of a high statistics study of the quenched spectrum
using Wilson fermions at $\beta=6.0$ on $32^3 \times 64$ lattices.  We
calculate the masses of mesons and baryons composed of both degenerate
and non-degenerate quarks.  Using non-degenerate quark combinations
allows us to study baryon mass splittings in detail. We find
significant deviations from the lowest order chiral expansion,
deviations that are consistent with the expectations of quenched
chiral perturbation theory.  We find that there is a $\sim 20\%$
systematic error in the extracted value of $m_s$, depending on the
meson mass ratio used to set its value.  Using the largest estimate of
$m_s$ we find that the extrapolated octet mass-splittings are in
agreement with the experimental values, as is $M_\Delta - M_N$, while
the decuplet splittings are $30\%$ smaller than experiment.  Combining
our results with data from the GF11 collaboration we find considerable
ambiguity in the extrapolation to the continuum limit.  Our preferred
values are $M_N / M_\rho = 1.38(7)$ and $M_\Delta / M_\rho =
1.73(10)$, suggesting that the quenched approximation is good to only
$\sim 10-15\%$.  We also analyze the $O(ma)$ discretization errors in
heavy quark masses.

\Date{20 DEC, 1995.}
\baselineskip=12pt 

\newsec{Introduction}

Precise measurements of the hadron spectrum using lattice QCD 
are crucial both to validate QCD as the correct theory of
strong interactions and to establish the reliability of numerical
simulations for extracting weak matrix elements.
Current lattice calculations
suffer from a variety of systematic errors,
most notably those due to quenching, discretization and the need to
extrapolate to light quark masses.
In this work we present a detailed study of these systematics for 
the hadron spectrum.

Such a study requires small statistical errors. We have reduced these
by using a moderately large ensemble, 170 lattices,
and working on a large lattice, $32^3\times 64$ at
$\beta = 6.0$ in the quenched approximation.
We use unimproved Wilson fermions.
Preliminary results from a subset of 100 lattices
were presented at the LATTICE94 meeting
\ref\TbRgBiel{T. Bhattacharya and R.~Gupta, 
\spireseprint{hep-lat}{9501016}{\bielefeld\ 935}.}.
Our lattices are large enough that we expect finite size effects to be small.
The major technical features of our work are
(a) using two kinds of sources that yield correlators that converge to
their asymptotic values from opposite directions, so as to improve
the reliability of the masses extracted;
(b) calculating hadron masses using non-degenerate light quarks,
which allows us to study the quark-mass dependence in detail; and
(c) using ratios of correlators to obtain accurate estimates of
mass differences.
We find that terms of higher order than linear in the quark mass
are very significant, and that their inclusion is essential for
the extrapolation to the physical light quark masses,
particularly for mass splittings amongst spin-1/2 baryons.
 Higher order terms are also important
for the vector mesons and the spin-3/2 baryons.

The outline of the paper is as follows.
In the following section we explain how we generate lattices and
calculate quark propagators. After a brief discussion
of fitting, and a summary of the expected chiral behavior of hadron masses
in the quenched approximation, we present our results for mesons and baryon
masses. We then extrapolate mass ratios to the continuum limit by
combining our results with those of the GF11 collaboration
\ref\weinhm{GF11 Collaboration, \spiresjournal{Nuc.+Phys.}{B430}{179}
{\NPB{430} (1994) 179}.}. 
There turns out to be considerable ambiguity in this extrapolation.
Our preferred values for the extrapolated ratios
are $M_N / M_\rho = 1.38(7)$ and $M_\Delta / M_\rho = 1.73(10)$.
This suggests that the quenched approximation is good to $\sim 10-15\%$,
less accurate than suggested in Ref.~\weinhm.
We close with some conclusions and suggestions for further work.

We use the following conventions throughout the paper. Hadron masses
are denoted by upper case $M$, while for quark masses we use lower
case $m$.  All masses are in lattice units unless explicitly 
expressed in $\MeV$ (or $\GeV$). 

\newsec{Details of simulations}

Gauge configurations are generated using a combination of over-relaxed
(OR), pseudo-heatbath (PHB) and Metropolis algorithms.  Typically, 5
OR sweeps are followed by a PHB update, with the latter consisting of
three hits, one in each of the SU(2) subgroups.  In some cases the PHB
update is replaced by a 20-hit Metropolis sweep.  Two independent
streams were generated, each starting from a lattice consisting of two
independently thermalized $32^4$ lattices joined together.  A further
$2000 \times (5 OR + 1 PHB)$ thermalization sweeps were then
performed.  Thereafter we analyze lattices separated by $400 \times (5
OR + 1 PHB)$ sweeps.

We calculate quark propagators using the simple Wilson action,
with periodic boundary conditions in all 4 directions. We use
two kinds of extended sources -- Wuppertal and Wall -- at each of the
five values of quark mass given by $\kappa = 0.135$ ($C$), $0.153$
($S$), $0.155$ ($U_1$), $0.1558$ ($U_2$), and $0.1563$ ($U_3$).  These
quarks correspond to pseudoscalar mesons of mass $2835$, $983$, $690$,
$545$ and $431$ $\MeV$ respectively, using $1/a=2.33\GeV$
for the lattice scale.  We use the three light quarks to extrapolate
the data to the physical isospin symmetric light quark mass $\mbar =
(m_u+m_d)/2$, while the $C$ and $S$ $\kappa$ values are selected to be
close to the physical charm and strange quark masses.  The physical
value of strange quark, in fact, lies between $S$ and $U_1$ and we use these
two points to interpolate to it.  
In most cases we find that the extrapolation to \mbar\ can be done
using the six combinations of light quarks $U_1 U_1, \ U_1 U_2, \ U_1
U_3,\ U_2 U_2,\ U_2 U_3,\ U_3 U_3$.  For brevity we use \lilj\ to
refer to this set of masses.

We analyze three types of hadron correlators distinguished by
the sources and sinks used to generate quark propagators.  These are
wall source and point sink (WL), Wuppertal source and point
sink (SL), and Wuppertal source and sink (SS).  The notation and 
details of the implementation of the Wuppertal source are as
in our previous work
\ref\rdynwfb{R.~Gupta,\etal, \spiresjournal{Phys.+Rev.}{D44}{3272}
{\PRD{44} (1991) 3272}.}.  The smearing 
parameter is set to $\kappa_{KG} = 0.181$, corresponding to a smearing size 
of $\Omega^2 \approx 3$. (In \rdynwfb\ this was mistakenly
written as $\Omega \approx 3$.) The generation of the 
Wuppertal sources is a negligible overhead on the inversion. 

The Dirac equation is solved using the over-relaxed preconditioned
(fourth order polynomial) minimal residue (MR4) algorithm described in
Ref.~\ref\rdynwfa{R.~Gupta,\etal,
\spiresjournal{Phys.+Rev.}{D40}{2072} {\PRD{40} (1989) 2072}.}.  The
convergence rate is quite insensitive to the over-relaxation parameter
$\omega$ as long as it is in the range $1.2 - 1.35$; we use
$\omega=1.3$.  We set the convergence criteria (for all values of
$\kappa$) to $ |r^2| / |\chi^2| < 10^{-14}$, where $r$ is the
remainder and $\chi$ is the solution.  This tolerance is as
small as we can demand, as we use IEEE single precision arithmetic.
To ensure convergence we run the MR4 inverter up to three times,
refreshing the starting remainder each time, and then switch to
conjugate gradient, and force it to run at least one cycle.
To date, we have observed no failures of MR4.  
We note that the simple MR algorithms are much less sensitive
to our use of 32-bit arithmetic than the CG-based algorithms, whose
convergence rate depends on maintaining orthogonality of a
sequence of vectors.
Indeed, we have found that MR4 is the most efficient and stable 
of the algorithms (MR, MR2, MR4, CG, BiCG5
\ref\bicgfive{A. Borici and P. deForcrand, private communications.}
and BiCGStab \ref\bicgstab{A. Frommer et al.,
\spiresjournal{Int.+J.+Mod.+Phys.}{C5}{1073}{{\it Int. J. Mod. Phys.}
{\bf C5} (1994) 1073.}}) 
we have implemented on the CM5.

A technical detail of our MR4 algorithm that makes it suitable for 32
bit precision is as follows. The $n^{th}$ iterate of the solution
$\chi$ (and similarly the remainder $r$) are given by $\vec \chi_n =
\vec \chi_{n-1} + \omega \alpha_n \vec r_{n-1}$.  The global sums needed 
in the calculation of $\alpha_n$ are done in double precision.  Any
residual errors, $\delta\alpha_n$, can be absorbed into $\omega$ and
do not adversely affect the convergence rate as long as $\omega_n =
\omega(1 + \delta\alpha_n)$ stays within the optimal range. Our tests 
also show that the calculation of $\alpha_n$ can be done in single
precision, and the convergence rate is not affected provided the
remainder is refreshed somewhat more often depending on the quark
mass.

The only place that round-off errors due to use of 32 bit precision
could affect the results is in the evaluation of the final
convergence.  For this purpose we make two checks. Firstly, we monitor
the final value of $ |r^2|$ and $|r^2| / |\chi^2| $ on each time-slice
in addition to the global value which could be biased by time-slices
closest to the source. We find that the values at the time-slice
farthest from the source are $|r^2| / |\chi^2| \approx 10^{-6}$ for the
$C$ quark and $\le 10^{-13}$ for other quarks. This means that, even
for the heaviest quark, the effect of incomplete convergence is
smaller than the statistical errors. Secondly, we double the number of
inversion sweeps on randomly selected lattices.  We find that the
relative change in the value of hadronic 2-point correlation functions
is $\lsim 10^{-5}$, which is negligible.  

\figure\fpicomp{\epsfysize=4in\epsfbox{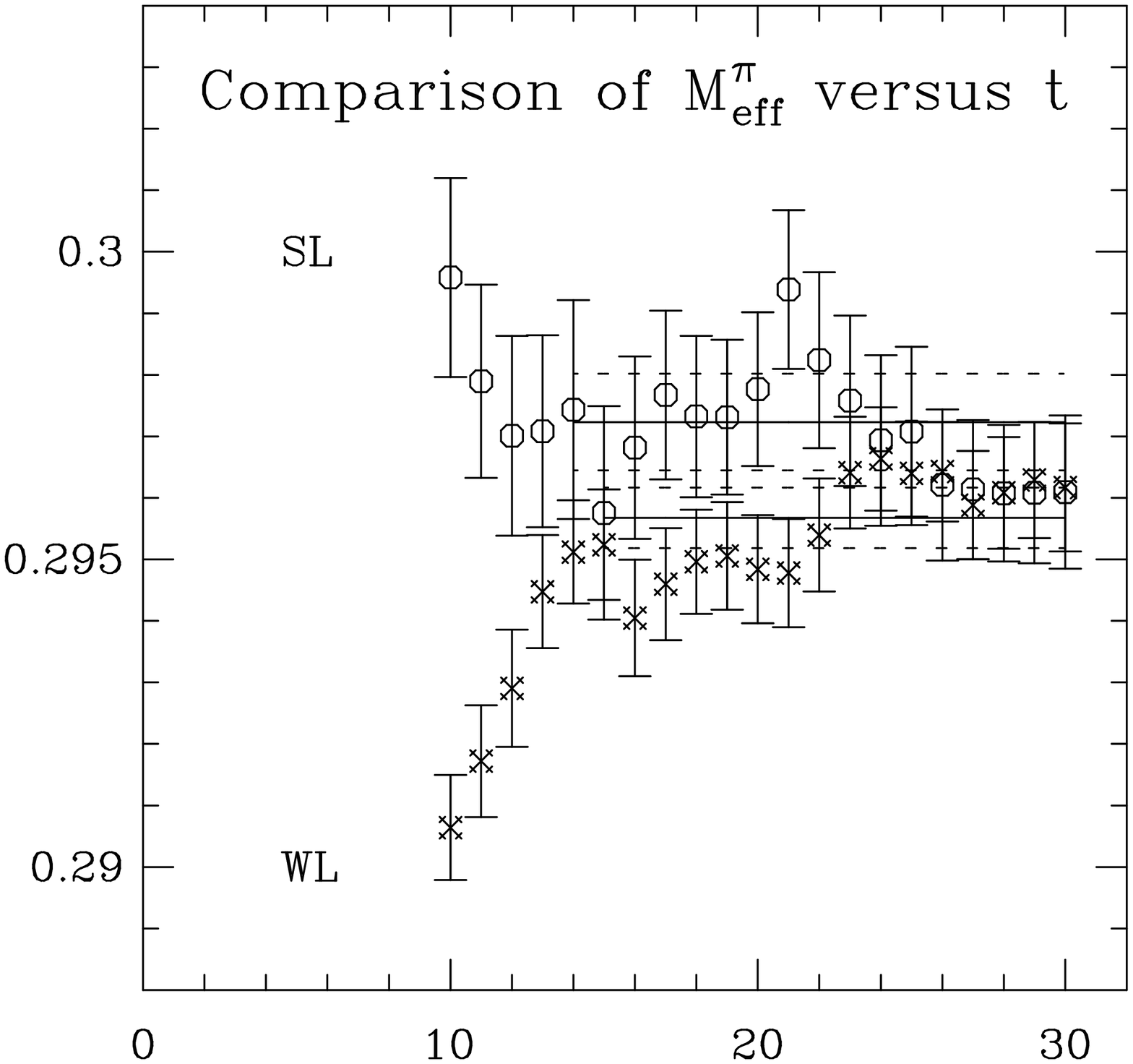}}{Comparison of 
$M_{\eff}(t)$ for $U_1U_1$ pion correlators with SL and WL sources.}

\figure\frhocomp{\epsfysize=4in\epsfbox{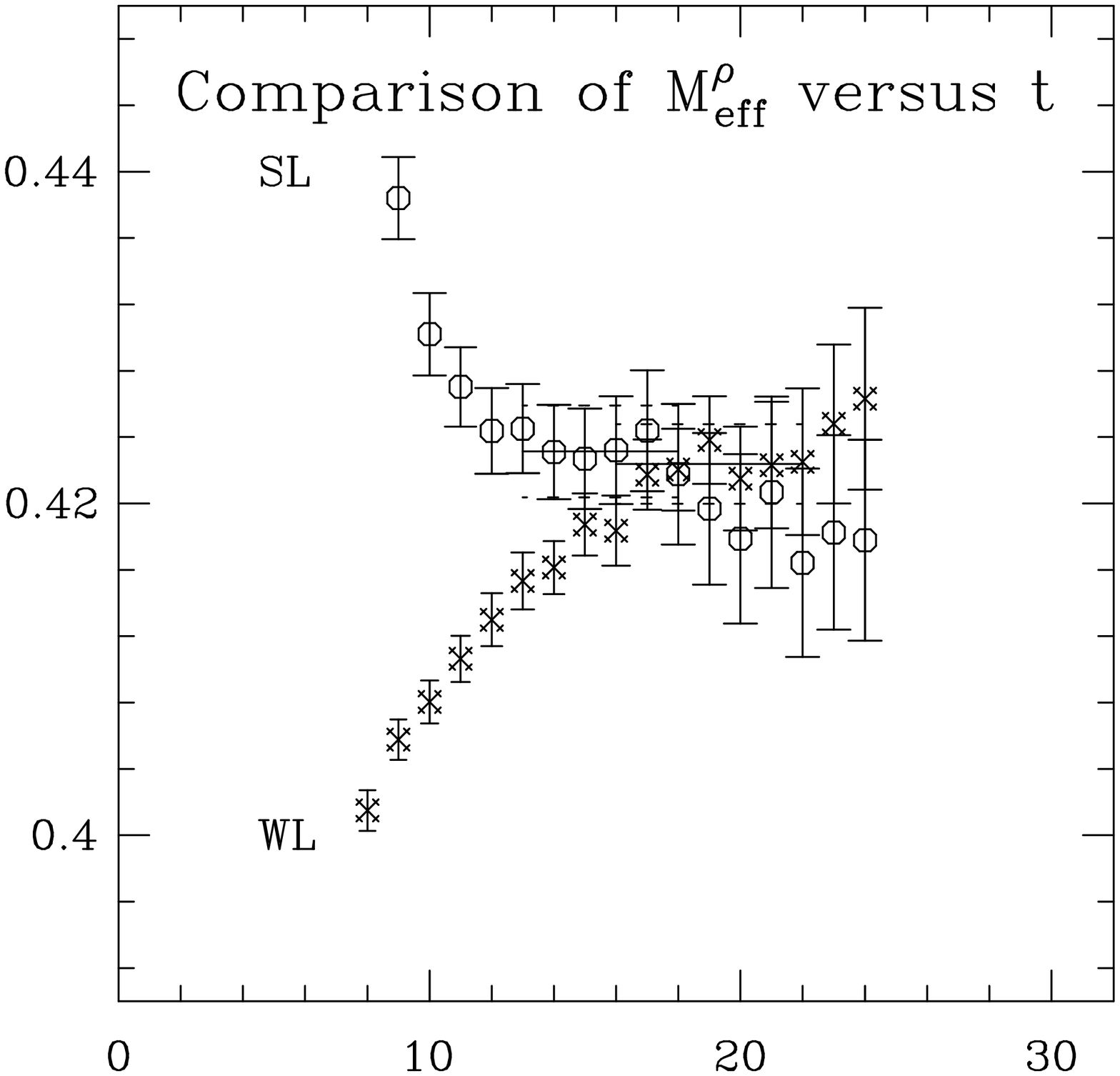}}{Comparison of 
$M_{\eff}(t)$ for $U_1U_1$ rho correlators with SL and WL sources.}

\figure\fnuccomp{\epsfysize=4in\epsfbox{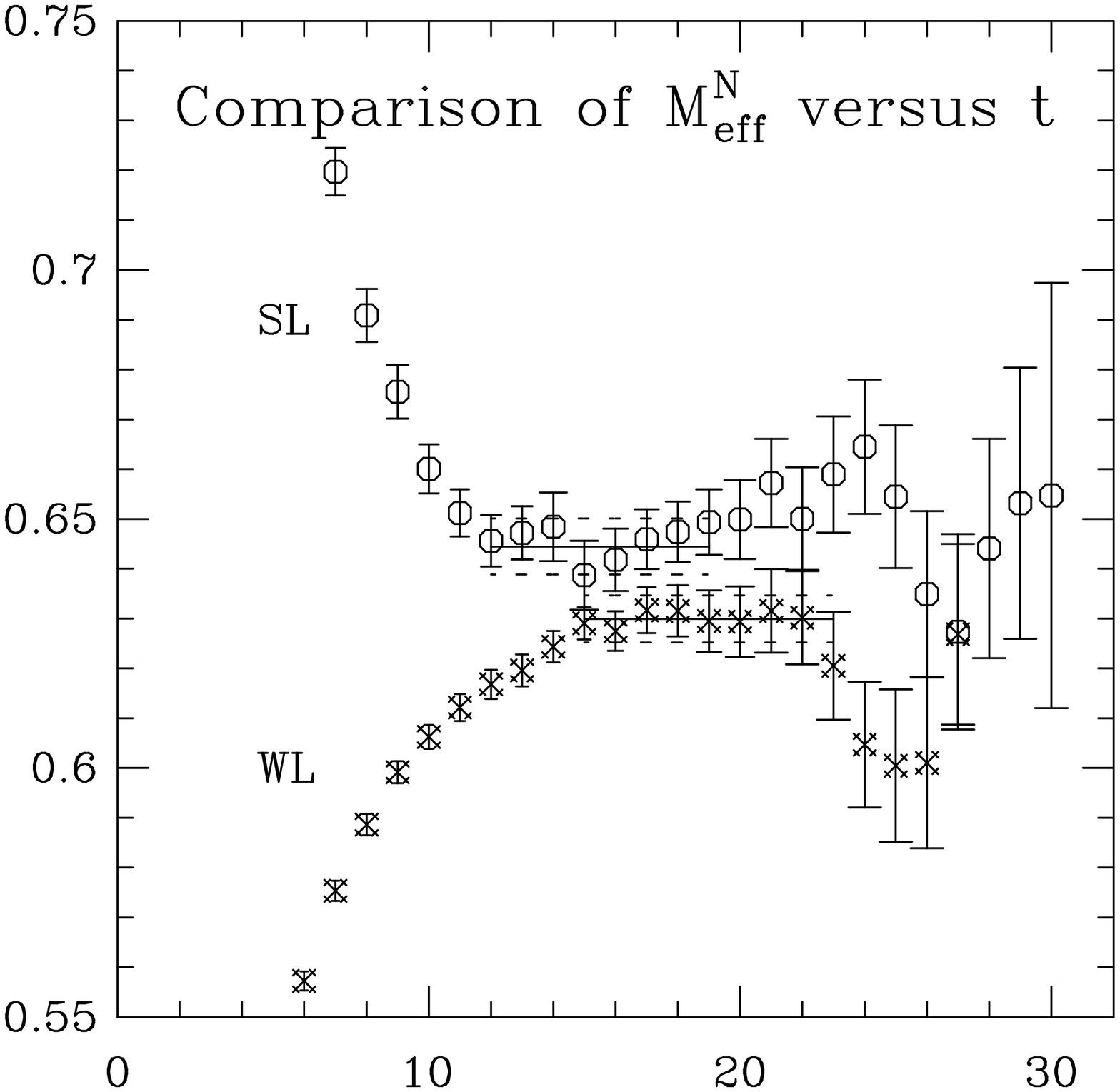}}{Comparison of 
$M_{\eff}(t)$ for $U_1U_1U_1$ nucleon correlators with SL and WL sources.}

\newsec{Fitting}

To illustrate some of the issues involved in fitting, we show, in
Figs.~\nameuse\fpicomp-\nameuse\fnuccomp, representative results for
the effective mass.
We define this as $M_{\eff}(t)=\ln[C(t-1)/C(t)]$, plus corrections due
to periodicity in the time direction.
The source lies at $t=0$ in all plots. 
It is a general feature that the WL effective masses
approach their asymptotic value from below, while those for SL and SS
correlators approach from above.  We consider in turn the problems we
face and the solutions we have adopted.

\item{1.}
A major problem is that, at the $2 \sigma$ level,
the convergence to the asymptotic value of $M_{\eff}$ is extremely slow.
The SL and WL effective masses do come together eventually
for pion and rho correlators, but for the nucleon the signal disappears 
into noise before convergence has occurred.
Masses extracted from WL correlators are systematically lower than
those from SL or SS correlators.
For the pion and rho channels, this difference is $\sim1\sigma$, 
while for the nucleons it is $\sim 2-3\sigma$.

\item{}
We think that this behavior is mainly due to limited statistics---a
recent JLQCD study \ref\jlqcd{S. Aoki {\it et al.}, \eprint{hep-lat}{9510013}.}
indicates that fluctuations at the $2\sigma$ level are the rule rather
than the exception.
What we have done in practice is to average the masses with weightings
$(2*WL + SL + SS)/4$. The equal weighting of wall and Wuppertal sources
ensures that the resulting mass lies between the two sets of data on
the effective mass plot.
Errors are obtained using single-elimination jack-knife,
performing a complete analysis (including forming the average
masses) on each jack-knife sample.

\item{2.}
The previous problem is exacerbated by the fact that we have
been unable to do stable fitting using the full correlation matrix.
When we do so, some fits lead to clearly unreasonable values for masses,
presumably because small errors in the correlation matrix are magnified
when this matrix is inverted to calculate $\chi^2$.
This problem is well known, and various remedies have been proposed 
\ref\reffixcorr{C. Michael, \spiresjournal{Phys.+Rev.}{D49}{2616}
{\PRD{49} (1994) 2616}, 
\semi G. Kilcup, \spireseprint{hep-lat}{9312081}{\dallas\ 350}.}.
For example, one can project against the eigenvectors of
the correlation matrix having small eigenvalues, and then invert.
Alternatively, one can reduce the range of times to which one fits.
We have tried both schemes, but find that the resulting mass estimates
are indistinguishable from those obtained fitting just with the
diagonal elements of the correlation matrix (``uncorrelated fits'').
Given that the schemes involve further subjective selections (e.g. how many
eigenvectors to discard) we choose to use uncorrelated fits for
our standard results. We do use the ``correlated fits'', however,
to check that our fits are reasonable.
Given that this is so, our jack-knife errors should be reliable.

\item{3.}
In a similar spirit, we have chosen to use single mass fits for
our final answers. We have also made fits with two masses, but find that
these fits require considerable tuning by hand, because our minimization
routines tend to give the same result for the two masses for a few of
the jack-knife samples. When we remove this problem, the resulting estimate
for the lightest mass is consistent with that from the single mass fit,
while the errors are 50\% larger. Given the large number of
correlators that we fit, it is impractical to do the tuning for each channel.

\item{4.}
Finally, we must choose a fit range.  
We choose the minimum time by inspecting the effective mass plots
and deciding where the ``plateau'' region begins.
The maximum time is then taken to be that at which the
diagonal errors have roughly doubled compared to the beginning of the
plateau, or that at which the signal shows a clear break.  
For the pion correlator,
where the errors do not grow with time, this means that we use all
points beyond the minimum time.

\smallskip
\noindent
Our overriding criterion is to include as many
time-slices as possible in the fits. 
We have not succeeded in developing a robust
automated procedure that meets this objective
when using two mass fits or incorporating the full covariance matrix.
Since we analyze $\sim 4000$ channels, it is not 
practical to tune the fitting of each by hand. 
Thus, for consistency, we use, for all channels, single mass fits keeping
only the diagonal elements of the correlation matrix. 
We feel confident in both our central values and our error estimates,
however, since in the many channels that we have fitted by hand, the results
do not change significantly when we use the
full correlation matrix and two mass fits.

\newsec{Quenching errors}
\seclab\secquench

In the last few years it has been argued that,
in the quenched approximation,
the $\eta'$ is a pseudo-Goldstone boson,
and that $\eta'$ loops give rise to chiral logarithms which make the
quenched approximation singular in the chiral limit
\ref\SRScpt{ S. Sharpe, \spiresjournal{Phys.+Rev.}{D41}{3233}
{\PRD{41} (1990) 3233}, \spiresjournal{Phys.+Rev.}{D46}{3146}
{PRD{46} (1992) 3146} \semi 
S.~Sharpe, 1994 TASI lectures, \eprint{hep-ph}{9412243}.}
\ref\CbMgcpt{C.~Bernard and M.~Golterman, \spiresjournal{Phys.+Rev.}{D46}{853}
{\PRD{46} (1992) 853} \semi M.~Golterman, \eprint{hep-lat}{9405002}.}.
The same methods (``quenched chiral perturbation theory'')
have been used to estimate the errors due to quenching.
It is important to test
the predictions of this theory against numerical results.
There is some supporting evidence---see Refs.
\ref\chiralrg{R.~Gupta, \spireseprint{hep-lat}{9412078}
{\NPBPS{42} (1995) 85}.} and
\ref\chiralmg{M.~Golterman, 
\eprint{hep-lat}{9411005}.}
for recent reviews---but more work is needed.

We collect here the results of quenched chiral perturbation
theory relevant for this work.
The expansion for the mass of a ``pion'' composed of quarks of
mass $m_1$ and $m_2$ is
\eqn\echptpi{
M_\pi^2 =   c_\pi  (m_1 + m_2) (1 - \delta \log[(m_1+m_2)]) + e_\pi m_q^2
+ \cdots
\,,}
with $c_\pi$, $e_\pi$
 and $\delta$ constants, and the ellipsis representing
higher order terms in the chiral expansion. 
The $m_q^2$ term is shorthand both for analytic terms, 
i.e. those proportional to $(m_1+m_2)^2$, $m_1^2$, and $m_2^2$, 
and for non-analytic chiral logarithms of general form $m_q^2 \log(m_q)$.
Such terms are present both in quenched and in full QCD, although the
constants multiplying them will be different in the two theories.
In contrast, the $\delta$ term is an artifact of quenching---it 
arises from $\eta'$ loops, and is divergent in the chiral limit.
Evidence for this divergence has been found with staggered fermions
\ref\kimsinclair{ S. Kim and D.K. Sinclair, 
\spiresjournal{Phys.+Rev.}{D52}{2614}{\PRD{52} (1995) 2614}.}, 
but the effect is small,
and becomes noticeable only at quark masses smaller than those we use.
Thus when fitting the pion masses we ignore the $\delta$ term.

The predicted form for baryon masses is, schematically
\ref\labrenzsharpe{J.~Labrenz and S. Sharpe, \spireseprint{hep-lat}{9312067}
{\dallas, 335}.},
\eqn\echptnuc{
M_{N} = a_N + \delta [b_N m_q^{1/2} + b_N' m_q \ln(m_q)] + 
c_N m_q + d_N m_q^{3/2} + e_N m_q^2 + \cdots
\,,}
where $\delta$ is the same constant as in Eq. \echptpi,
while $a_N-e_N$ are additional constants.
The expansion has the same form in full QCD, 
except that the $\delta$ term, which again comes from $\eta'$ loops, 
is absent. As for the pion, we ignore the $\delta$ term
in almost all fits.
It is small (because $\delta$ is small), and, furthermore,
is numerically indistinguishable from the higher order terms
within our range of light quark masses.
What we can test in some detail, however, 
is the expected form of the $m_q$, $m_q^{3/2}$, and $m_q^{2}$ terms.
Here we benefit greatly from our use of non-degenerate quarks.

Similar comments apply to the vector meson masses. 
Although these have not been discussed in the quenched chiral perturbation
theory literature, it is straightforward to extend the work done in
QCD \ref\vectorchpt{E. Jenkins, A. Manohar and M. Wise,
\spiresjournal{Phys.+Rev.+Lett.}{75}{2272}{\PRL{75} (1995) 2272}.}, and 
arrive at the prediction 
\eqn\echptrho{
M_{\rho} = a_\rho + \delta b_\rho m_q^{1/2} + 
c_\rho (m_1 + m_2) + d_\rho m_q^{3/2} + e_\rho m_q^{2} + \cdots
\,.}
Unlike for baryons, the detailed expressions for $b_\rho$, $c_\rho$ and
$d_\rho$ in terms of constants in the quenched chiral Lagrangian are not known.
Nevertheless, we expect that the $\delta$ term will be small,
and again ignore it in our fitting.

\bigskip

\table\tmesonwaL{
\let\iffitrange=\iffalse
  \def\myskip{\omit&height1.5pt&\omit&&\omit&&\omit&&\omit&&\omit&&\omit&&\omit
&&\omit&\cr}
  \def\tlr{\noalign{\hrule}}
  \vbox{\hbox{\vbox{\tabskip=0pt\offinterlineskip
  \halign{\strut#&\vrule\vrule#\tabskip=3pt&\hfil$#$\hfil&\vrule#\vrule%
           &\hfil$#$\hfil&\vrule#%
           &\hfil$#$\hfil&\vrule#%
           &\hfil$#$\hfil&\vrule#%
           &\hfil$#$\hfil&\vrule#\vrule%
           &\hfil$#$\hfil&\vrule#\vrule%
           &\hfil$#$\hfil&\vrule#\vrule%
           &\hfil$#$\hfil&\vrule#\vrule%
\tabskip=0pt\cr\tlr\tlr
%
\myskip
&&&&\pi&&A_4/A_4&&A_4/\pi&&\pi/A_4&&\rho&&a_0&&a_1&\cr\myskip\tlr\tlr
\myskip
&& CC&&1.217( 01)&&1.217( 00)&&1.217( 01)&&1.217( 01)&&1.229( 01)&&1.432( 06)&&1.439( 05)&\cr
\iffitrange\myskip&&&&\hbox{23 -- 31}&&\hbox{20 -- 28}&&\hbox{20 -- 29}&&\hbox{23 -- 29}&&\hbox{24 -- 30}&&\hbox{12 -- 16}&&\hbox{12 -- 15}&\cr\fi
\myskip\tlr
\myskip
&& CS&&0.853( 01)&&0.854( 01)&&0.853( 01)&&0.854( 01)&&0.878( 01)&&          &&1.104( 04)&\cr
\iffitrange\myskip&&&&\hbox{18 -- 28}&&\hbox{20 -- 28}&&\hbox{20 -- 29}&&\hbox{23 -- 29}&&\hbox{16 -- 22}&&\hbox{ 8 -- 10}&&\hbox{ 8 -- 10}&\cr\fi
\myskip\tlr
\myskip
&& CU_1&&0.813( 01)&&0.814( 01)&&0.813( 01)&&0.814( 01)&&0.841( 01)&&          &&1.081( 07)&\cr
\iffitrange\myskip&&&&\hbox{18 -- 28}&&\hbox{20 -- 28}&&\hbox{20 -- 29}&&\hbox{23 -- 29}&&\hbox{16 -- 22}&&\hbox{ 8 -- 10}&&\hbox{ 8 -- 10}&\cr\fi
\myskip\tlr
\myskip
&& CU_2&&0.797( 01)&&0.798( 01)&&0.798( 01)&&0.798( 01)&&0.826( 01)&&          &&1.080( 10)&\cr
\iffitrange\myskip&&&&\hbox{18 -- 28}&&\hbox{20 -- 28}&&\hbox{20 -- 29}&&\hbox{23 -- 29}&&\hbox{16 -- 22}&&\hbox{ 8 -- 10}&&\hbox{ 8 -- 10}&\cr\fi
\myskip\tlr
\myskip
&& CU_3&&0.787( 01)&&0.789( 02)&&0.789( 02)&&0.789( 02)&&0.816( 02)&&          &&1.087( 13)&\cr
\iffitrange\myskip&&&&\hbox{18 -- 28}&&\hbox{20 -- 28}&&\hbox{20 -- 29}&&\hbox{23 -- 29}&&\hbox{16 -- 22}&&\hbox{ 8 -- 10}&&\hbox{ 7 -- 10}&\cr\fi
\myskip\tlr
\tlr
\myskip
&& SS&&0.421( 00)&&0.421( 01)&&0.422( 01)&&0.421( 01)&&0.504( 01)&&0.743( 15)&&0.753( 05)&\cr
\iffitrange\myskip&&&&\hbox{15 -- 28}&&\hbox{15 -- 28}&&\hbox{15 -- 26}&&\hbox{15 -- 26}&&\hbox{15 -- 22}&&\hbox{ 8 -- 11}&&\hbox{ 6 -- 10}&\cr\fi
\myskip\tlr
\myskip
&& SU_1&&0.362( 00)&&0.363( 01)&&0.363( 01)&&0.362( 01)&&0.463( 01)&&0.720( 18)&&0.721( 07)&\cr
\iffitrange\myskip&&&&\hbox{15 -- 28}&&\hbox{15 -- 28}&&\hbox{15 -- 26}&&\hbox{15 -- 26}&&\hbox{15 -- 22}&&\hbox{ 7 -- 10}&&\hbox{ 6 -- 10}&\cr\fi
\myskip\tlr
\myskip
&& SU_2&&0.338( 01)&&0.338( 01)&&0.338( 01)&&0.337( 01)&&0.447( 02)&&0.743( 31)&&0.712( 09)&\cr
\iffitrange\myskip&&&&\hbox{15 -- 28}&&\hbox{15 -- 28}&&\hbox{15 -- 26}&&\hbox{15 -- 26}&&\hbox{15 -- 22}&&\hbox{ 7 -- 10}&&\hbox{ 6 -- 10}&\cr\fi
\myskip\tlr
\myskip
&& SU_3&&0.322( 01)&&0.322( 01)&&0.323( 01)&&0.321( 01)&&0.438( 02)&&          &&0.714( 12)&\cr
\iffitrange\myskip&&&&\hbox{15 -- 28}&&\hbox{15 -- 28}&&\hbox{15 -- 26}&&\hbox{15 -- 26}&&\hbox{15 -- 22}&&\hbox{ 7 -- 10}&&\hbox{ 6 -- 10}&\cr\fi
\myskip\tlr
\tlr
\myskip
&& U_1U_1&&0.296( 01)&&0.296( 01)&&0.296( 01)&&0.295( 01)&&0.422( 02)&&0.724( 53)&&0.685( 09)&\cr
\iffitrange\myskip&&&&\hbox{15 -- 28}&&\hbox{15 -- 28}&&\hbox{15 -- 26}&&\hbox{15 -- 26}&&\hbox{15 -- 21}&&\hbox{ 8 -- 10}&&\hbox{ 6 -- 10}&\cr\fi
\myskip\tlr
\myskip
&& U_1U_2&&0.266( 01)&&0.266( 01)&&0.267( 01)&&0.265( 01)&&0.404( 03)&&          &&0.674( 12)&\cr
\iffitrange\myskip&&&&\hbox{15 -- 28}&&\hbox{15 -- 28}&&\hbox{15 -- 26}&&\hbox{15 -- 26}&&\hbox{15 -- 21}&&\hbox{ 8 -- 10}&&\hbox{ 6 -- 10}&\cr\fi
\myskip\tlr
\myskip
&& U_1U_3&&0.246( 01)&&0.247( 01)&&0.247( 01)&&0.245( 01)&&0.393( 03)&&          &&0.672( 15)&\cr
\iffitrange\myskip&&&&\hbox{15 -- 28}&&\hbox{15 -- 28}&&\hbox{15 -- 26}&&\hbox{15 -- 26}&&\hbox{15 -- 21}&&\hbox{ 7 --  9}&&\hbox{ 6 -- 10}&\cr\fi
\myskip\tlr
\myskip
&& U_2U_2&&0.233( 01)&&0.233( 01)&&0.234( 01)&&0.232( 01)&&0.386( 03)&&          &&0.660( 14)&\cr
\iffitrange\myskip&&&&\hbox{15 -- 28}&&\hbox{15 -- 28}&&\hbox{15 -- 26}&&\hbox{15 -- 26}&&\hbox{14 -- 21}&&\hbox{ 7 --  9}&&\hbox{ 6 -- 10}&\cr\fi
\myskip\tlr
\myskip
&& U_2U_3&&0.210( 01)&&0.210( 01)&&0.211( 01)&&0.209( 01)&&0.371( 03)&&          &&0.656( 18)&\cr
\iffitrange\myskip&&&&\hbox{15 -- 28}&&\hbox{15 -- 28}&&\hbox{15 -- 26}&&\hbox{15 -- 26}&&\hbox{12 -- 16}&&\hbox{ 3 --  5}&&\hbox{ 6 -- 10}&\cr\fi
\myskip\tlr
\myskip
&& U_3U_3&&0.184( 01)&&0.184( 01)&&0.186( 01)&&0.183( 01)&&0.359( 04)&&          &&0.646( 22)&\cr
\iffitrange\myskip&&&&\hbox{15 -- 28}&&\hbox{15 -- 28}&&\hbox{15 -- 26}&&\hbox{15 -- 26}&&\hbox{12 -- 16}&&\hbox{ 2 --  4}&&\hbox{ 6 -- 10}&\cr\fi
\myskip\tlr
}}}}
 
 }
{Meson masses from $WL$ correlators at $\vec p = 0$.}

\table\tmesonwuL{
\let\iffitrange=\iffalse
  \def\myskip{\omit&height1.5pt&\omit&&\omit&&\omit&&\omit&&\omit&&\omit&&\omit
&&\omit&\cr}
  \def\tlr{\noalign{\hrule}}
  \vbox{\hbox{\vbox{\tabskip=0pt\offinterlineskip
  \halign{\strut#&\vrule\vrule#\tabskip=3pt&\hfil$#$\hfil&\vrule#\vrule%
           &\hfil$#$\hfil&\vrule#%
           &\hfil$#$\hfil&\vrule#%
           &\hfil$#$\hfil&\vrule#%
           &\hfil$#$\hfil&\vrule#\vrule%
           &\hfil$#$\hfil&\vrule#\vrule%
           &\hfil$#$\hfil&\vrule#\vrule%
           &\hfil$#$\hfil&\vrule#\vrule%
\tabskip=0pt\cr\tlr\tlr
%
\myskip
&&&&\pi&&A_4/A_4&&A_4/\pi&&\pi/A_4&&\rho&&a_0&&a_1&\cr\myskip\tlr\tlr
\myskip
&& CC&&1.217( 01)&&1.217( 01)&&1.217( 01)&&1.217( 01)&&1.230( 01)&&1.421( 11)&&1.448( 09)&\cr
\iffitrange\myskip&&&&\hbox{16 -- 30}&&\hbox{18 -- 30}&&\hbox{20 -- 30}&&\hbox{20 -- 30}&&\hbox{16 -- 26}&&\hbox{11 -- 14}&&\hbox{ 9 -- 13}&\cr\fi
\myskip\tlr
\myskip
&& CS&&0.854( 01)&&0.854( 01)&&0.855( 01)&&0.854( 01)&&0.881( 01)&&1.079( 17)&&1.126( 12)&\cr
\iffitrange\myskip&&&&\hbox{15 -- 30}&&\hbox{15 -- 28}&&\hbox{15 -- 28}&&\hbox{15 -- 28}&&\hbox{15 -- 26}&&\hbox{11 -- 14}&&\hbox{ 9 -- 13}&\cr\fi
\myskip\tlr
\myskip
&& CU_1&&0.815( 01)&&0.815( 01)&&0.815( 01)&&0.814( 01)&&0.844( 02)&&1.047( 22)&&1.097( 15)&\cr
\iffitrange\myskip&&&&\hbox{15 -- 30}&&\hbox{15 -- 28}&&\hbox{15 -- 28}&&\hbox{15 -- 28}&&\hbox{16 -- 26}&&\hbox{11 -- 14}&&\hbox{ 9 -- 13}&\cr\fi
\myskip\tlr
\myskip
&& CU_2&&0.800( 02)&&0.800( 02)&&0.800( 02)&&0.800( 02)&&0.829( 02)&&1.040( 28)&&1.089( 18)&\cr
\iffitrange\myskip&&&&\hbox{15 -- 30}&&\hbox{15 -- 28}&&\hbox{15 -- 28}&&\hbox{15 -- 28}&&\hbox{16 -- 24}&&\hbox{11 -- 14}&&\hbox{ 9 -- 13}&\cr\fi
\myskip\tlr
\myskip
&& CU_3&&0.791( 02)&&0.792( 03)&&0.791( 02)&&0.792( 02)&&0.821( 03)&&1.048( 38)&&1.090( 23)&\cr
\iffitrange\myskip&&&&\hbox{16 -- 28}&&\hbox{15 -- 28}&&\hbox{15 -- 28}&&\hbox{15 -- 28}&&\hbox{16 -- 24}&&\hbox{11 -- 14}&&\hbox{ 9 -- 13}&\cr\fi
\myskip\tlr
\tlr
\myskip
&& SS&&0.423( 01)&&0.422( 01)&&0.423( 01)&&0.422( 01)&&0.506( 02)&&0.716( 16)&&0.777( 13)&\cr
\iffitrange\myskip&&&&\hbox{18 -- 28}&&\hbox{17 -- 28}&&\hbox{17 -- 28}&&\hbox{15 -- 28}&&\hbox{16 -- 22}&&\hbox{ 8 -- 11}&&\hbox{ 8 -- 11}&\cr\fi
\myskip\tlr
\myskip
&& SU_1&&0.364( 01)&&0.363( 01)&&0.364( 01)&&0.364( 01)&&0.464( 03)&&0.684( 24)&&0.741( 16)&\cr
\iffitrange\myskip&&&&\hbox{18 -- 28}&&\hbox{17 -- 30}&&\hbox{17 -- 28}&&\hbox{15 -- 28}&&\hbox{16 -- 22}&&\hbox{ 8 -- 11}&&\hbox{ 8 -- 11}&\cr\fi
\myskip\tlr
\myskip
&& SU_2&&0.340( 01)&&0.339( 01)&&0.340( 01)&&0.340( 01)&&0.450( 03)&&0.697( 29)&&0.731( 19)&\cr
\iffitrange\myskip&&&&\hbox{18 -- 28}&&\hbox{17 -- 30}&&\hbox{17 -- 28}&&\hbox{15 -- 28}&&\hbox{14 -- 19}&&\hbox{ 7 -- 11}&&\hbox{ 8 -- 11}&\cr\fi
\myskip\tlr
\myskip
&& SU_3&&0.324( 01)&&0.323( 02)&&0.324( 01)&&0.324( 01)&&0.440( 03)&&0.733( 48)&&0.729( 22)&\cr
\iffitrange\myskip&&&&\hbox{18 -- 28}&&\hbox{17 -- 30}&&\hbox{17 -- 28}&&\hbox{15 -- 28}&&\hbox{14 -- 19}&&\hbox{ 7 -- 11}&&\hbox{ 8 -- 11}&\cr\fi
\myskip\tlr
\tlr
\myskip
&& U_1U_1&&0.297( 01)&&0.296( 01)&&0.297( 01)&&0.297( 01)&&0.423( 03)&&0.677( 37)&&0.705( 20)&\cr
\iffitrange\myskip&&&&\hbox{14 -- 30}&&\hbox{17 -- 28}&&\hbox{17 -- 28}&&\hbox{15 -- 28}&&\hbox{13 -- 18}&&\hbox{ 7 -- 11}&&\hbox{ 8 -- 11}&\cr\fi
\myskip\tlr
\myskip
&& U_1U_2&&0.268( 01)&&0.267( 02)&&0.268( 01)&&0.268( 01)&&0.406( 04)&&0.704( 42)&&0.695( 25)&\cr
\iffitrange\myskip&&&&\hbox{18 -- 28}&&\hbox{17 -- 28}&&\hbox{17 -- 28}&&\hbox{15 -- 28}&&\hbox{13 -- 18}&&\hbox{ 6 -- 10}&&\hbox{ 8 -- 11}&\cr\fi
\myskip\tlr
\myskip
&& U_1U_3&&0.248( 01)&&0.247( 02)&&0.248( 02)&&0.248( 01)&&0.396( 05)&&0.736( 39)&&0.695( 29)&\cr
\iffitrange\myskip&&&&\hbox{18 -- 28}&&\hbox{17 -- 28}&&\hbox{17 -- 28}&&\hbox{15 -- 28}&&\hbox{13 -- 18}&&\hbox{ 5 --  8}&&\hbox{ 8 -- 11}&\cr\fi
\myskip\tlr
\myskip
&& U_2U_2&&0.235( 01)&&0.233( 02)&&0.235( 02)&&0.235( 01)&&0.389( 05)&&          &&0.687( 30)&\cr
\iffitrange\myskip&&&&\hbox{18 -- 28}&&\hbox{17 -- 28}&&\hbox{17 -- 28}&&\hbox{15 -- 28}&&\hbox{13 -- 18}&&\hbox{ 5 --  8}&&\hbox{ 8 -- 11}&\cr\fi
\myskip\tlr
\myskip
&& U_2U_3&&0.212( 01)&&0.210( 02)&&0.212( 02)&&0.212( 02)&&0.377( 06)&&          &&0.688( 36)&\cr
\iffitrange\myskip&&&&\hbox{18 -- 28}&&\hbox{17 -- 28}&&\hbox{17 -- 28}&&\hbox{15 -- 28}&&\hbox{13 -- 18}&&\hbox{ 4 --  6}&&\hbox{ 8 -- 11}&\cr\fi
\myskip\tlr
\myskip
&& U_3U_3&&0.186( 01)&&0.184( 02)&&0.186( 02)&&0.185( 02)&&0.363( 08)&&          &&0.691( 45)&\cr
\iffitrange\myskip&&&&\hbox{18 -- 28}&&\hbox{17 -- 28}&&\hbox{17 -- 28}&&\hbox{15 -- 28}&&\hbox{13 -- 18}&&\hbox{ 3 --  6}&&\hbox{ 8 -- 11}&\cr\fi
\myskip\tlr
}}}}
 
 }
{Meson masses from $SL$ correlators at $\vec p = 0$.}

\table\tmesonwuS{
\let\iffitrange=\iffalse
  \def\myskip{\omit&height1.5pt&\omit&&\omit&&\omit&&\omit&&\omit&&\omit&&\omit&&\omit&\cr}
  \def\tlr{\noalign{\hrule}}
  \vbox{\hbox{\vbox{\tabskip=0pt\offinterlineskip
  \halign{\strut#&\vrule\vrule#\tabskip=3pt&\hfil$#$\hfil&\vrule#\vrule%
           &\hfil$#$\hfil&\vrule#%
           &\hfil$#$\hfil&\vrule#%
           &\hfil$#$\hfil&\vrule#%
           &\hfil$#$\hfil&\vrule#\vrule%
           &\hfil$#$\hfil&\vrule#\vrule%
           &\hfil$#$\hfil&\vrule#\vrule%
           &\hfil$#$\hfil&\vrule#\vrule%
\tabskip=0pt\cr\tlr\tlr
%
\myskip
&&&&\pi&&A_4/A_4&&A_4/\pi&&\pi/A_4&&\rho&&a_0&&a_1&\cr\myskip\tlr\tlr
\myskip
&& CC&&1.218( 01)&&1.218( 01)&&1.218( 01)&&1.217( 01)&&1.230( 01)&&1.421( 12)&&1.438( 11)&\cr
\iffitrange\myskip&&&&\hbox{15 -- 28}&&\hbox{15 -- 28}&&\hbox{15 -- 28}&&\hbox{15 -- 28}&&\hbox{14 -- 24}&&\hbox{10 -- 14}&&\hbox{ 9 -- 14}&\cr\fi
\myskip\tlr
\myskip
&& CS&&0.854( 01)&&0.855( 01)&&0.855( 01)&&0.854( 01)&&0.880( 01)&&1.071( 21)&&1.118( 15)&\cr
\iffitrange\myskip&&&&\hbox{15 -- 28}&&\hbox{15 -- 28}&&\hbox{15 -- 28}&&\hbox{15 -- 28}&&\hbox{14 -- 24}&&\hbox{11 -- 15}&&\hbox{ 9 -- 12}&\cr\fi
\myskip\tlr
\myskip
&& CU_1&&0.814( 01)&&0.815( 02)&&0.815( 01)&&0.815( 02)&&0.843( 02)&&1.043( 29)&&1.084( 19)&\cr
\iffitrange\myskip&&&&\hbox{15 -- 28}&&\hbox{15 -- 28}&&\hbox{15 -- 28}&&\hbox{15 -- 28}&&\hbox{14 -- 24}&&\hbox{11 -- 15}&&\hbox{ 9 -- 12}&\cr\fi
\myskip\tlr
\myskip
&& CU_2&&0.799( 02)&&0.800( 02)&&0.800( 02)&&0.800( 02)&&0.829( 02)&&1.040( 38)&&1.074( 23)&\cr
\iffitrange\myskip&&&&\hbox{15 -- 28}&&\hbox{15 -- 28}&&\hbox{15 -- 28}&&\hbox{15 -- 28}&&\hbox{15 -- 25}&&\hbox{11 -- 15}&&\hbox{ 9 -- 12}&\cr\fi
\myskip\tlr
\myskip
&& CU_3&&0.791( 02)&&0.792( 03)&&0.791( 03)&&0.792( 03)&&0.820( 03)&&1.045( 42)&&1.072( 28)&\cr
\iffitrange\myskip&&&&\hbox{16 -- 28}&&\hbox{15 -- 28}&&\hbox{15 -- 28}&&\hbox{15 -- 28}&&\hbox{15 -- 25}&&\hbox{10 -- 15}&&\hbox{ 9 -- 12}&\cr\fi
\myskip\tlr
\tlr
\myskip
&& SS&&0.422( 01)&&0.422( 01)&&0.422( 01)&&0.422( 01)&&0.507( 02)&&0.707( 18)&&0.770( 14)&\cr
\iffitrange\myskip&&&&\hbox{15 -- 28}&&\hbox{15 -- 28}&&\hbox{15 -- 28}&&\hbox{15 -- 28}&&\hbox{12 -- 21}&&\hbox{ 8 -- 11}&&\hbox{ 7 -- 12}&\cr\fi
\myskip\tlr
\myskip
&& SU_1&&0.364( 01)&&0.364( 01)&&0.364( 01)&&0.364( 01)&&0.465( 02)&&0.677( 27)&&0.738( 13)&\cr
\iffitrange\myskip&&&&\hbox{15 -- 28}&&\hbox{15 -- 28}&&\hbox{15 -- 28}&&\hbox{15 -- 28}&&\hbox{12 -- 21}&&\hbox{ 8 -- 11}&&\hbox{ 6 -- 11}&\cr\fi
\myskip\tlr
\myskip
&& SU_2&&0.340( 01)&&0.339( 01)&&0.340( 01)&&0.339( 01)&&0.449( 03)&&0.691( 31)&&0.725( 19)&\cr
\iffitrange\myskip&&&&\hbox{15 -- 28}&&\hbox{15 -- 28}&&\hbox{15 -- 28}&&\hbox{15 -- 28}&&\hbox{12 -- 21}&&\hbox{ 7 -- 10}&&\hbox{ 7 -- 11}&\cr\fi
\myskip\tlr
\myskip
&& SU_3&&0.324( 01)&&0.324( 02)&&0.324( 01)&&0.324( 01)&&0.440( 03)&&0.724( 29)&&0.726( 16)&\cr
\iffitrange\myskip&&&&\hbox{15 -- 28}&&\hbox{15 -- 28}&&\hbox{15 -- 28}&&\hbox{15 -- 28}&&\hbox{12 -- 21}&&\hbox{ 5 --  9}&&\hbox{ 6 -- 10}&\cr\fi
\myskip\tlr
\tlr
\myskip
&& U_1U_1&&0.297( 01)&&0.296( 01)&&0.297( 01)&&0.297( 01)&&0.422( 03)&&0.672( 39)&&0.702( 15)&\cr
\iffitrange\myskip&&&&\hbox{15 -- 28}&&\hbox{15 -- 28}&&\hbox{15 -- 28}&&\hbox{15 -- 28}&&\hbox{12 -- 21}&&\hbox{ 7 -- 10}&&\hbox{ 6 -- 10}&\cr\fi
\myskip\tlr
\myskip
&& U_1U_2&&0.268( 01)&&0.267( 01)&&0.268( 01)&&0.267( 01)&&0.405( 04)&&0.692( 30)&&0.691( 17)&\cr
\iffitrange\myskip&&&&\hbox{15 -- 28}&&\hbox{15 -- 28}&&\hbox{15 -- 28}&&\hbox{15 -- 28}&&\hbox{12 -- 21}&&\hbox{ 5 --  8}&&\hbox{ 6 -- 10}&\cr\fi
\myskip\tlr
\myskip
&& U_1U_3&&0.248( 01)&&0.247( 02)&&0.248( 01)&&0.248( 02)&&0.395( 05)&&0.726( 32)&&0.687( 20)&\cr
\iffitrange\myskip&&&&\hbox{15 -- 28}&&\hbox{15 -- 28}&&\hbox{15 -- 28}&&\hbox{15 -- 28}&&\hbox{12 -- 21}&&\hbox{ 4 --  7}&&\hbox{ 6 -- 10}&\cr\fi
\myskip\tlr
\myskip
&& U_2U_2&&0.235( 01)&&0.233( 02)&&0.235( 01)&&0.234( 01)&&0.387( 05)&&          &&0.679( 20)&\cr
\iffitrange\myskip&&&&\hbox{15 -- 28}&&\hbox{15 -- 28}&&\hbox{15 -- 28}&&\hbox{15 -- 28}&&\hbox{12 -- 21}&&\hbox{ 4 --  7}&&\hbox{ 6 -- 10}&\cr\fi
\myskip\tlr
\myskip
&& U_2U_3&&0.212( 01)&&0.210( 02)&&0.212( 02)&&0.211( 02)&&0.375( 07)&&          &&0.675( 24)&\cr
\iffitrange\myskip&&&&\hbox{15 -- 28}&&\hbox{15 -- 28}&&\hbox{15 -- 28}&&\hbox{15 -- 28}&&\hbox{12 -- 21}&&\hbox{ 3 --  6}&&\hbox{ 6 -- 10}&\cr\fi
\myskip\tlr
\myskip
&& U_3U_3&&0.186( 01)&&0.184( 02)&&0.186( 02)&&0.184( 02)&&0.361( 08)&&          &&0.671( 28)&\cr
\iffitrange\myskip&&&&\hbox{15 -- 28}&&\hbox{15 -- 28}&&\hbox{15 -- 28}&&\hbox{15 -- 28}&&\hbox{12 -- 20}&&\hbox{ 2 --  4}&&\hbox{ 6 -- 10}&\cr\fi
\myskip\tlr
}}}}
 
 }
{Meson masses from $SS$ correlators at $\vec p = 0$.}

\table\tmesonAV{
\let\iffitrange=\iffalse
  \def\myskip{\omit&height1.5pt&\omit&&\omit&&\omit&&\omit&&\omit&&\omit&&\omit
&&\omit&\cr}
  \def\tlr{\noalign{\hrule}}
  \vbox{\hbox{\vbox{\tabskip=0pt\offinterlineskip
  \halign{\strut#&\vrule\vrule#\tabskip=3pt&\hfil$#$\hfil&\vrule#\vrule%
           &\hfil$#$\hfil&\vrule#%
           &\hfil$#$\hfil&\vrule#%
           &\hfil$#$\hfil&\vrule#%
           &\hfil$#$\hfil&\vrule#\vrule%
           &\hfil$#$\hfil&\vrule#\vrule%
           &\hfil$#$\hfil&\vrule#\vrule%
           &\hfil$#$\hfil&\vrule#\vrule%
\tabskip=0pt\cr\tlr\tlr
%
%
\myskip
&&&&\pi&&A_4/A_4&&A_4/\pi&&\pi/A_4&&\rho&&a_0&&a_1&\cr\myskip\tlr\tlr
\myskip
&& CC&&1.217( 00)&&1.217( 01)&&1.217( 01)&&1.217( 01)&&1.229( 01)&&1.426( 07)&&1.441( 06)&\cr
\iffitrange\myskip&&&&\hbox{16 -- 30}&&\hbox{18 -- 30}&&\hbox{20 -- 30}&&\hbox{20 -- 30}&&\hbox{16 -- 26}&&\hbox{11 -- 14}&&\hbox{ 9 -- 13}&\cr\fi
\myskip\tlr
\myskip
&& CS&&0.854( 01)&&0.854( 01)&&0.854( 01)&&0.854( 01)&&0.879( 01)&&          &&1.113( 08)&\cr
\iffitrange\myskip&&&&\hbox{15 -- 30}&&\hbox{15 -- 28}&&\hbox{15 -- 28}&&\hbox{15 -- 28}&&\hbox{15 -- 26}&&\hbox{11 -- 14}&&\hbox{ 9 -- 13}&\cr\fi
\myskip\tlr
\myskip
&& CU_1&&0.814( 01)&&0.814( 01)&&0.814( 01)&&0.814( 01)&&0.842( 01)&&          &&1.086( 10)&\cr
\iffitrange\myskip&&&&\hbox{15 -- 30}&&\hbox{15 -- 28}&&\hbox{15 -- 28}&&\hbox{15 -- 28}&&\hbox{16 -- 26}&&\hbox{11 -- 14}&&\hbox{ 9 -- 13}&\cr\fi
\myskip\tlr
\myskip
&& CU_2&&0.798( 01)&&0.799( 01)&&0.799( 01)&&0.799( 01)&&0.828( 01)&&          &&1.081( 12)&\cr
\iffitrange\myskip&&&&\hbox{15 -- 30}&&\hbox{15 -- 28}&&\hbox{15 -- 28}&&\hbox{15 -- 28}&&\hbox{16 -- 24}&&\hbox{11 -- 14}&&\hbox{ 9 -- 13}&\cr\fi
\myskip\tlr
\myskip
&& CU_3&&0.789( 01)&&0.791( 02)&&0.790( 01)&&0.790( 01)&&0.819( 02)&&          &&1.084( 15)&\cr
\iffitrange\myskip&&&&\hbox{16 -- 28}&&\hbox{15 -- 28}&&\hbox{15 -- 28}&&\hbox{15 -- 28}&&\hbox{16 -- 24}&&\hbox{11 -- 14}&&\hbox{ 9 -- 13}&\cr\fi
\myskip\tlr
\tlr
\myskip
&& SS&&0.422( 01)&&0.422( 01)&&0.422( 01)&&0.422( 01)&&0.505( 01)&&0.727( 12)&&0.763( 08)&\cr
\iffitrange\myskip&&&&\hbox{18 -- 28}&&\hbox{17 -- 28}&&\hbox{17 -- 28}&&\hbox{15 -- 28}&&\hbox{16 -- 22}&&\hbox{ 8 -- 11}&&\hbox{ 8 -- 11}&\cr\fi
\myskip\tlr
\myskip
&& SU_1&&0.363( 01)&&0.363( 01)&&0.364( 01)&&0.363( 01)&&0.464( 01)&&0.700( 16)&&0.730( 08)&\cr
\iffitrange\myskip&&&&\hbox{18 -- 28}&&\hbox{17 -- 30}&&\hbox{17 -- 28}&&\hbox{15 -- 28}&&\hbox{16 -- 22}&&\hbox{ 8 -- 11}&&\hbox{ 8 -- 11}&\cr\fi
\myskip\tlr
\myskip
&& SU_2&&0.339( 01)&&0.339( 01)&&0.339( 01)&&0.338( 01)&&0.448( 02)&&0.718( 21)&&0.720( 11)&\cr
\iffitrange\myskip&&&&\hbox{18 -- 28}&&\hbox{17 -- 30}&&\hbox{17 -- 28}&&\hbox{15 -- 28}&&\hbox{14 -- 19}&&\hbox{ 7 -- 11}&&\hbox{ 8 -- 11}&\cr\fi
\myskip\tlr
\myskip
&& SU_3&&0.323( 01)&&0.323( 01)&&0.323( 01)&&0.323( 01)&&0.439( 02)&&          &&0.720( 12)&\cr
\iffitrange\myskip&&&&\hbox{18 -- 28}&&\hbox{17 -- 30}&&\hbox{17 -- 28}&&\hbox{15 -- 28}&&\hbox{14 -- 19}&&\hbox{ 7 -- 11}&&\hbox{ 8 -- 11}&\cr\fi
\myskip\tlr
\tlr
\myskip
&& U_1U_1&&0.296( 01)&&0.296( 01)&&0.297( 01)&&0.296( 01)&&0.422( 02)&&0.699( 31)&&0.694( 10)&\cr
\iffitrange\myskip&&&&\hbox{14 -- 30}&&\hbox{17 -- 28}&&\hbox{17 -- 28}&&\hbox{15 -- 28}&&\hbox{13 -- 18}&&\hbox{ 7 -- 11}&&\hbox{ 8 -- 11}&\cr\fi
\myskip\tlr
\myskip
&& U_1U_2&&0.267( 01)&&0.266( 01)&&0.267( 01)&&0.266( 01)&&0.405( 02)&&          &&0.683( 12)&\cr
\iffitrange\myskip&&&&\hbox{18 -- 28}&&\hbox{17 -- 28}&&\hbox{17 -- 28}&&\hbox{15 -- 28}&&\hbox{13 -- 18}&&\hbox{ 6 -- 10}&&\hbox{ 8 -- 11}&\cr\fi
\myskip\tlr
\myskip
&& U_1U_3&&0.247( 01)&&0.247( 01)&&0.248( 01)&&0.246( 01)&&0.394( 03)&&          &&0.682( 15)&\cr
\iffitrange\myskip&&&&\hbox{18 -- 28}&&\hbox{17 -- 28}&&\hbox{17 -- 28}&&\hbox{15 -- 28}&&\hbox{13 -- 18}&&\hbox{ 5 --  8}&&\hbox{ 8 -- 11}&\cr\fi
\myskip\tlr
\myskip
&& U_2U_2&&0.234( 01)&&0.233( 01)&&0.234( 01)&&0.233( 01)&&0.387( 03)&&          &&0.672( 14)&\cr
\iffitrange\myskip&&&&\hbox{18 -- 28}&&\hbox{17 -- 28}&&\hbox{17 -- 28}&&\hbox{15 -- 28}&&\hbox{13 -- 18}&&\hbox{ 5 --  8}&&\hbox{ 8 -- 11}&\cr\fi
\myskip\tlr
\myskip
&& U_2U_3&&0.211( 01)&&0.210( 01)&&0.211( 01)&&0.210( 01)&&0.373( 03)&&          &&0.669( 17)&\cr
\iffitrange\myskip&&&&\hbox{18 -- 28}&&\hbox{17 -- 28}&&\hbox{17 -- 28}&&\hbox{15 -- 28}&&\hbox{13 -- 18}&&\hbox{ 4 --  6}&&\hbox{ 8 -- 11}&\cr\fi
\myskip\tlr
\myskip
&& U_3U_3&&0.185( 01)&&0.184( 01)&&0.186( 01)&&0.184( 01)&&0.361( 05)&&          &&0.664( 20)&\cr
\iffitrange\myskip&&&&\hbox{18 -- 28}&&\hbox{17 -- 28}&&\hbox{17 -- 28}&&\hbox{15 -- 28}&&\hbox{13 -- 18}&&\hbox{ 3 --  6}&&\hbox{ 8 -- 11}&\cr\fi
\myskip\tlr
}}}}
 
 }
{\vtop{\advance\hsize by -2\parindent 
\noindent 
Best estimates of meson masses (given by the average $(2*WL+SL+SS)/4$ as 
explained in the text) at $\vec p = 0$.}}

\newsec{Meson masses}
\penalty10000
We give our results for meson masses
in Tables~\nameuse\tmesonwaL$-$ \nameuse\tmesonAV.
We quote separately the results for the three source-sink combinations
and the average $(2*WL+SL+SS)/4$.
There are four columns for the pion in each table because
we use Lorentz structures $\gamma_5=P$ and $\gamma_4\gamma_5=A_4$
for both source and sink. Thus $PP$ has a pseudoscalar interpolating
field at both source and sink, while $PA_4$ has an axial interpolating
field at the sink. The four possibilities yield consistent results
for masses, and we use the average of the four (inside our jack-knife loop)
to give our best estimate of $M_\pi$.

States at $\vec p = 0$ created by $\bar \psi \psi $ are labeled $a_0$
and those by $\bar \psi \gamma_i \gamma_5 \psi $ are called $a_1$. The
signal in these channels is not good.  We only present data for those
mass combinations where there is a ``plateau'' over at least three
time-slices.  Even in the best case the signal dies out by $t=14$, so
contamination from higher states is likely.  We do not consider the
data good enough to warrant further analysis.

The tables show in detail how all the masses are systematically
lower for the WL correlators than for the SL or SS correlators.
The effect is, however, less than $1 \sigma$ for the pion and rho. 
As discussed above, we take the
results of Table~\nameuse\tmesonAV\ as our best estimates.

\figure\fmnpvskap{\epsfysize=4in\epsfbox{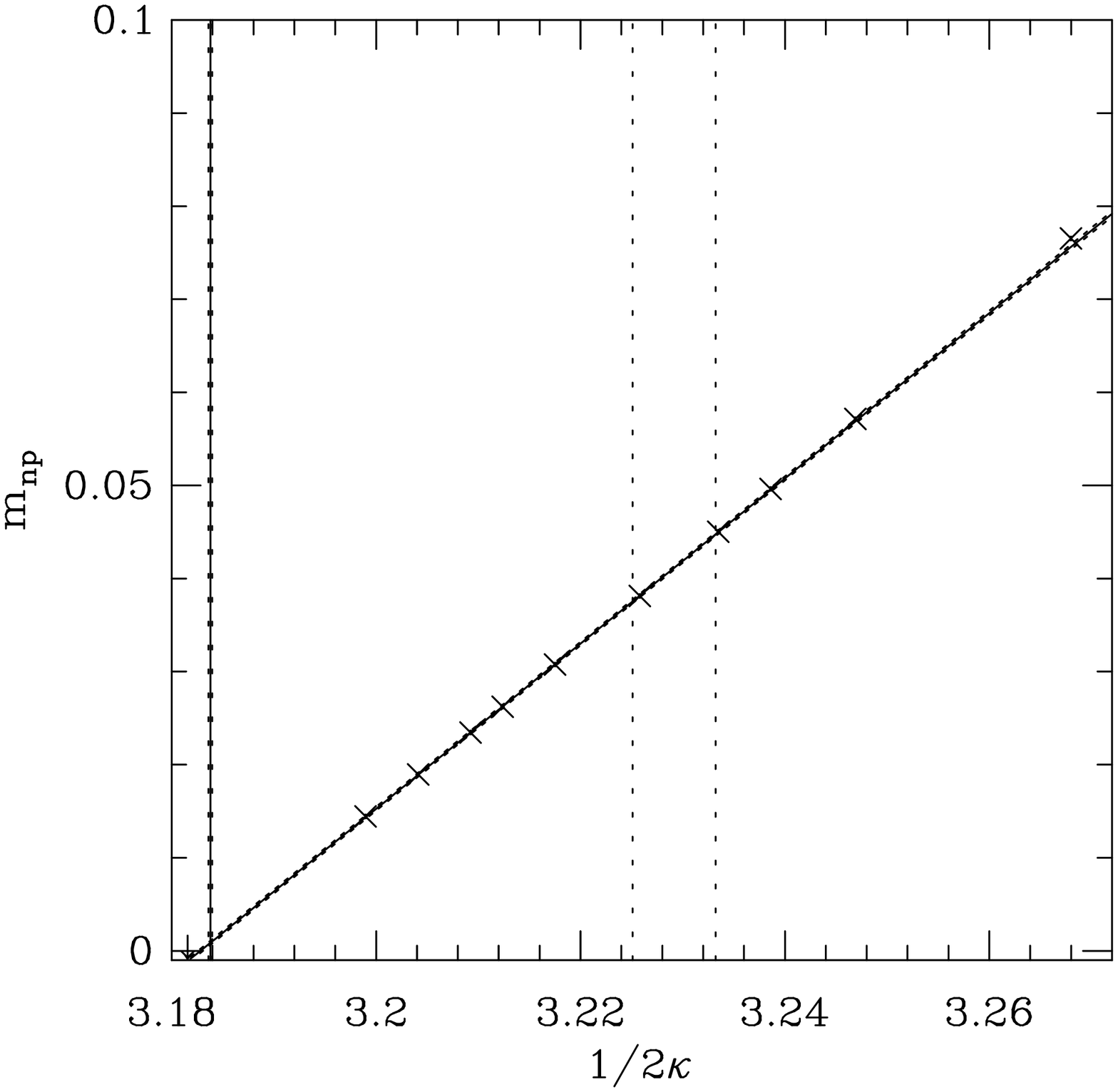}}
{\vtop{\advance\hsize by -2\parindent \noindent Plot of 
data for $m_{np}$ versus $1/2\kappa$.  The linear fit is to the 
six \lilj\ (lightest) points. The vertical lines show \mbar\ and the range 
of estimates of $m_s$.}}

To extrapolate the hadron masses towards the chiral limit,
and to test the forms predicted by quenched chiral perturbation theory,
we need to choose a definition of quark mass.
We consider two possibilities.
The first is the standard perturbative definition
\eqn\defmsbar{
m_q^\MSbar = Z_m (1/2\kappa - 1/2\kappa_c) \,, 
}
while the second is non-perturbative 
\ref\USpheno{R.~Gupta \etal, \spiresjournal{Phys.+Rev.}{D46}{3130}
{\PRD{46} (1992) 3130}.}
\eqn\defnmq{
        - { M_\pi } \
        { \vev{ 0 | A_4(t) J(0)| 0 }
          \over
          \vev{ 0 | P(t)   J(0)| 0 } }
         \ {\buildrel {\longrightarrow} \over {\scriptstyle t \to \infty } } \
        {Z_P \over Z_A} \ {(m_{1,np}+m_{2,np})  \over 2} \,.
}
Here $P$ and $A_4$ are local operators and $J$ (which has the same
Lorentz structure as either $P$ or $A_4$) is constructed from the
smeared Wuppertal source propagators.  The data for the two choices of $J$ 
are consistent, so we use the average as the best estimate. 
We use tadpole improved
renormalization constants, defined by
\ref\rlepagemackenzie{P.~Lepage, P.~Mackenzie, \spiresjournal{Phys.+Rev.}
{D48}{2250}{\PRD{48} (1993) 2250}}
\eqn\eZArenorm{\eqalign{
{Z_m \over 8 \kappa_c } \ &= \ 1 -  \alpha_{\MSbar}(q^*) 
                    \left( {2 \over \pi} \log(\mu a) - 0.01 \right)\,, \cr
\sqrt{Z_\psi^1 Z_\psi^2} Z_A \ &= \ 
                        \sqrt{1 - 3 \kappa_1 / 4 \kappa_c} 
		        \sqrt{1 - 3 \kappa_2 / 4 \kappa_c}  \ 
			\big( 1 -  0.316 \alpha_{\MSbar}(q^*) \big)\,, \cr
\sqrt{Z_\psi^1 Z_\psi^2} Z_P \ &= \ 
                        \sqrt{1 - 3 \kappa_1 / 4 \kappa_c} 
		        \sqrt{1 - 3 \kappa_2 / 4 \kappa_c}  \ 
			\big( 1 +  \alpha_{\MSbar}(q^*) ({2 \over \pi} 
                                   \log(\mu a) - 1.0335) \big) \,.  \cr
}}
$Z_\psi^{1,2}$ are wavefunction renormalizations for the two quarks
coupling to the bilinears, which have hopping parameters
$\kappa_{1,2}$ respectively. The $Z_\psi$'s cancel in the ratio
$Z_P/Z_A$.  $\mu$ is the scale at which we match to the continuum
renormalization scheme (here $\MSbar$), while $q^*$ is a typical
momentum in the 1-loop integral.  We choose $\mu=1/a$, and assume
$q^*=1/a$.  The masses can be run to other scales using the two-loop
anomalous dimension relation
\ref\mqAPE{C.~Allton, \etal, \spiresjournal{Nuc.+Phys.}{B431}{667}
{\NPB{431} (1994) 667}.} 
\eqn\mqrunning{
{m(Q) \over m(\mu)}  = 
        \bigg({g^2(Q) \over g^2(\mu)}\bigg)^{\gamma_0 \over 2 \beta_0} 
         \ \ \bigg( 1 + {g^2(Q) - g^2(\mu) \over 16\pi^2} \big( 
        {\gamma_1 \beta_0 - \gamma_0 \beta_1 \over 2 \beta_0^2} \big) \bigg)\ .
}
We mostly quote results for the scale $1/a$,
but also give some results for the scale 2 GeV, 
to allow comparison with other work.

It is important to note that,
although we do not label $m_{np}$ with $\MSbar$ explicitly,
both definitions yield estimates for 
the continuum mass in the $\MSbar$ scheme,
and should agree in the continuum limit.
We give our results for both definitions in Table~\nameuse\tmonetwo,
and find a substantial difference between them.
Most likely this is due to the failure of the perturbative expression 
for $Z_P$ \mqAPE\ \chiralrg. Details of this analysis will be presented 
elsewhere.

This difference is not, however, important for our study,
since the two definitions of mass turn out to proportional
to very high accuracy for the 6 cases \lilj.
This is shown in Fig.~\nameuse\fmnpvskap, in which we plot
the average $m_{np}$ of the quark and anti-quark against the
average value of $1/2\kappa$.
The fit is given by $ m_{np} = 0.867(4) (1/2\kappa) - 2.76(1) $.
Thus, when extrapolating, 
it does not matter which definition of mass one uses.
We choose to use $m_{np}$ for most purposes in this paper---and we will
drop the subscript henceforth. The perturbative mass, when it appears,
will be distinguished by the superscript $\MSbar$.

\figure\fpionvsmnp{\epsfysize=4in\epsfbox{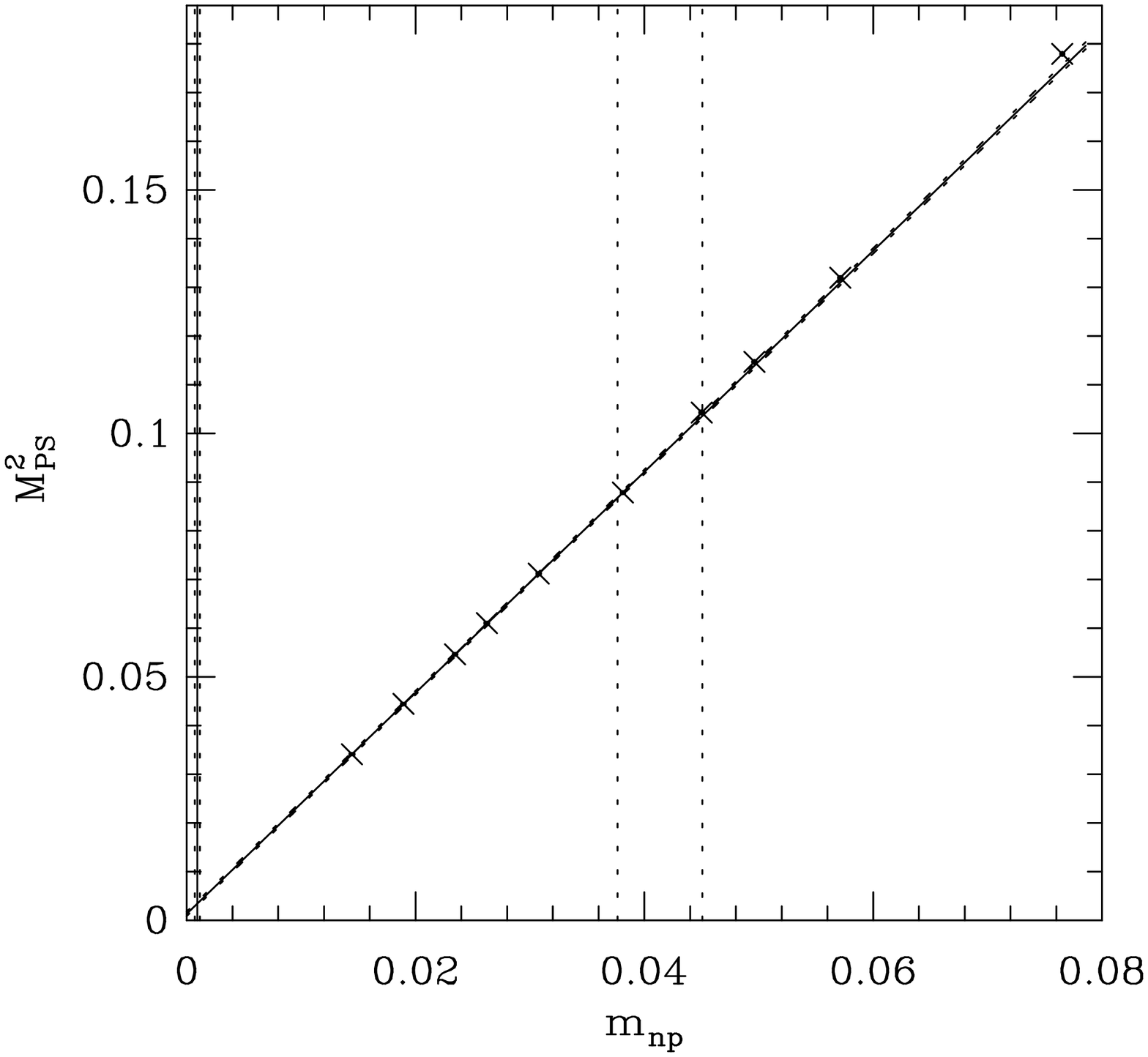}}{%
$M_\pi^2 $ versus $m_{np}$.  The linear fit is to the 
six \lilj\ points.}

\figure\frhovsmnp{\epsfysize=4in\epsfbox{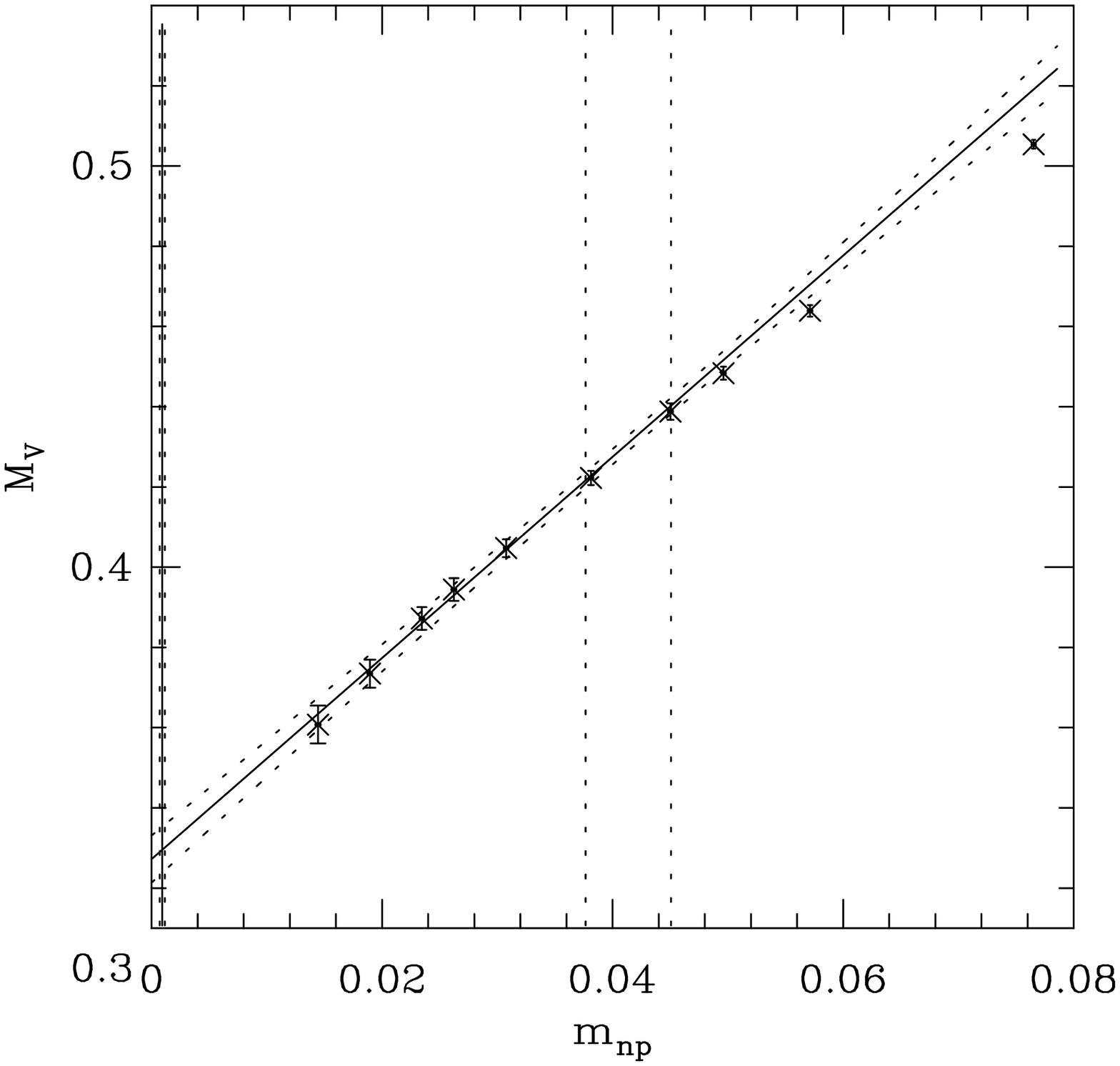}}{%
$M_\rho$ versus $m_{np}$.  The linear fit is to the 
six \lilj\ points.}

We now consider extrapolations to the chiral limit.
Figures~\nameuse\fpionvsmnp\ and \nameuse\frhovsmnp, show,
respectively, $M_{\pi}^2$ and $M_\rho$ as a function of the average
$m_{np}$.  Only data from the combinations \lilj\ (the six points with
the smallest masses) and \sslj\ (the rightmost four points) are
included. 

We first consider linear fits to the data.
These work well for the \lilj\ points, and are the fits shown in the figures.
The fits give
\eqn\epionmass{\eqalign{
M_\pi^2  &= a_\pi  + c_\pi  (m_1 + m_2) \qquad\ = \ 
0.0013(5) + 2.296(11) (m_1+m_2)/2 \,, \cr
M_{\rho} &= a_\rho + c_\rho (m_1 + m_2) \qquad\ = \ 
0.327(6) + 2.54(14)  (m_1+m_2)/2 \,. \cr
}}
One interesting feature is that the pion intercept is slightly
inconsistent with zero, whereas in the continuum limit of full QCD the
intercept should vanish. This discrepancy could be due to
discretization errors: for $M_\pi^2$ to be proportional to $m_{np}$ it
must be true that $\vev{ 0 | A_4 | P} \propto M_\pi$ (see Eq. \defnmq)
which is guaranteed only in the continuum limit. A violation of this
proportionality at $O(a)$ would lead to an
additional term in Eq. \epionmass\ proportional to $\sqrt{a m_{np}}$,
where $a$ is the lattice spacing. This could give rise to a non-zero
intercept.  Other possible culprits are quenched chiral
logarithms---the terms proportional to $\delta$ in Eq. \echptpi---and
finite size effects.  We expect the latter to be insignificant,
however, since $M_\pi L \ge 6$.  Our data is not good enough to
investigate this effect further. As a result of the non-zero
intercept, the chiral limit (defined by $M_\pi=0$) is at
$m_{np}^{chiral}=-0.00058(21)$.
In the following, when quoting quark masses in physical units
we use $m_{np}-m_{np}^{chiral}$, i.e. we offset the zero of the mass scale.
This offset is only significant for $\mbar$.
When we use masses in lattice units, as we do in all our fits and
extrapolations, we do not include this offset.

Using our linear fits we determine $\mbar_{np}$ by extrapolating to
the point where $M_\pi^2/M_\rho^2 = (137/768)^2$, finding
$\mbar_{np}=0.00093(21)$.  The resulting value of $M_\rho a$ gives us
an estimate of the scale
\eqn\scalea{
a^{-1}(M_\rho) = 2.330(41) {\GeV} \,.
}
Using this value, and including the offset in the mass scale,
we find $\mbar(1/a)= 3.51(6) \MeV$ ($\mbar(2 \GeV) = 3.60(5)$).
To facilitate comparison with earlier calculations it is useful to
give results for the critical and light quark hopping parameters.
Using linear fits versus $1/\kappa$ we find
\eqn\kappac{
\kappa_c = 0.157131(9)\,, \qquad
\kappa_l = 0.157046(9)\,.
}
We note that linear fits work well for the subsample {\sslj} (and for
\clj, not shown in the figures) as well, and we use them below when 
calculating the masses of strange and charm mesons.

The figures show that linear fits are inadequate 
for the combined \lilj\ and \sslj\ data.
This is particularly striking for $M_\rho$, where there is a definite
negative curvature, which, if ignored,
could lead to an overestimate of $a_\rho$.
Thus we have tried fits to the combined data involving higher order terms 
in the chiral expansion. 
A quadratic fit to $M_\pi^2$ works well,
but has little impact on extrapolations,
increasing $a_\pi $ by about $1\sigma$. 
For the $\rho$, motivated by chiral perturbation theory, 
we have used two forms of higher order terms
\eqn\erhochiral{\eqalign{
M_{\rho}&= a_\rho + c_\rho (m_1 + m_2) + d_\rho (m_1 + m_2)^{3/2} \,, \cr
M_{\rho}&= a_\rho + c_\rho (m_1 + m_2) + e_\rho (m_1 + m_2)^{2} \,. \cr
}}
These fits, when extrapolated to {\mbar} (calculated self-consistently), yield
smaller values of $M_\rho(\mbar)$ and thus larger scales than the linear fit:
\eqn\scaleaH{\eqalign{
a^{-1}(M_\rho) &= 2.365(48) {\GeV} \qquad (m^{3/2} \hbox{ fit}) \,, \cr 
a^{-1}(M_\rho) &= 2.344(42) {\GeV} \ \ \qquad (m^2     \hbox{ fit}) \,, \cr 
}}
The difference from the result of the linear fit is, however, within the
statistical errors. Our data is good enough to see the curvature, but not
precise enough to study it in detail. In particular, we cannot distinguish
between the two forms in Eq. \erhochiral, both doing a good job of fitting
the combined \lilj\ and \sslj\ data sets.
In view of this, we chose to use the scale from the linear
fit for most of our subsequent analysis.
For comparison, we note that the scale we find from 
$f_\pi$ is $2265(57)\MeV$ 
\ref\DCfinal{T. Bhattacharya and R.~Gupta, \eprint{hep-lat}{9510044}.}, 
while the NRQCD collaboration reports a value $2.4(1) \GeV$ from a mean of
charmonium and $\Upsilon$ spectroscopy
\ref\rNRQCDscale{C.~Davies, \etal, \spiresjournal{Phys.+Rev.}{D50}{6963}
{\PRD{50} (1994) 6963}.}.  
Thus, different estimates based on meson correlators are consistent.
In the remainder of the paper we shall take the scale from $M_\rho$
and assign a $3\%$ systematic uncertainty to cover the spread in $a^{-1}$
obtained from different mesonic observables and different types of
extrapolations.

Using the linear fits defined in Eq.~\epionmass\ we determine the
lattice strange quark mass by first extrapolating $M_K^2/M_\pi^2$,
$M_{K^\ast} / M_\rho$, or $M_{\phi} / M_\rho$ to $\mbar$, and then
linearly interpolating between $U_1$ and $S$ until these quantities
match their physical values.  (We use $M_K=495\ \MeV$, $M_{K^\ast}=894 \MeV$, 
and $M_\phi=1019$ MeV.) 
The results are given in Table~\nameuse\tmstab\ in terms of $\kappa$ and
the two definitions of quark mass discussed above,
except that we have here run the masses to a scale of $2 \GeV$ to
facilitate comparison with other work.

\table\tmstab{
\let\ifspace=\iffalse
\def\myskip{\omit&height1.5pt&%
\omit&&%
\omit&&%
\omit&&%
 &\cr}
\vbox{\hbox{\vbox{
\tabskip=0pt\offinterlineskip
\def\tlr{\noalign{\hrule}}
\halign {\strut#& \vrule\vrule#\tabskip=3pt&
  \hfil$#$\hfil&\vrule#&
  \hfil$#$\hfil&\vrule#&
  \hfil$#$\hfil&\vrule#&
  \hfil$#$\hfil&\vrule#&
  \hfil$#$\hfil&\vrule#\tabskip=0pt\cr\tlr
\omit&height1.5pt&\multispan{9   }&\cr
\myskip
&& {\rm Qty.}
&& \kappa_s
&& m_{s,np}(1/a)
&& m_{s,np}(2 \GeV)
&& m_s^\MSbar(2 \GeV)
  &\cr
\myskip\tlr
\omit&height0.5pt&\multispan{9   }&\cr\tlr
\myskip
&& M_K^2 / M_\rho^2
&& 0.15503(7)
&& 87(2)
&& 89(2)
&& 129(2)
  &\cr\ifspace\myskip&&
&& 
&&
&& 
&& 
  &\cr\fi\myskip\tlr
\myskip
&& M_{K^*} / M_\rho
&& 0.15479(19)
&& 98(7)
&& 100(7)
&& 145(9)
  &\cr\ifspace\myskip&&
&& 
&&
&& 
&& 
  &\cr\fi\myskip\tlr
\myskip
&& M_\phi / M_\rho                     
&& 0.15464(17)
&& 104(6)
&& 106(6)
&& 154(8)
  &\cr\ifspace\myskip&&
&& 
&&
&& 
&& 
  &\cr\fi\myskip\tlr
\cr}}}}

}
{\vtop{\advance\hsize by -2\parindent 
\noindent
Estimates of the strange quark mass obtained by matching
different quantities with their physical values.
Results are given for (1) $\kappa_s$
(2) the lattice non-perturbative mass $m_{np}/a$ evaluated at $1/a$
and $2 \GeV$,
and (3) the lattice perturbative mass $m_s^\MSbar$ 
evaluated at $2 \GeV$.
The latter three masses are in MeV.
}}

The three ratios lead to significantly different results for $m_s$,
presumably because of a combination of quenching and discretization errors.
Using $M_K^2/M_\pi^2$ to fix $m_s$ implies $m_s \equiv 25 \mbar$,
since we use the lowest order chiral expansion to fit the data.  
On the other hand, the estimates using $M_{\phi} / M_\rho$ and
$M_{K^*}/M_\rho$ are not constrained by the chiral expansion,
and give $m_s / \mbar \approx 30$, 
in surprisingly good agreement with the next-to-leading chiral result
\ref\Donoghue{ J. Donoghue, B. Holstein, D. Wyler, 
\spiresjournal{Phys.+Rev.+Lett.}{69}{3444}{\PRL{69} (1992) 3444}.}. 
In this paper we quote all results using $m_{\rm s}(M_{\phi})$.

Recently Lacock and Michael 
\ref\rJmichael{P.~Lacock and C.~Michael, \eprint{hep-lat}{9506009}.} 
suggested using the dimensionless quantity 
$J_V = M_{V} \partial M_V / \partial M_{\pi}^2$ to test the quenched 
approximation. Using a linear fit we find 
\eqn\eJparam{\eqalign{
J_{K^\ast} \ &= \ 0.41(1) , \cr
J_\rho  \ &= \ 0.36(1) , \cr
}}
to be compared to the experimental values of $\approx 0.48$ and $\approx 0.41$ 
respectively. Again, the discrepancy is presumably due to a combination
of discretization and quenching errors.

\newsec{$O(ma)$ discretization errors}
\seclab\secdisp

At $\beta=6$, the charm mass is such that $m_c a \sim 1$,
so there are potentially large $O(m_c a)$ discretization errors
in all quantities involving charm quarks.
These are in addition to errors of $O(\Lambda_{\rm QCD} a)$, which are
common to all quantities.
One effect of $O(a)$ errors is that there is a difference
between the ``static'' mass $M_1=E(\vec p=0)$,
and the ``kinetic'' mass 
$M_2 \equiv (\partial^2 E / \partial p^2 |_{p=0})^{-1}$.
Here the energy is determined from the rate of 
exponential decay of the correlator, $C(t) \propto \exp(-E(\vec p) t)$.
$M_1$ and $M_2$ agree in the continuum limit, whatever the mass of the state.

\table\tmesonMOM{
\let\ifaverage=\iftrue
  \def\myskip{\omit&height1.5pt&\omit&&\omit&&\omit&&\omit&&\omit&&\omit&&\omit
&&\omit&&\omit&\cr}
  \def\tlr{\noalign{\hrule}}
  \vbox{\hbox{\vbox{\tabskip=0pt\offinterlineskip
  \halign{\strut#&\vrule\vrule#\tabskip=1.5pt&\hfil$#$\hfil&\vrule#\vrule%
           &\hfil$#$\hfil&\vrule#%
           &\hfil$#$\hfil&\vrule#%
           &\hfil$#$\hfil&\vrule#%
           &\hfil$#$\hfil&\vrule#\vrule%
           &\hfil$#$\hfil&\vrule#%
           &\hfil$#$\hfil&\vrule#%
           &\hfil$#$\hfil&\vrule#%
           &\hfil$#$\hfil&\vrule#%
\vrule\tabskip=0pt\cr\tlr\tlr
%
\myskip
&&&&p^2=  1&&p^2=  2&&p^2=  3&&p^2=  4&&p^2=  1&&p^2=  2&&p^2=  3&&p^2=  4
&\cr\myskip\tlr\tlr
\myskip
&&CC&%
\ifaverage
&1.231( 01)&&1.244( 01)&&1.255( 02)&&1.268( 01)&%
\else
&1.231( 01)&&1.244( 01)&&1.255( 02)&&1.268( 02)&%
\fi
&1.243( 01)&&1.255( 01)&&1.267( 02)&&1.278( 02)&%
\cr
\myskip\tlr
\myskip
&&CS&%
\ifaverage
&0.874( 01)&&0.893( 02)&&0.913( 03)&&0.930( 02)&%
\else
&0.875( 01)&&0.894( 02)&&0.912( 03)&&0.929( 02)&%
\fi
&0.899( 01)&&0.917( 02)&&0.937( 03)&&0.951( 03)&%
\cr
\myskip\tlr
\myskip
&&CU_1&%
\ifaverage
&0.836( 02)&&0.855( 02)&&0.876( 03)&&0.895( 02)&%
\else
&0.836( 02)&&0.856( 02)&&0.875( 03)&&0.892( 03)&%
\fi
&0.863( 02)&&0.881( 02)&&0.901( 03)&&0.917( 04)&%
\cr
\myskip\tlr
\myskip
&&CU_2&%
\ifaverage
&0.821( 02)&&0.841( 02)&&0.861( 04)&&0.881( 03)&%
\else
&0.822( 02)&&0.842( 03)&&0.860( 04)&&0.879( 03)&%
\fi
&0.849( 02)&&0.868( 03)&&0.888( 04)&&0.905( 04)&%
\cr
\myskip\tlr
\myskip
&&CU_3&%
\ifaverage
&0.814( 03)&&0.833( 03)&&0.853( 04)&&0.873( 03)&%
\else
&0.813( 03)&&0.834( 03)&&0.852( 04)&&0.872( 04)&%
\fi
&0.841( 03)&&0.860( 04)&&0.882( 04)&&0.898( 05)&%
\cr
\myskip\tlr
\tlr
\myskip
&&SS&%
\ifaverage
&0.466( 01)&&0.504( 02)&&0.547( 04)&&0.578( 04)&%
\else
&0.466( 01)&&0.504( 02)&&0.547( 06)&&0.576( 04)&%
\fi
&0.541( 02)&&0.575( 03)&&0.614( 05)&&0.642( 04)&%
\cr
\myskip\tlr
\myskip
&&SU_1&%
\ifaverage
&0.414( 02)&&0.456( 03)&&0.505( 06)&&0.539( 04)&%
\else
&0.414( 01)&&0.456( 03)&&0.504( 07)&&0.538( 05)&%
\fi
&0.503( 03)&&0.540( 03)&&0.583( 06)&&0.613( 05)&%
\cr
\myskip\tlr
\myskip
&&SU_2&%
\ifaverage
&0.393( 02)&&0.438( 03)&&0.487( 07)&&0.522( 05)&%
\else
&0.393( 02)&&0.437( 03)&&0.486( 08)&&0.521( 06)&%
\fi
&0.489( 03)&&0.528( 04)&&0.571( 07)&&0.602( 06)&%
\cr
\myskip\tlr
\myskip
&&SU_3&%
\ifaverage
&0.380( 02)&&0.427( 03)&&0.474( 08)&&0.511( 06)&%
\else
&0.380( 02)&&0.426( 04)&&0.474( 09)&&0.509( 07)&%
\fi
&0.482( 04)&&0.522( 05)&&0.565( 09)&&0.596( 07)&%
\cr
\myskip\tlr
\tlr
\myskip
&&U_1U_1&%
\ifaverage
&0.357( 02)&&0.406( 03)&&0.460( 08)&&0.497( 06)&%
\else
&0.357( 02)&&0.406( 04)&&0.460( 10)&&0.495( 08)&%
\fi
&0.465( 04)&&0.508( 05)&&0.553( 08)&&0.583( 06)&%
\cr
\myskip\tlr
\myskip
&&U_1U_2&%
\ifaverage
&0.333( 02)&&0.386( 03)&&0.438( 09)&&0.480( 07)&%
\else
&0.333( 02)&&0.386( 04)&&0.439( 12)&&0.477( 10)&%
\fi
&0.452( 05)&&0.496( 06)&&0.541( 10)&&0.572( 07)&%
\cr
\myskip\tlr
\myskip
&&U_1U_3&%
\ifaverage
&0.317( 02)&&0.372( 04)&&0.423( 11)&&0.469( 08)&%
\else
&0.318( 02)&&0.372( 05)&&0.424( 15)&&0.465( 12)&%
\fi
&0.445( 06)&&0.490( 07)&&0.534( 12)&&0.566( 08)&%
\cr
\myskip\tlr
\myskip
&&U_2U_3&%
\ifaverage
&0.307( 02)&&0.363( 04)&&0.416( 11)&&0.464( 09)&%
\else
&0.308( 02)&&0.363( 05)&&0.418( 15)&&0.459( 13)&%
\fi
&0.439( 06)&&0.484( 08)&&0.529( 12)&&0.561( 09)&%
\cr
\myskip\tlr
\myskip
&&U_2U_3&%
\ifaverage
&0.290( 03)&&0.349( 05)&&0.399( 13)&&0.454( 11)&%
\else
&0.291( 03)&&0.348( 06)&&0.400( 18)&&0.448( 16)&%
\fi
&0.432( 08)&&0.478( 10)&&0.520( 15)&&0.554( 10)&%
\cr
\myskip\tlr
\myskip
&&U_3U_3&%
\ifaverage
&0.272( 03)&&0.333( 05)&&0.379( 16)&&0.445( 13)&%
\else
&0.274( 04)&&0.331( 07)&&0.378( 23)&&0.438( 21)&%
\fi
&0.424( 10)&&0.471( 13)&&0.509( 19)&&0.546( 13)&%
\cr
\myskip\tlr
}}}}
 
 }
{\vtop{\advance\hsize by -2\parindent \noindent 
Pseudoscalar (first set) and vector (second set) meson energies 
as a function of momentum. The data are 
the average of $SL$ and $SS$ estimates.}}

To evaluate $M_2$, we have tested four forms of dispersion relation
\eqn\elatdispersion{\eqalign{
(A)\quad 
E^2 = p^2 + M^2        \aftergroup{\hfill}      &{}\Rightarrow{}
\quad  M_2 =  M \,; \cr
(B)\quad 
\sinh^2E = \sin^2p + \sinh^2M \aftergroup{\hfill} &{}\Rightarrow{} 
\quad M_2 = {1\over 2}  \sinh2M \,; \cr
(C)\quad 
\sinh^2{E\over2} = \sin^2{p\over 2} + \sinh^2{M\over 2} &{}\Rightarrow{} 
\quad M_2 =  \sinh M \,;\cr
(D)\quad 
\sinh^2{E\over4} = \sin^2{p\over 4} + \sinh^2{M\over 4} &{}\Rightarrow{} 
\quad M_2 =  2 \sinh {M \over 2} \,. \cr
}}
(A) is the continuum relation,
while (B), (C) and (D) are lattice forms following from different
choices of lattice action.
In particular, (C) follows from the nearest neighbor symmetric difference
discretization of the action for a scalar.
Our results for $E(\vec p)$ are collected in Table~\nameuse\tmesonMOM, 
and in Fig.~\nameuse\fdispersion\ we show how the various dispersion
relations fare for the $CU_1$ meson.
For all heavy-heavy and heavy-light mesons, our results turn out to be
consistent with dispersion relation (C), but not with the other forms.

\figure\fdispersion{\epsfysize=4in\epsfbox{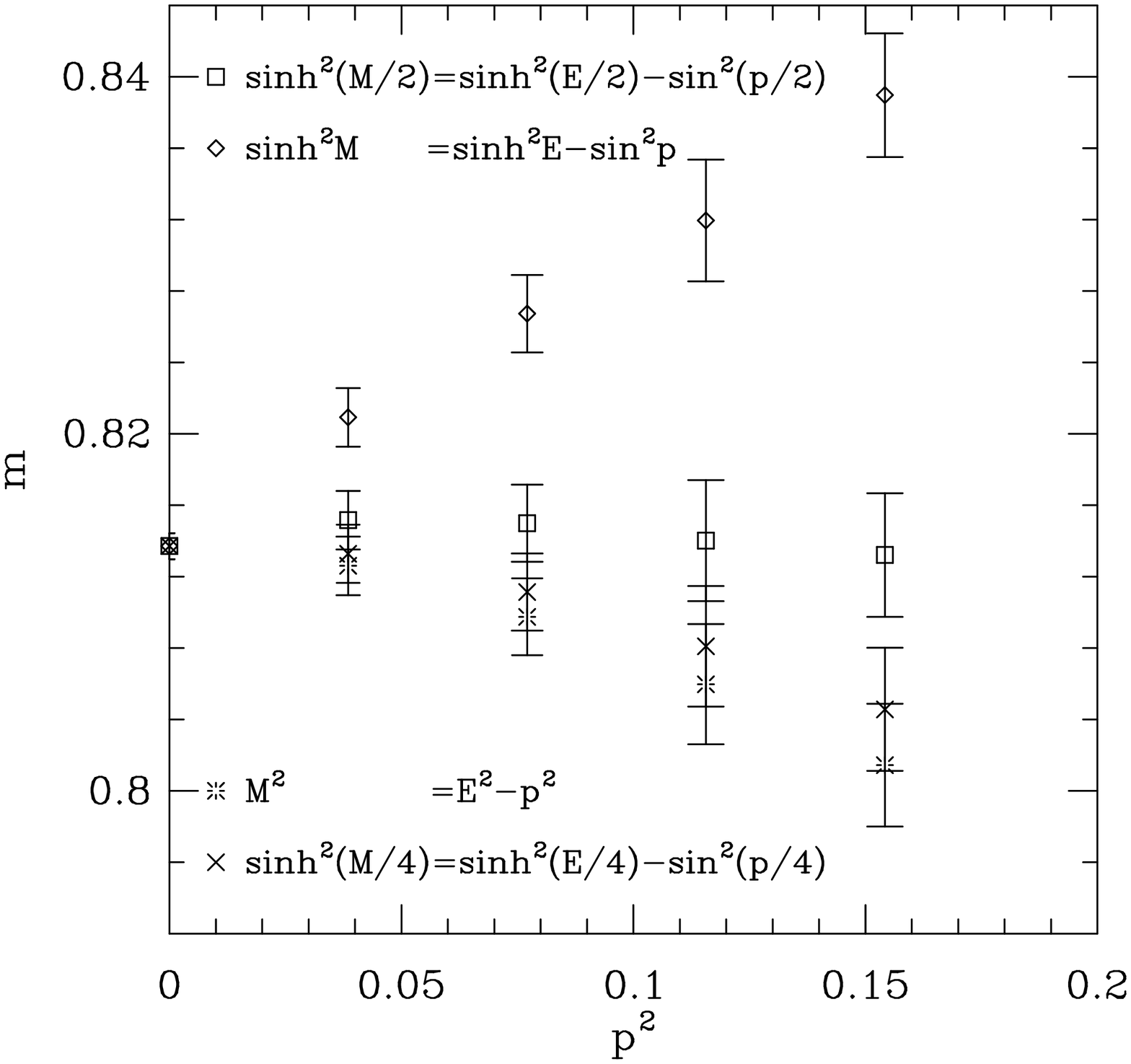}}
{\vtop{\advance\hsize by -2\parindent 
\noindent 
Test of four 
lattice dispersion relations (see Eq.~\elatdispersion) using the lattice 
pion data at various momenta for the case $CU_1$.  The data favor 
the nearest-neighbor symmetric-difference relativistic dispersion relation
$\sinh^2(E/2) - \sin^2(p/2) = \sinh^2(M/2)$, as shown by the square symbols.}}

\table\tmonetwo{
\def\dhr{\noalign{\hrule}}
\vbox{\hbox{\indent\vbox{\tabskip=0pt\offinterlineskip
\halign {\strut#& \vrule#\tabskip=2pt&
\hfil#\hfil&\vrule#&
#\hfil&\vrule#&
#\hfil&\vrule#&
#\hfil&\vrule#&
#\hfil&\vrule#&
#\hfil&\vrule#&
#\hfil&\vrule#\tabskip=0pt\cr\noalign{\hrule}
%
%
\omit&height2pt& && \multispan{3} && \multispan{3} && && &\cr
&& \hbox{}
&& \multispan{3} {\hfil $M_\pi$  \hfil}
&& \multispan{3} {\hfil $M_\rho$ \hfil}
&& \hfil $m_{np}$
&& \hfil $m_q$
& \cr\dhr
\omit&height2pt& && && && && && && &\cr
&& \hbox{State}
&& \hfil $M_1$
&& \hfil $M_2$
&& \hfil $M_1$
&& \hfil $M_2$
&&
&&
& \cr
\omit&height2pt& && && && && && && &\cr\dhr
%
%
&&$CC    $&&$1.217(0)$&&$1.541(1)$&&$1.229(1)$&&$1.564(2)$&&$0.4569(2)$&&$0.6563(2)$&\cr
&&$CS    $&&$0.854(1)$&&$0.962(1)$&&$0.879(1)$&&$0.999(2)$&&$0.2625(2)$&&$0.3822(2)$&\cr
&&$CU_1  $&&$0.814(1)$&&$0.908(2)$&&$0.842(1)$&&$0.947(2)$&&$0.2419(3)$&&$0.3557(2)$&\cr
&&$CU_2  $&&$0.799(1)$&&$0.888(2)$&&$0.828(1)$&&$0.927(3)$&&$0.2339(3)$&&$0.3453(2)$&\cr
&&$CU_3  $&&$0.790(1)$&&$0.877(3)$&&$0.819(2)$&&$0.916(4)$&&$0.2291(4)$&&$0.3388(2)$&\cr
&&$SS    $&&$0.422(1)$&&$0.435(1)$&&$0.505(1)$&&$0.528(2)$&&$0.0756(2)$&&$0.1081(2)$&\cr
&&$SU_1  $&&$0.363(1)$&&$0.372(1)$&&$0.464(1)$&&$0.481(3)$&&$0.0565(2)$&&$0.0816(2)$&\cr
&&$SU_2  $&&$0.339(1)$&&$0.346(1)$&&$0.448(2)$&&$0.465(3)$&&$0.0490(2)$&&$0.0711(2)$&\cr
&&$SU_3  $&&$0.323(1)$&&$0.330(1)$&&$0.439(2)$&&$0.454(4)$&&$0.0445(2)$&&$0.0647(2)$&\cr
&&$U_1U_1$&&$0.296(1)$&&$0.301(1)$&&$0.422(2)$&&$0.436(3)$&&$0.0377(2)$&&$0.0550(2)$&\cr
&&$U_1U_2$&&$0.267(1)$&&$0.271(1)$&&$0.405(2)$&&$0.417(4)$&&$0.0304(2)$&&$0.0446(2)$&\cr
&&$U_1U_3$&&$0.247(1)$&&$0.250(1)$&&$0.394(3)$&&$0.406(5)$&&$0.0259(2)$&&$0.0382(2)$&\cr
&&$U_2U_2$&&$0.234(1)$&&$0.236(1)$&&$0.387(3)$&&$0.398(5)$&&$0.0232(2)$&&$0.0342(2)$&\cr
&&$U_2U_3$&&$0.211(1)$&&$0.213(1)$&&$0.373(3)$&&$0.385(7)$&&$0.0187(2)$&&$0.0277(2)$&\cr
&&$U_3U_3$&&$0.185(1)$&&$0.186(1)$&&$0.361(5)$&&$0.370(9)$&&$0.0143(2)$&&$0.0213(2)$&\cr
%
\dhr
%
%
\crcr}}}}
 
 }
{\vtop{\advance\hsize by -2\parindent 
\noindent 
Comparison of $M_1=E(\vec p=0)$ and $M_2=\sinh M_1$.  We also give the
values of the average non-perturbative $m_{np}$ and perturbative
definitions of the quark mass in the $\MSbar$ scheme as defined in
Eqs.~\nameuse\defnmq\ and \nameuse\eZArenorm. Both are in lattice units,
evaluated at the scale $1/a$.}}

\table\tDmass{
\let\ifspace=\iffalse
\def\myskip{\omit&height1.5pt&%
\omit&&%
\omit&&%
\omit&&%
 &\cr}
\vbox{\hbox{\vbox{
\tabskip=0pt\offinterlineskip
\def\tlr{\noalign{\hrule}}
\halign {\strut#& \vrule\vrule#\tabskip=3pt&
  \hfil$#$\hfil&\vrule#&
  \hfil$#$\hfil&\vrule#&
  \hfil$#$\hfil&\vrule#&
  \hfil$#$\hfil&\vrule#\tabskip=0pt\cr\tlr
\omit&height1.5pt&\multispan{7   }&\cr
&&\multispan{7   }\hfil
{\bf D meson masses in MeV} \hfil&\cr
\omit&height1.5pt&\multispan{7   }&\cr\tlr
\omit&height1.5pt&\multispan{7   }&\cr\tlr
\myskip
&& 
&& M_1
&& M_2
&& Expt.
  &\cr
\myskip\tlr
\omit&height0.5pt&\multispan{7   }&\cr\tlr
\myskip
&& M_D
&& 1805(31)
&& 1990(34)
&& 1869
  &\cr\ifspace\myskip&&
&& 
&& 
&& 
  &\cr\fi\myskip\tlr
\myskip
&& M_{D^*}
&& 1876(32)
&& 2085(35)
&& 2008
  &\cr\ifspace\myskip&&
&& 
&& 
&& 
  &\cr\fi\myskip\tlr
\myskip
&& M_{D_s}(m_s(M_K))
&& 1896(30)
&& 2112(32)
&& 1969
  &\cr\ifspace\myskip&&
&& 
&& 
&& 
  &\cr\fi\myskip\tlr
\myskip
&& M_{D_s}(m_s(M_\phi))
&& 1914(26)
&& 2137(27)
&& 1969
  &\cr\ifspace\myskip&&
&& 
&& 
&& 
  &\cr\fi\myskip\tlr
\myskip
&& M_{D_s^*}(m_s(M_K))
&& 1961(31)
&& 2201(34)
&& 2110?
  &\cr\ifspace\myskip&&
&& 
&& 
&& 
  &\cr\fi\myskip\tlr
\myskip
&& M_{D_s^*}(m_s(M_\phi))
&& 1978(27)
&& 2224(29)
&& 2110?
  &\cr\ifspace\myskip&&
&& 
&& 
&& 
  &\cr\fi\myskip\tlr
\myskip
&& M_{\eta_c}({}^1S_0)
&& 2836(50)
&& 3590(64)
&& 2980
  &\cr\ifspace\myskip&&
&& 
&& 
&& 
  &\cr\fi\myskip\tlr
\myskip
&& M_{J/\psi}({}^3S_1)
&& 2865(51)
&& 3643(65)
&& 3097
  &\cr\ifspace\myskip&&
&& 
&& 
&& 
  &\cr\fi\myskip\tlr
\myskip
&& M_{\chi_{c0}}({}^3P_0)
&& 3324(60)
&& 4572(86)
&& 3415
  &\cr\ifspace\myskip&&
&& 
&& 
&& 
  &\cr\fi\myskip\tlr
\myskip
&& M_{\chi_{c1}}({}^3P_1)
&& 3357(60)
&& 4646(86)
&& 3510
  &\cr\ifspace\myskip&&
&& 
&& 
&& 
  &\cr\fi\myskip\tlr
\myskip
&& \Delta M ({}^3S_1 - {}^1S_0)
&& 29(1)
&& 53(2)
&& 117
  &\cr\ifspace\myskip&&
&& 
&& 
&& 
  &\cr\fi\myskip\tlr
\myskip
&& \Delta M ({}^3P_1 - {}^3P_0)
&& 33(9)
&& 74(18)
&& 95
  &\cr\ifspace\myskip&&
&& 
&& 
&& 
  &\cr\fi\myskip\tlr
\myskip
&& \Delta M ({}^3P_0 - {}^3S_1)
&& 459(17)
&& 929(36)
&& 318
  &\cr\ifspace\myskip&&
&& 
&& 
&& 
  &\cr\fi\myskip\tlr
\cr}}}}

 }
{\vtop{\advance\hsize by -2\parindent 
\noindent
A comparison of lattice estimates of $D$ meson and charmonium masses with the 
experimental data.  We show results for $M_1$ and $M_2$ and for 
the two different ways of setting $m_s$ described in the text.}}

Using this result, we have that $M_2 = \sinh M_1$.  In
Table~\nameuse\tmonetwo\ we compare $M_1$ and $M_2$ for the
pseudoscalar and vector mesons.  The difference is tiny for the smallest
quark masses, but substantial for charmed mesons. Our results for
charmed meson masses (obtained by linear extrapolation in the light quark
mass and with $m_c$ chosen to be $\kappa=0.135$) are given in
Table~\nameuse\tDmass. There is a significant difference between the
estimates using $M_1$ and $M_2$: for the $D$ mesons the difference is
$\sim 10\%$, while for the charmonium system it is $25-30\%$.
This suggests that $O(ma)$ corrections in the Wilson
action are already large in the charm region. 
This is also apparent from the mass-splittings---the spin-spin 
and spin-orbit interactions are underestimated by the Wilson action,
as has been previously observed
\ref\rsplittings{J. Sloan, \spireseprint{hep-lat}{9412095}{\bielefeld, 171}.}.

\newsec{Baryon masses}

Our baryon mass analysis is based on two overlapping data sets. For
the complete data set (170 configurations) we have results for only $WL$ and
$SL$ correlators, and for only a subset of possible quark combinations.
On the last 110 configurations we also calculate
$SS$ correlators, and use all degenerate and non-degenerate
combinations made up of $S$ and $U_i$ quarks.  
Our results for the masses are
given in Tables~\nameuse\tlambda, \nameuse\tnucleon, and \nameuse\tdecuplet. 
The sample size is indicated by the subscript in the table headings.
In the following, some analyses are only possible on the smaller sample,
and we label such results by an asterisk next to the error estimate.

\subsec{Correlators}

It is necessary to introduce some notation to explain the 
correlation functions we calculate. 
We adapt that used in quenched chiral perturbation
theory for baryons \labrenzsharpe.
Assume that we have three flavors of quark, labeled by $u$, $d$ and $s$.
To create spin-1/2 baryons we can use the interpolating operators
\eqn\ebarydef{
{\cal O}_{(ij)k} = 
(\psi_{a,i}^T C\gamma_5 \psi_{b,j})\psi_{c,k} \epsilon^{abc} 
\,, }
where $a$, $b$ and $c$ label color, while $i$, $j$ and $k$ label flavor.
It is simple to show that ${\cal O}_{(ij)k} = -{\cal O}_{(ji)k}$,
so that there are only nine independent operators---eight $SU(3)$ octets
and the singlet $\sum_{ijk} \epsilon^{ijk} {\cal O}_{ijk}$.
One way to project against the singlet is to form
\eqn\ebarydefn{
{\cal B}_{ijk} = {\cal O}_{(ij)k} + {\cal O}_{(ik)j} = {\cal B}_{ikj} \,.
}
There are eight independent $B_{ijk}$'s, the relation to the
usual states being exemplified by
\eqn\ebaryrel{\eqalign{
\sqrt{2} p	  &=
-{\cal B}_{duu}=2{\cal B}_{uud}= 2{\cal B}_{udu} \,,\cr
\sqrt{2}\Sigma^+&=
-{\cal B}_{suu}=2{\cal B}_{uus}= 2{\cal B}_{usu} \,,\cr
\Sigma^0 &= {\cal B}_{uds}+{\cal B}_{dsu} = - {\cal B}_{sud} \,.\cr
\sqrt{3} \Lambda^0 	  &= {\cal B}_{uds}-{\cal B}_{dsu} \,.\cr
}}
The overall factor in these equations is arbitrary,
while the relative normalization is fixed by $SU(3)$ symmetry.

\figure\fcontractions{\epsfysize=3in\epsfbox{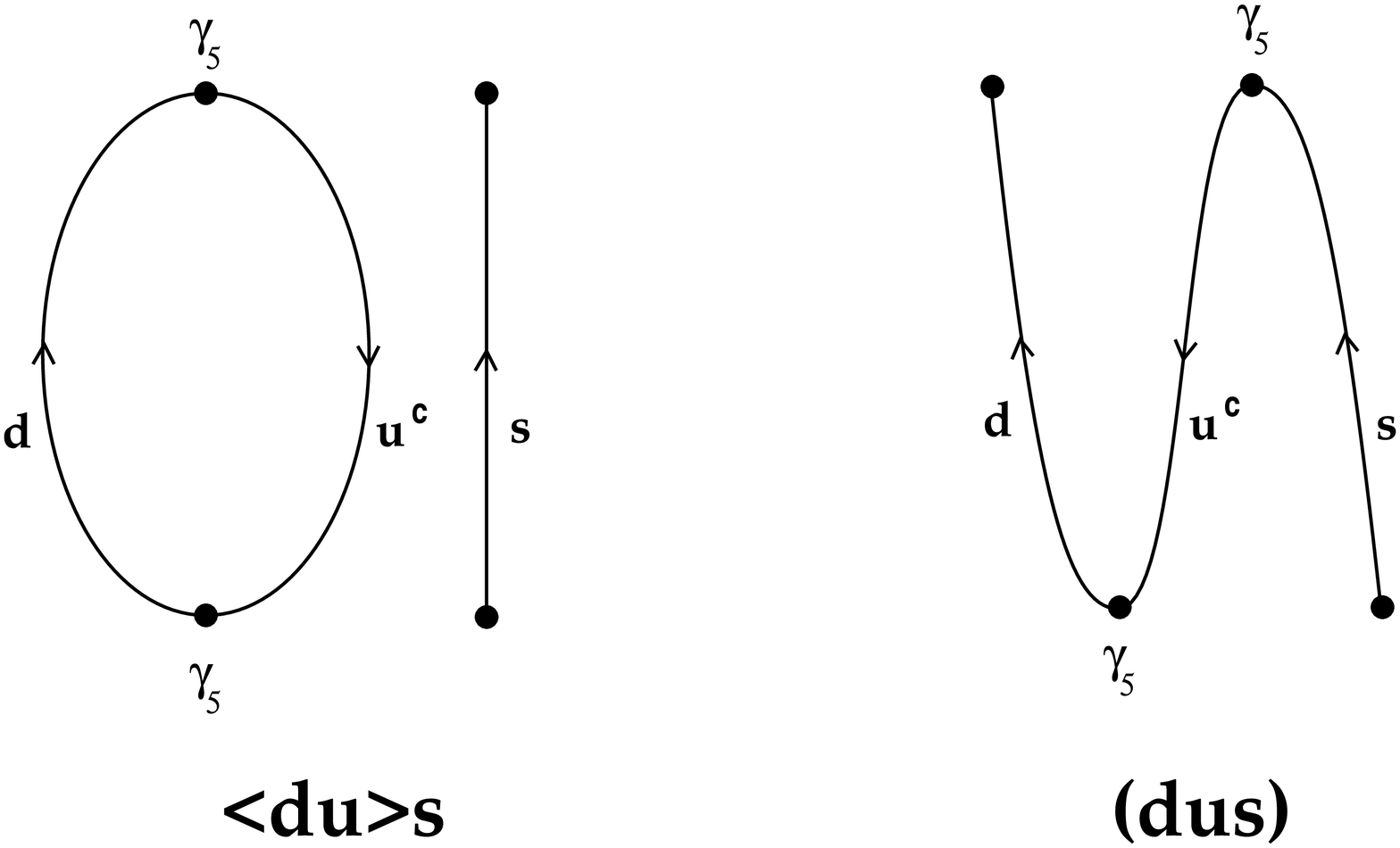}}
{\vtop{\advance\hsize by -2\parindent 
\noindent 
The two different types of contractions for the baryon states.
}}

\table\tlambda{
%
\let\ifspace=\iffalse
\def\myskip{\omit&height1pt&%
\omit&&%
\omit&&%
\omit&&%
\omit&&%
 &\cr}
\vbox{\hbox{\vbox{
\tabskip=0pt\offinterlineskip
\def\tlr{\noalign{\hrule}}

\halign {#& \vrule\vrule#\tabskip=3pt&
  \hfil$#$\hfil&\vrule#&
  \hfil$#$\hfil&\vrule#&
  \hfil$#$\hfil&\vrule#&
  \hfil$#$\hfil&\vrule#&
  \hfil$#$\hfil&\vrule#\tabskip=0pt\cr\tlr
%
\myskip
&& 
&& WL_{110}             
&& SL_{110}             
&& SS_{110}             
&& AV_{110}             
  &\cr
\myskip\tlr
\omit&height0.5pt&\multispan{ 9   }&\cr\tlr
\myskip
&& S[SS]             
&& 0.786( 03)
&& 0.791( 04)
&& 0.793( 04)
&& 0.789( 02)
  &\cr\ifspace\myskip&&
&& 
&& 
&& 
&& 
  &\cr\fi\myskip\tlr
\myskip
&& S[SU_1]           
&& 0.734( 04)
&& 0.740( 04)
&& 0.742( 04)
&& 0.738( 02)
  &\cr\ifspace\myskip&&
&& 
&& 
&& 
&& 
  &\cr\fi\myskip\tlr
\myskip
&& S[SU_2]           
&& 0.712( 04)
&& 0.720( 04)
&& 0.721( 04)
&& 0.716( 03)
  &\cr\ifspace\myskip&&
&& 
&& 
&& 
&& 
  &\cr\fi\myskip\tlr
\myskip
&& S[SU_3]           
&& 0.698( 04)
&& 0.707( 05)
&& 0.708( 04)
&& 0.703( 03)
  &\cr\ifspace\myskip&&
&& 
&& 
&& 
&& 
  &\cr\fi\myskip\tlr
\myskip
&& S[U_1U_1]         
&& 0.678( 04)
&& 0.687( 04)
&& 0.688( 04)
&& 0.683( 03)
  &\cr\ifspace\myskip&&
&& 
&& 
&& 
&& 
  &\cr\fi\myskip\tlr
\myskip
&& S[U_1U_2]         
&& 0.655( 04)
&& 0.665( 05)
&& 0.665( 04)
&& 0.660( 03)
  &\cr\ifspace\myskip&&
&& 
&& 
&& 
&& 
  &\cr\fi\myskip\tlr
\myskip
&& S[U_1U_3]         
&& 0.640( 05)
&& 0.652( 05)
&& 0.651( 05)
&& 0.646( 03)
  &\cr\ifspace\myskip&&
&& 
&& 
&& 
&& 
  &\cr\fi\myskip\tlr
\myskip
&& S[U_2U_2]         
&& 0.630( 05)
&& 0.642( 05)
&& 0.642( 05)
&& 0.636( 03)
  &\cr\ifspace\myskip&&
&& 
&& 
&& 
&& 
  &\cr\fi\myskip\tlr
\myskip
&& S[U_2U_3]         
&& 0.614( 06)
&& 0.627( 06)
&& 0.627( 05)
&& 0.620( 04)
  &\cr\ifspace\myskip&&
&& 
&& 
&& 
&& 
  &\cr\fi\myskip\tlr
\myskip
&& S[U_3U_3]         
&& 0.597( 07)
&& 0.611( 07)
&& 0.610( 06)
&& 0.604( 04)
  &\cr\ifspace\myskip&&
&& 
&& 
&& 
&& 
  &\cr\fi\myskip\tlr
\myskip
&& U_1[SS]           
&& 0.740( 04)
&& 0.747( 04)
&& 0.749( 04)
&& 0.744( 03)
  &\cr\ifspace\myskip&&
&& 
&& 
&& 
&& 
  &\cr\fi\myskip\tlr
\myskip
&& U_1[SU_1]         
&& 0.687( 05)
&& 0.696( 05)
&& 0.697( 04)
&& 0.691( 03)
  &\cr\ifspace\myskip&&
&& 
&& 
&& 
&& 
  &\cr\fi\myskip\tlr
\myskip
&& U_1[SU_2]         
&& 0.663( 04)
&& 0.675( 05)
&& 0.675( 05)
&& 0.669( 03)
  &\cr\ifspace\myskip&&
&& 
&& 
&& 
&& 
  &\cr\fi\myskip\tlr
\myskip
&& U_1[SU_3]         
&& 0.648( 05)
&& 0.661( 05)
&& 0.661( 05)
&& 0.655( 03)
  &\cr\ifspace\myskip&&
&& 
&& 
&& 
&& 
  &\cr\fi\myskip\tlr
\myskip
&& U_1[U_1U_1]       
&& 0.630( 05)
&& 0.643( 05)
&& 0.642( 05)
&& 0.636( 03)
  &\cr\ifspace\myskip&&
&& 
&& 
&& 
&& 
  &\cr\fi\myskip\tlr
\myskip
&& U_1[U_1U_2]       
&& 0.605( 05)
&& 0.620( 06)
&& 0.619( 05)
&& 0.612( 03)
  &\cr\ifspace\myskip&&
&& 
&& 
&& 
&& 
  &\cr\fi\myskip\tlr
\myskip
&& U_1[U_1U_3]       
&& 0.589( 06)
&& 0.605( 06)
&& 0.604( 06)
&& 0.597( 04)
  &\cr\ifspace\myskip&&
&& 
&& 
&& 
&& 
  &\cr\fi\myskip\tlr
\myskip
&& U_1[U_2U_2]       
&& 0.579( 06)
&& 0.596( 06)
&& 0.595( 06)
&& 0.587( 04)
  &\cr\ifspace\myskip&&
&& 
&& 
&& 
&& 
  &\cr\fi\myskip\tlr
\myskip
&& U_1[U_2U_3]       
&& 0.562( 07)
&& 0.580( 07)
&& 0.579( 06)
&& 0.571( 04)
  &\cr\ifspace\myskip&&
&& 
&& 
&& 
&& 
  &\cr\fi\myskip\tlr
\myskip
&& U_1[U_3U_3]       
&& 0.544( 08)
&& 0.563( 09)
&& 0.562( 07)
&& 0.553( 05)
  &\cr\ifspace\myskip&&
&& 
&& 
&& 
&& 
  &\cr\fi\myskip\tlr
\myskip
&& U_2[SS]           
&& 0.723( 04)
&& 0.731( 05)
&& 0.732( 04)
&& 0.727( 03)
  &\cr\ifspace\myskip&&
&& 
&& 
&& 
&& 
  &\cr\fi\myskip\tlr
\myskip
&& U_2[SU_1]         
&& 0.668( 05)
&& 0.679( 05)
&& 0.679( 05)
&& 0.674( 03)
  &\cr\ifspace\myskip&&
&& 
&& 
&& 
&& 
  &\cr\fi\myskip\tlr
\myskip
&& U_2[SU_2]         
&& 0.643( 05)
&& 0.657( 06)
&& 0.657( 05)
&& 0.650( 03)
  &\cr\ifspace\myskip&&
&& 
&& 
&& 
&& 
  &\cr\fi\myskip\tlr
\myskip
&& U_2[SU_3]         
&& 0.627( 06)
&& 0.643( 06)
&& 0.643( 06)
&& 0.635( 04)
  &\cr\ifspace\myskip&&
&& 
&& 
&& 
&& 
  &\cr\fi\myskip\tlr
\myskip
&& U_2[U_1U_1]       
&& 0.610( 06)
&& 0.626( 06)
&& 0.625( 05)
&& 0.617( 04)
  &\cr\ifspace\myskip&&
&& 
&& 
&& 
&& 
  &\cr\fi\myskip\tlr
\myskip
&& U_2[U_1U_2]       
&& 0.584( 06)
&& 0.602( 07)
&& 0.601( 06)
&& 0.593( 04)
  &\cr\ifspace\myskip&&
&& 
&& 
&& 
&& 
  &\cr\fi\myskip\tlr
\myskip
&& U_2[U_1U_3]       
&& 0.567( 07)
&& 0.587( 08)
&& 0.586( 07)
&& 0.577( 05)
  &\cr\ifspace\myskip&&
&& 
&& 
&& 
&& 
  &\cr\fi\myskip\tlr
\myskip
&& U_2[U_2U_2]       
&& 0.557( 07)
&& 0.578( 08)
&& 0.577( 07)
&& 0.567( 05)
  &\cr\ifspace\myskip&&
&& 
&& 
&& 
&& 
  &\cr\fi\myskip\tlr
\myskip
&& U_2[U_2U_3]       
&& 0.539( 08)
&& 0.561( 09)
&& 0.560( 08)
&& 0.550( 05)
  &\cr\ifspace\myskip&&
&& 
&& 
&& 
&& 
  &\cr\fi\myskip\tlr
\myskip
&& U_2[U_3U_3]       
&& 0.520( 10)
&& 0.543( 11)
&& 0.543( 09)
&& 0.532( 06)
  &\cr\ifspace\myskip&&
&& 
&& 
&& 
&& 
  &\cr\fi\myskip\tlr
\myskip
&& U_3[SS]           
&& 0.712( 05)
&& 0.722( 05)
&& 0.723( 05)
&& 0.717( 03)
  &\cr\ifspace\myskip&&
&& 
&& 
&& 
&& 
  &\cr\fi\myskip\tlr
\myskip
&& U_3[SU_1]         
&& 0.656( 06)
&& 0.670( 06)
&& 0.669( 05)
&& 0.663( 04)
  &\cr\ifspace\myskip&&
&& 
&& 
&& 
&& 
  &\cr\fi\myskip\tlr
\myskip
&& U_3[SU_2]         
&& 0.631( 06)
&& 0.647( 07)
&& 0.646( 06)
&& 0.639( 04)
  &\cr\ifspace\myskip&&
&& 
&& 
&& 
&& 
  &\cr\fi\myskip\tlr
\myskip
&& U_3[SU_3]         
&& 0.614( 07)
&& 0.632( 08)
&& 0.631( 07)
&& 0.623( 04)
  &\cr\ifspace\myskip&&
&& 
&& 
&& 
&& 
  &\cr\fi\myskip\tlr
\myskip
&& U_3[U_1U_1]       
&& 0.598( 06)
&& 0.616( 07)
&& 0.615( 06)
&& 0.607( 04)
  &\cr\ifspace\myskip&&
&& 
&& 
&& 
&& 
  &\cr\fi\myskip\tlr
\myskip
&& U_3[U_1U_2]       
&& 0.572( 07)
&& 0.592( 08)
&& 0.590( 07)
&& 0.581( 05)
  &\cr\ifspace\myskip&&
&& 
&& 
&& 
&& 
  &\cr\fi\myskip\tlr
\myskip
&& U_3[U_1U_3]       
&& 0.554( 08)
&& 0.576( 09)
&& 0.574( 08)
&& 0.564( 05)
  &\cr\ifspace\myskip&&
&& 
&& 
&& 
&& 
  &\cr\fi\myskip\tlr
\myskip
&& U_3[U_2U_2]       
&& 0.544( 08)
&& 0.567( 09)
&& 0.565( 08)
&& 0.555( 05)
  &\cr\ifspace\myskip&&
&& 
&& 
&& 
&& 
  &\cr\fi\myskip\tlr
\myskip
&& U_3[U_2U_3]       
&& 0.525( 09)
&& 0.550( 11)
&& 0.549( 09)
&& 0.537( 06)
  &\cr\ifspace\myskip&&
&& 
&& 
&& 
&& 
  &\cr\fi\myskip\tlr
\myskip
&& U_3[U_3U_3]       
&& 0.506( 11)
&& 0.531( 13)
&& 0.531( 10)
&& 0.518( 07)
  &\cr\ifspace\myskip&&
&& 
&& 
&& 
&& 
  &\cr\fi\myskip\tlr
\cr}}}}

 }
{Mass estimates for $\Lambda$-like baryons.}

All spin-1/2 baryon correlators are built out of the two contractions
shown in Fig.~\nameuse\fcontractions.
The notation $\langle DU \rangle S = \langle UD \rangle S$ 
corresponds to quarks of flavors $U$ and $D$ contracted into
a closed loop, while the propagator for $S$ carries the spin quantum
numbers of the baryon.  
The notation $(DUS)$ corresponds
to a single ordered contraction of the three quarks.
We consider two types of correlator, ``$\Sigma$-like'' and ``$\Lambda$-like''.
The former is exemplified by that of the $\Sigma^0$
\eqn\ebarysig{\eqalign{
S\{UD\} = S \{DU\} &\equiv 
\vev{{\cal B}_{sud}(x) \ \overline{{\cal B}_{sud}(0)}} \cr
&= \langle US \rangle D + \langle DS \rangle U + (USD) + (DSU) \,. \cr
}}
(This equation defines our sign conventions for the contractions.)
The proton, neutron, $\Sigma^+$, $\Sigma^-$, $\Xi^0$ and $\Xi^-$
correlators are also of this type:
they are, respectively, $D\{UU\}$, $U\{DD\}$, $S\{UU\}$, $S\{DD\}$,
$U\{SS\}$ and $D\{SS\}$.
The second type of correlator is that of the $\Lambda^0$
\eqn\ebarylam{\eqalign{
S[UD] = S[DU] &\equiv (1/3) \left[
\langle US \rangle D + \langle DS \rangle U + 4 \langle UD \rangle S 
-  (USD) - (DSU)  \right.\cr 
&\quad \left. + 2(SUD) + 2(SDU) + 2(UDS) + 2(DUS) \right]
\,. \cr
}}
When $m_u\ne m_d$, there is also a non-vanishing
$\Lambda^0-\Sigma^0$ cross correlator, but
we have not found this to give useful results.

\table\tnucleon{
\let\ifspace=\iffalse
\def\myskip{\omit&height1pt&%
\omit&&%
\omit&&%
\omit&&%
\omit&&%
\omit&&%
\omit&&%
\omit&&%
 &\cr}
\vbox{\hbox{\vbox{
\tabskip=0pt\offinterlineskip
\def\tlr{\noalign{\hrule}}

\halign {#& \vrule\vrule#\tabskip=3pt&
  \hfil$#$\hfil&\vrule#&
  \hfil$#$\hfil&\vrule#&
  \hfil$#$\hfil&\vrule#&
  \hfil$#$\hfil&\vrule#&
  \hfil$#$\hfil&\vrule#&
  \hfil$#$\hfil&\vrule#&
  \hfil$#$\hfil&\vrule#&
  \hfil$#$\hfil&\vrule#\tabskip=0pt\cr\tlr
%
\myskip
&& 
&& WL_{110}             
&& SL_{110}             
&& SS_{110}             
&& AV_{110}             
&& WL_{170}            
&& SL_{170}            
&& AV_{170}            
  &\cr
\myskip\tlr
\omit&height0.5pt&\multispan{15   }&\cr\tlr
%
\myskip
&& S\{SS\}             
&& 0.786( 03)
&& 0.791( 04)
&& 0.793( 04)
&& 0.789( 02)
&& 0.786( 03)
&& 0.794( 03)
&& 0.790( 02)
  &\cr\ifspace\myskip&&
&& 
&& 
&& 
&& 
&& 
&& 
&& 
  &\cr\fi\myskip\tlr
\myskip
&& S\{SU_1\}           
&& 0.738( 04)
&& 0.745( 04)
&& 0.746( 04)
&& 0.742( 03)
&& 
&& 
&& 
  &\cr\ifspace\myskip&&
&& 
&& 
&& 
&& 
&& 
&& 
&& 
  &\cr\fi\myskip\tlr
\myskip
&& S\{SU_2\}           
&& 0.718( 04)
&& 0.727( 04)
&& 0.728( 04)
&& 0.723( 03)
&& 
&& 
&& 
  &\cr\ifspace\myskip&&
&& 
&& 
&& 
&& 
&& 
&& 
&& 
  &\cr\fi\myskip\tlr
\myskip
&& S\{SU_3\}           
&& 0.706( 04)
&& 0.716( 05)
&& 0.717( 05)
&& 0.711( 03)
&& 
&& 
&& 
  &\cr\ifspace\myskip&&
&& 
&& 
&& 
&& 
&& 
&& 
&& 
  &\cr\fi\myskip\tlr
\myskip
&& S\{U_1U_1\}         
&& 0.689( 04)
&& 0.699( 05)
&& 0.700( 04)
&& 0.695( 03)
&& 0.689( 04)
&& 0.702( 04)
&& 0.695( 03)
  &\cr\ifspace\myskip&&
&& 
&& 
&& 
&& 
&& 
&& 
&& 
  &\cr\fi\myskip\tlr
\myskip
&& S\{U_1U_2\}         
&& 0.669( 05)
&& 0.682( 05)
&& 0.682( 05)
&& 0.676( 03)
&& 
&& 
&& 
  &\cr\ifspace\myskip&&
&& 
&& 
&& 
&& 
&& 
&& 
&& 
  &\cr\fi\myskip\tlr
\myskip
&& S\{U_1U_3\}         
&& 0.657( 05)
&& 0.672( 06)
&& 0.671( 05)
&& 0.664( 04)
&& 
&& 
&& 
  &\cr\ifspace\myskip&&
&& 
&& 
&& 
&& 
&& 
&& 
&& 
  &\cr\fi\myskip\tlr
\myskip
&& S\{U_2U_2\}         
&& 0.649( 06)
&& 0.665( 06)
&& 0.664( 06)
&& 0.657( 04)
&& 0.649( 04)
&& 0.667( 05)
&& 0.658( 03)
  &\cr\ifspace\myskip&&
&& 
&& 
&& 
&& 
&& 
&& 
&& 
  &\cr\fi\myskip\tlr
\myskip
&& S\{U_2U_3\}         
&& 0.636( 07)
&& 0.654( 07)
&& 0.654( 06)
&& 0.645( 04)
&& 
&& 
&& 
  &\cr\ifspace\myskip&&
&& 
&& 
&& 
&& 
&& 
&& 
&& 
  &\cr\fi\myskip\tlr
\myskip
&& S\{U_3U_3\}         
&& 0.622( 08)
&& 0.644( 09)
&& 0.643( 08)
&& 0.633( 05)
&& 0.625( 06)
&& 0.647( 07)
&& 0.636( 04)
  &\cr\ifspace\myskip&&
&& 
&& 
&& 
&& 
&& 
&& 
&& 
  &\cr\fi\myskip\tlr
\myskip
&& U_1\{SS\}           
&& 0.732( 04)
&& 0.738( 04)
&& 0.739( 04)
&& 0.736( 02)
&& 0.732( 03)
&& 0.744( 04)
&& 0.738( 02)
  &\cr\ifspace\myskip&&
&& 
&& 
&& 
&& 
&& 
&& 
&& 
  &\cr\fi\myskip\tlr
\myskip
&& U_1\{SU_1\}         
&& 0.681( 05)
&& 0.690( 04)
&& 0.690( 04)
&& 0.686( 03)
&& 
&& 
&& 
  &\cr\ifspace\myskip&&
&& 
&& 
&& 
&& 
&& 
&& 
&& 
  &\cr\fi\myskip\tlr
\myskip
&& U_1\{SU_2\}         
&& 0.659( 04)
&& 0.671( 05)
&& 0.671( 04)
&& 0.665( 03)
&& 
&& 
&& 
  &\cr\ifspace\myskip&&
&& 
&& 
&& 
&& 
&& 
&& 
&& 
  &\cr\fi\myskip\tlr
\myskip
&& U_1\{SU_3\}         
&& 0.646( 05)
&& 0.659( 05)
&& 0.658( 05)
&& 0.652( 03)
&& 
&& 
&& 
  &\cr\ifspace\myskip&&
&& 
&& 
&& 
&& 
&& 
&& 
&& 
  &\cr\fi\myskip\tlr
\myskip
&& U_1\{U_1U_1\}       
&& 0.630( 05)
&& 0.643( 05)
&& 0.642( 05)
&& 0.636( 03)
&& 0.630( 05)
&& 0.645( 05)
&& 0.638( 03)
  &\cr\ifspace\myskip&&
&& 
&& 
&& 
&& 
&& 
&& 
&& 
  &\cr\fi\myskip\tlr
\myskip
&& U_1\{U_1U_2\}       
&& 0.608( 05)
&& 0.624( 06)
&& 0.623( 05)
&& 0.616( 04)
&& 
&& 
&& 
  &\cr\ifspace\myskip&&
&& 
&& 
&& 
&& 
&& 
&& 
&& 
  &\cr\fi\myskip\tlr
\myskip
&& U_1\{U_1U_3\}       
&& 0.594( 06)
&& 0.612( 07)
&& 0.611( 06)
&& 0.603( 04)
&& 
&& 
&& 
  &\cr\ifspace\myskip&&
&& 
&& 
&& 
&& 
&& 
&& 
&& 
  &\cr\fi\myskip\tlr
\myskip
&& U_1\{U_2U_2\}       
&& 0.586( 06)
&& 0.605( 07)
&& 0.604( 06)
&& 0.595( 04)
&& 0.588( 05)
&& 0.607( 06)
&& 0.598( 04)
  &\cr\ifspace\myskip&&
&& 
&& 
&& 
&& 
&& 
&& 
&& 
  &\cr\fi\myskip\tlr
\myskip
&& U_1\{U_2U_3\}       
&& 0.572( 07)
&& 0.593( 08)
&& 0.592( 07)
&& 0.582( 05)
&& 
&& 
&& 
  &\cr\ifspace\myskip&&
&& 
&& 
&& 
&& 
&& 
&& 
&& 
  &\cr\fi\myskip\tlr
\myskip
&& U_1\{U_3U_3\}       
&& 0.557( 09)
&& 0.581( 10)
&& 0.580( 08)
&& 0.569( 06)
&& 0.562( 07)
&& 0.583( 08)
&& 0.573( 05)
  &\cr\ifspace\myskip&&
&& 
&& 
&& 
&& 
&& 
&& 
&& 
  &\cr\fi\myskip\tlr
\myskip
&& U_2\{SS\}           
&& 0.710( 04)
&& 0.717( 04)
&& 0.718( 04)
&& 0.714( 03)
&& 0.708( 04)
&& 0.720( 04)
&& 0.714( 03)
  &\cr\ifspace\myskip&&
&& 
&& 
&& 
&& 
&& 
&& 
&& 
  &\cr\fi\myskip\tlr
\myskip
&& U_2\{SU_1\}         
&& 0.657( 05)
&& 0.667( 05)
&& 0.667( 04)
&& 0.662( 03)
&& 
&& 
&& 
  &\cr\ifspace\myskip&&
&& 
&& 
&& 
&& 
&& 
&& 
&& 
  &\cr\fi\myskip\tlr
\myskip
&& U_2\{SU_2\}         
&& 0.634( 05)
&& 0.646( 05)
&& 0.646( 05)
&& 0.640( 03)
&& 
&& 
&& 
  &\cr\ifspace\myskip&&
&& 
&& 
&& 
&& 
&& 
&& 
&& 
  &\cr\fi\myskip\tlr
\myskip
&& U_2\{SU_3\}         
&& 0.619( 06)
&& 0.633( 06)
&& 0.633( 05)
&& 0.626( 04)
&& 
&& 
&& 
  &\cr\ifspace\myskip&&
&& 
&& 
&& 
&& 
&& 
&& 
&& 
  &\cr\fi\myskip\tlr
\myskip
&& U_2\{U_1U_1\}       
&& 0.603( 05)
&& 0.618( 06)
&& 0.618( 05)
&& 0.611( 03)
&& 0.605( 06)
&& 0.621( 06)
&& 0.613( 03)
  &\cr\ifspace\myskip&&
&& 
&& 
&& 
&& 
&& 
&& 
&& 
  &\cr\fi\myskip\tlr
\myskip
&& U_2\{U_1U_2\}       
&& 0.580( 06)
&& 0.598( 07)
&& 0.597( 06)
&& 0.589( 04)
&& 
&& 
&& 
  &\cr\ifspace\myskip&&
&& 
&& 
&& 
&& 
&& 
&& 
&& 
  &\cr\fi\myskip\tlr
\myskip
&& U_2\{U_1U_3\}       
&& 0.565( 07)
&& 0.585( 08)
&& 0.584( 06)
&& 0.575( 04)
&& 
&& 
&& 
  &\cr\ifspace\myskip&&
&& 
&& 
&& 
&& 
&& 
&& 
&& 
  &\cr\fi\myskip\tlr
\myskip
&& U_2\{U_2U_2\}       
&& 0.557( 07)
&& 0.578( 08)
&& 0.577( 07)
&& 0.567( 05)
&& 0.561( 06)
&& 0.581( 07)
&& 0.571( 04)
  &\cr\ifspace\myskip&&
&& 
&& 
&& 
&& 
&& 
&& 
&& 
  &\cr\fi\myskip\tlr
\myskip
&& U_2\{U_2U_3\}       
&& 0.542( 08)
&& 0.565( 09)
&& 0.564( 08)
&& 0.553( 05)
&& 
&& 
&& 
  &\cr\ifspace\myskip&&
&& 
&& 
&& 
&& 
&& 
&& 
&& 
  &\cr\fi\myskip\tlr
\myskip
&& U_2\{U_3U_3\}       
&& 0.526( 10)
&& 0.552( 11)
&& 0.551( 09)
&& 0.538( 06)
&& 0.534( 08)
&& 0.555( 09)
&& 0.544( 06)
  &\cr\ifspace\myskip&&
&& 
&& 
&& 
&& 
&& 
&& 
&& 
  &\cr\fi\myskip\tlr
\myskip
&& U_3\{SS\}           
&& 0.696( 05)
&& 0.704( 05)
&& 0.704( 04)
&& 0.700( 03)
&& 0.694( 04)
&& 0.708( 05)
&& 0.701( 03)
  &\cr\ifspace\myskip&&
&& 
&& 
&& 
&& 
&& 
&& 
&& 
  &\cr\fi\myskip\tlr
\myskip
&& U_3\{SU_1\}         
&& 0.642( 06)
&& 0.653( 05)
&& 0.652( 05)
&& 0.647( 03)
&& 
&& 
&& 
  &\cr\ifspace\myskip&&
&& 
&& 
&& 
&& 
&& 
&& 
&& 
  &\cr\fi\myskip\tlr
\myskip
&& U_3\{SU_2\}         
&& 0.617( 06)
&& 0.631( 06)
&& 0.630( 05)
&& 0.624( 04)
&& 
&& 
&& 
  &\cr\ifspace\myskip&&
&& 
&& 
&& 
&& 
&& 
&& 
&& 
  &\cr\fi\myskip\tlr
\myskip
&& U_3\{SU_3\}         
&& 0.602( 07)
&& 0.616( 07)
&& 0.616( 06)
&& 0.609( 04)
&& 
&& 
&& 
  &\cr\ifspace\myskip&&
&& 
&& 
&& 
&& 
&& 
&& 
&& 
  &\cr\fi\myskip\tlr
\myskip
&& U_3\{U_1U_1\}       
&& 0.586( 06)
&& 0.602( 06)
&& 0.602( 06)
&& 0.594( 04)
&& 0.589( 06)
&& 0.606( 06)
&& 0.597( 04)
  &\cr\ifspace\myskip&&
&& 
&& 
&& 
&& 
&& 
&& 
&& 
  &\cr\fi\myskip\tlr
\myskip
&& U_3\{U_1U_2\}       
&& 0.563( 07)
&& 0.581( 08)
&& 0.580( 06)
&& 0.572( 04)
&& 
&& 
&& 
  &\cr\ifspace\myskip&&
&& 
&& 
&& 
&& 
&& 
&& 
&& 
  &\cr\fi\myskip\tlr
\myskip
&& U_3\{U_1U_3\}       
&& 0.547( 08)
&& 0.567( 09)
&& 0.566( 07)
&& 0.557( 05)
&& 
&& 
&& 
  &\cr\ifspace\myskip&&
&& 
&& 
&& 
&& 
&& 
&& 
&& 
  &\cr\fi\myskip\tlr
\myskip
&& U_3\{U_2U_2\}       
&& 0.539( 08)
&& 0.560( 09)
&& 0.559( 07)
&& 0.549( 05)
&& 0.543( 08)
&& 0.564( 08)
&& 0.553( 05)
  &\cr\ifspace\myskip&&
&& 
&& 
&& 
&& 
&& 
&& 
&& 
  &\cr\fi\myskip\tlr
\myskip
&& U_3\{U_2U_3\}       
&& 0.523( 09)
&& 0.545( 11)
&& 0.545( 09)
&& 0.534( 06)
&& 
&& 
&& 
  &\cr\ifspace\myskip&&
&& 
&& 
&& 
&& 
&& 
&& 
&& 
  &\cr\fi\myskip\tlr
\myskip
&& U_3\{U_3U_3\}       
&& 0.506( 11)
&& 0.531( 13)
&& 0.531( 10)
&& 0.518( 07)
&& 0.515( 10)
&& 0.536( 10)
&& 0.525( 07)
  &\cr\ifspace\myskip&&
&& 
&& 
&& 
&& 
&& 
&& 
&& 
  &\cr\fi\myskip\tlr
\cr}}}}
 
 }
{Mass estimates for $\Sigma$-like baryons.}

We have calculated the two types of spin-1/2 baryon
correlator for all independent
mass combinations involving $U_i$ and $S$ quarks.
The results are given in Tables~\nameuse\tlambda\ and \nameuse\tnucleon\ 
respectively.
Masses from correlators of the form $A[AB]$ and $A\{AB\}$ 
are also included even though they are not independent---the correlators are related by
\eqn\ebarynotindep{\eqalign{
A[AB] 	&= {1\over4} \left( 3 B[AA] + B\{AA\} \right) \,, \cr
A\{AB\}	&= {1\over4} \left( B[AA] + 3 B\{AA\} \right) \,. \cr
}}
One can think of the results for the $A\{BB\}$ and $A[BB]$ masses
as being those for the $\Sigma$ and $\Lambda$, respectively,
with $m_s=m_A$ and $m_u=m_d=m_B$.
Unlike in the real world, there is nothing to stop these two
masses being the same, i.e. $m_A=m_B$, in the quenched approximation.
Note, however, that in this case the $\Sigma$ and
$\Lambda$ are also degenerate, i.e. $M(A\{AA\})=M(A[AA])$.
Indeed, the contractions in the two cases are identical.

The interpretation of the results for
the completely non-degenerate correlators, $A[BC]$ and $A\{BC\}$, 
is more complicated.
Because isospin is broken,
the $\Sigma^0$- and $\Lambda$-like states mix,
with both correlators containing contributions from both
physical states.
Let $M_+$ and $M_-$ be the masses of the heavier and lighter states,
respectively, and $\delta\!M$ the mass difference.
At long times, the effective mass for both correlators 
will asymptote to $M_-$.
However, at times short compared to the inverse mass difference,
i.e. $\delta\!M t_{\rm max} \ll 1$, 
there will be an approximate plateau at a value which is
a weighted average of the two masses.
To see this, we pick the $\Lambda$ correlator and write it as
\eqn\etwostateC{
C_\Lambda(t) = 
A e^{-M_- t} (\cos^2\!\theta + \sin^2\!\theta e^{-\delta\!M t}) + \cdots \,, 
}
where $\tan\theta$ is the ratio of the amplitudes to create the
two mixed states, and the ellipsis represents excited states. 
The effective mass is
\eqn\etwostateM{
m(\Lambda)_\eff(t) = - {d \ln C_\Lambda(t) \over dt} = 
M_- + \sin^2\!\theta\, \delta\!M (1 + O(\delta\!M t)) 
\approx \cos^2\!\theta\, M_- + \sin^2\!\theta\, M_+ \,. 
}
Thus the effective mass is almost constant, and given our errors,
we cannot distinguish it from a plateau.
We discuss below the interpretation of the resulting ``mass''.

\table\tdecuplet{
\let\ifspace=\iffalse
\def\myskip{\omit&height1.5pt&%
\omit&&%
\omit&&%
\omit&&%
\omit&&%
\omit&&%
\omit&&%
\omit&&%
 &\cr}
\vbox{\hbox{\vbox{
\tabskip=0pt\offinterlineskip
\def\tlr{\noalign{\hrule}}

\halign {\strut#& \vrule\vrule#\tabskip=3pt&
  \hfil$#$\hfil&\vrule#&
  \hfil$#$\hfil&\vrule#&
  \hfil$#$\hfil&\vrule#&
  \hfil$#$\hfil&\vrule#&
  \hfil$#$\hfil&\vrule#&
  \hfil$#$\hfil&\vrule#&
  \hfil$#$\hfil&\vrule#&
  \hfil$#$\hfil&\vrule#\tabskip=0pt\cr\tlr
%
%
\myskip
&& 
&& WL_{110}             
&& SL_{110}             
&& SS_{110}             
&& AV_{110}             
&& WL_{170}            
&& SL_{170}            
&& AV_{170}            
  &\cr
\myskip\tlr
\omit&height0.5pt&\multispan{15   }&\cr\tlr
\myskip
&& \{SSS\}             
&& 0.832( 05)
&& 0.843( 06)
&& 0.843( 07)
&& 0.837( 04)
&& 0.831( 04)
&& 0.845( 05)
&& 0.838( 03)
  &\cr\ifspace\myskip&&
&& 
&& 
&& 
&& 
&& 
&& 
&& 
  &\cr\fi\myskip\tlr
\myskip
&& \{SSU_1\}           
&& 0.788( 06)
&& 0.803( 07)
&& 0.802( 07)
&& 0.795( 05)
&& 
&& 
&& 
  &\cr\ifspace\myskip&&
&& 
&& 
&& 
&& 
&& 
&& 
&& 
  &\cr\fi\myskip\tlr
\myskip
&& \{SSU_2\}           
&& 0.770( 07)
&& 0.789( 07)
&& 0.786( 08)
&& 0.779( 06)
&& 
&& 
&& 
  &\cr\ifspace\myskip&&
&& 
&& 
&& 
&& 
&& 
&& 
&& 
  &\cr\fi\myskip\tlr
\myskip
&& \{SSU_3\}           
&& 0.760( 08)
&& 0.781( 09)
&& 0.778( 08)
&& 0.770( 06)
&& 
&& 
&& 
  &\cr\ifspace\myskip&&
&& 
&& 
&& 
&& 
&& 
&& 
&& 
  &\cr\fi\myskip\tlr
\myskip
&& \{SU_1U_1\}         
&& 0.744( 08)
&& 0.764( 08)
&& 0.761( 08)
&& 0.753( 06)
&& 
&& 
&& 
  &\cr\ifspace\myskip&&
&& 
&& 
&& 
&& 
&& 
&& 
&& 
  &\cr\fi\myskip\tlr
\myskip
&& \{SU_1U_2\}         
&& 0.725( 10)
&& 0.750( 09)
&& 0.747( 08)
&& 0.736( 07)
&& 
&& 
&& 
  &\cr\ifspace\myskip&&
&& 
&& 
&& 
&& 
&& 
&& 
&& 
  &\cr\fi\myskip\tlr
\myskip
&& \{SU_1U_3\}         
&& 0.710( 11)
&& 0.741( 11)
&& 0.738( 09)
&& 0.725( 08)
&& 
&& 
&& 
  &\cr\ifspace\myskip&&
&& 
&& 
&& 
&& 
&& 
&& 
&& 
  &\cr\fi\myskip\tlr
\myskip
&& \{SU_2U_2\}         
&& 0.705( 10)
&& 0.736( 11)
&& 0.732( 09)
&& 0.719( 07)
&& 
&& 
&& 
  &\cr\ifspace\myskip&&
&& 
&& 
&& 
&& 
&& 
&& 
&& 
  &\cr\fi\myskip\tlr
\myskip
&& \{SU_2U_3\}         
&& 0.690( 11)
&& 0.727( 13)
&& 0.724( 10)
&& 0.708( 08)
&& 
&& 
&& 
  &\cr\ifspace\myskip&&
&& 
&& 
&& 
&& 
&& 
&& 
&& 
  &\cr\fi\myskip\tlr
\myskip
&& \{SU_3U_3\}         
&& 0.674( 09)
&& 0.718( 15)
&& 0.716( 11)
&& 0.696( 08)
&& 
&& 
&& 
  &\cr\ifspace\myskip&&
&& 
&& 
&& 
&& 
&& 
&& 
&& 
  &\cr\fi\myskip\tlr
\myskip
&& \{U_1U_1U_1\}       
&& 0.697( 08)
&& 0.726( 10)
&& 0.721( 09)
&& 0.710( 06)
&& 0.699( 08)
&& 0.721( 09)
&& 0.710( 06)
  &\cr\ifspace\myskip&&
&& 
&& 
&& 
&& 
&& 
&& 
&& 
  &\cr\fi\myskip\tlr
\myskip
&& \{U_1U_1U_2\}       
&& 0.677( 09)
&& 0.712( 12)
&& 0.706( 10)
&& 0.693( 07)
&& 
&& 
&& 
  &\cr\ifspace\myskip&&
&& 
&& 
&& 
&& 
&& 
&& 
&& 
  &\cr\fi\myskip\tlr
\myskip
&& \{U_1U_1U_3\}       
&& 0.663( 11)
&& 0.704( 14)
&& 0.699( 11)
&& 0.682( 08)
&& 
&& 
&& 
  &\cr\ifspace\myskip&&
&& 
&& 
&& 
&& 
&& 
&& 
&& 
  &\cr\fi\myskip\tlr
\myskip
&& \{U_1U_2U_2\}       
&& 0.656( 09)
&& 0.699( 14)
&& 0.693( 11)
&& 0.676( 08)
&& 
&& 
&& 
  &\cr\ifspace\myskip&&
&& 
&& 
&& 
&& 
&& 
&& 
&& 
  &\cr\fi\myskip\tlr
\myskip
&& \{U_1U_2U_3\}       
&& 0.642( 10)
&& 0.690( 17)
&& 0.685( 12)
&& 0.665( 09)
&& 
&& 
&& 
  &\cr\ifspace\myskip&&
&& 
&& 
&& 
&& 
&& 
&& 
&& 
  &\cr\fi\myskip\tlr
\myskip
&& \{U_1U_3U_3\}       
&& 0.629( 09)
&& 0.682( 21)
&& 0.678( 13)
&& 0.654( 09)
&& 
&& 
&& 
  &\cr\ifspace\myskip&&
&& 
&& 
&& 
&& 
&& 
&& 
&& 
  &\cr\fi\myskip\tlr
\myskip
&& \{U_2U_2U_2\}       
&& 0.635( 08)
&& 0.686( 18)
&& 0.680( 12)
&& 0.659( 08)
&& 0.644( 11)
&& 0.683( 12)
&& 0.664( 08)
  &\cr\ifspace\myskip&&
&& 
&& 
&& 
&& 
&& 
&& 
&& 
  &\cr\fi\myskip\tlr
\myskip
&& \{U_2U_2U_3\}       
&& 0.622( 10)
&& 0.678( 22)
&& 0.673( 14)
&& 0.649( 10)
&& 
&& 
&& 
  &\cr\ifspace\myskip&&
&& 
&& 
&& 
&& 
&& 
&& 
&& 
  &\cr\fi\myskip\tlr
\myskip
&& \{U_2U_3U_3\}       
&& 0.609( 11)
&& 0.670( 27)
&& 0.667( 16)
&& 0.639( 11)
&& 
&& 
&& 
  &\cr\ifspace\myskip&&
&& 
&& 
&& 
&& 
&& 
&& 
&& 
  &\cr\fi\myskip\tlr
\myskip
&& \{U_3U_3U_3\}       
&& 0.596( 13)
&& 0.661( 33)
&& 0.662( 18)
&& 0.628( 14)
&& 0.612( 13)
&& 0.660( 22)
&& 0.636( 13)
  &\cr\ifspace\myskip&&
&& 
&& 
&& 
&& 
&& 
&& 
&& 
  &\cr\fi\myskip\tlr
\cr}}}}
 
 }
{Mass estimates for the decuplet baryons.}

We have also calculated the masses of spin-3/2 baryons.
Here there is only one type of operator, which is completely
symmetric in flavor. Our results for all 20 mass combinations
built out of $U_i$ and $S$ quarks are given in Table \nameuse\tdecuplet.


We analyze the baryon masses using the chiral expansion.
We first consider the spin-1/2 baryons, returning to spin-3/2 baryons later
in this section.
We couch the discussion in terms similar to those used in full QCD.
If we keep only constants and terms linear in quark masses,
(and drop non-analytic terms of the form $\delta m_q^{1/2}$
as discussed in section \nameuse\secquench),
then it is straightforward to show, using quenched chiral perturbation
theory \labrenzsharpe, that
\eqn\ebarymassquenched{\eqalign{
M(\Sigma^+) = M(S\{UU\}) 
	&= M_0 + 4 F m_u + 2 (F-D) m_s \,, \cr
M(\Lambda) = M(S[UU])  
	&= M_0 + 4(F-{2D \over 3})m_u + 2 (F+{D\over 3}) m_s \,, \cr
}}
Here $M_0$ is spin-1/2 baryon mass in the chiral limit,
and $F$ and $D$ are the usual reduced matrix elements of scalar densities.
Note that there is no dependence on $m_d$, since the $d$ quark does
not enter the correlators.
These results can also be obtained using mass perturbation theory in
full QCD, and then deleting terms proportional to $m_u+m_d+m_s$
which correspond to contributions from internal quark loops.
We stress that these formulae apply to all the states we consider---for
example, the proton mass is $M(D\{UU\})$, and is obtained by
replacing $m_s$ with $m_d$ in the formula for $M(\Sigma^+)$.

At this order in the chiral expansion, it is simple to extend the
results to baryons composed of three non-degenerate quarks.
The mass matrix in the $(\Sigma^0, \Lambda)=(S\{UD\},S[UD])$ basis is
\eqn\emassmatrix{ 
\left( \matrix{
\alpha & \gamma \cr
\gamma & \beta  \cr
}\right) 
 =
\left( \matrix{
M_0 + 4 F \mbar + 2 (F-D) m_s 	& 
		{2D \over \sqrt3} (m_u-m_d)   \cr
{2D \over \sqrt3} (m_u-m_d)  	& 
		M_0 + 4(F-{2D \over 3})\mbar + 2 (F+{D\over 3}) m_s \cr
}\right) 
}
Diagonalizing this matrix gives the eigenvalues $M_\pm$ 
with a mixing angle $\theta$.
If we assume that the same mixing angle applies for the interpolating
fields, which we think is true up to corrections of $O(m_q^{3/2})$,
then $\theta$ is the angle appearing in our previous
expressions for the $\Lambda$ correlator (Eq.~\etwostateC). 
The ``short-time effective mass'', Eq.~\etwostateM, is then
\eqn\emixdeflam{
M(\Lambda)_\eff \approx \cos^2\theta M_- + \sin^2\theta M_+ = \beta \,.
}
A similar argument shows that
\eqn\emixdefsig{
M(\Sigma)_\eff \approx \sin^2\theta M_- + \cos^2\theta M_+ = \alpha \,.
}
Thus we find the surprising result that the short-time effective
masses are insensitive to the isospin breaking term $\gamma$.
Furthermore, the expressions for $\alpha$ and $\beta$ are exactly
the same as the formulae applicable when isospin is unbroken, 
Eqs. \ebarymassquenched,
except that $m_u$ is replaced by the average mass, $\mbar$.

We have not extended this analysis to higher order in the chiral expansion.
Nevertheless, we use it as motivation for including our results for
baryons composed of completely non-degenerate quarks by assuming
that the effective masses satisfy
\eqn\enondegenfix{
M(A\{BC\}) = M(A\{DD\})\ {\rm and}\ M(A[BC]) = M(A[DD]) \,,
}
where $m_D = (m_B+m_C)/ 2$.
We make clear in the following where we are using this assumption 
and where not.

A large part of our analysis concerns mass splittings between,
for example, spin-1/2 baryons.
In most cases, we have extracted these differences by directly
fitting to the ratio of the appropriate correlators.
Thus, for example,
$$
{\Gamma_\Sigma(t) \over \Gamma_N(t) } \sim e^{-(M_\Sigma - M_N) t}, 
$$
and so the mass difference is obtained from a single fit. This has the
advantage of both reducing some of the systematic errors 
(e.g. those arising from excited state contamination)
and of improving the statistical errors. We find that
estimates obtained in this manner are consistent with those obtained 
from individual fits but have errors which are $3-5$ times smaller.  

\subsec{Spin-1/2 baryon mass splittings}

In this subsection we use our results for mass splittings between
spin-1/2 baryons to extract predictions for the physical mass splittings,
and to study the chiral behavior of baryon masses.
It turns out that the leading chiral prediction,
Eq.~\ebarymassquenched, gives a poor description of our data.
Higher order terms are essential in order to extrapolate reliably
from our quark masses, and have a significant impact on the final result.
The higher order terms we use are motivated by quenched chiral perturbation
theory---which predicts non-analytic terms of $O(m_q^{3/2})$ 
and analytic terms proportional to $m_q^2$.
The form of these terms, and some predictions for their coefficients, 
have been worked out in
Ref.~\ref\rSRSchiral{J. Labrenz and S. Sharpe, in preparation.}. 
We attempt only a qualitative comparison with these predictions---though
we provide the results of our fits so that others can pursue this further.

%
%

We first consider the octet hyperfine splitting,
``$\Sigma-\Lambda$'', 
\eqn\esigmlam{\eqalign{
{M_{A\{BB\}} - M_{A[BB]}   \over m_A - m_B} &=
(-8D/3) + d_1 { M_{AB}^3 - M_{BB}^3  \over m_A - m_B}
+ d'_1 {M_{AA}^3 - M_{BB}^3 \over m_A-m_B} \cr
&\ + e_1 m_B + e'_1 (m_A+m_B) \,, \cr
}}
where $M_{AB}$ is the mass of the pion with flavor $\overline{A}B$, etc.
The constants $d_i$ and $e_i$ can be expressed in terms of parameters of
the quenched chiral Lagrangian.
For reasonable choices of these parameters, one expects $|d'_1|\ll |d_1|$.
There is no useful information concerning $e_1$ or $e'_1$.

\figure\fsiglamchiral{\epsfysize=4in\epsfbox{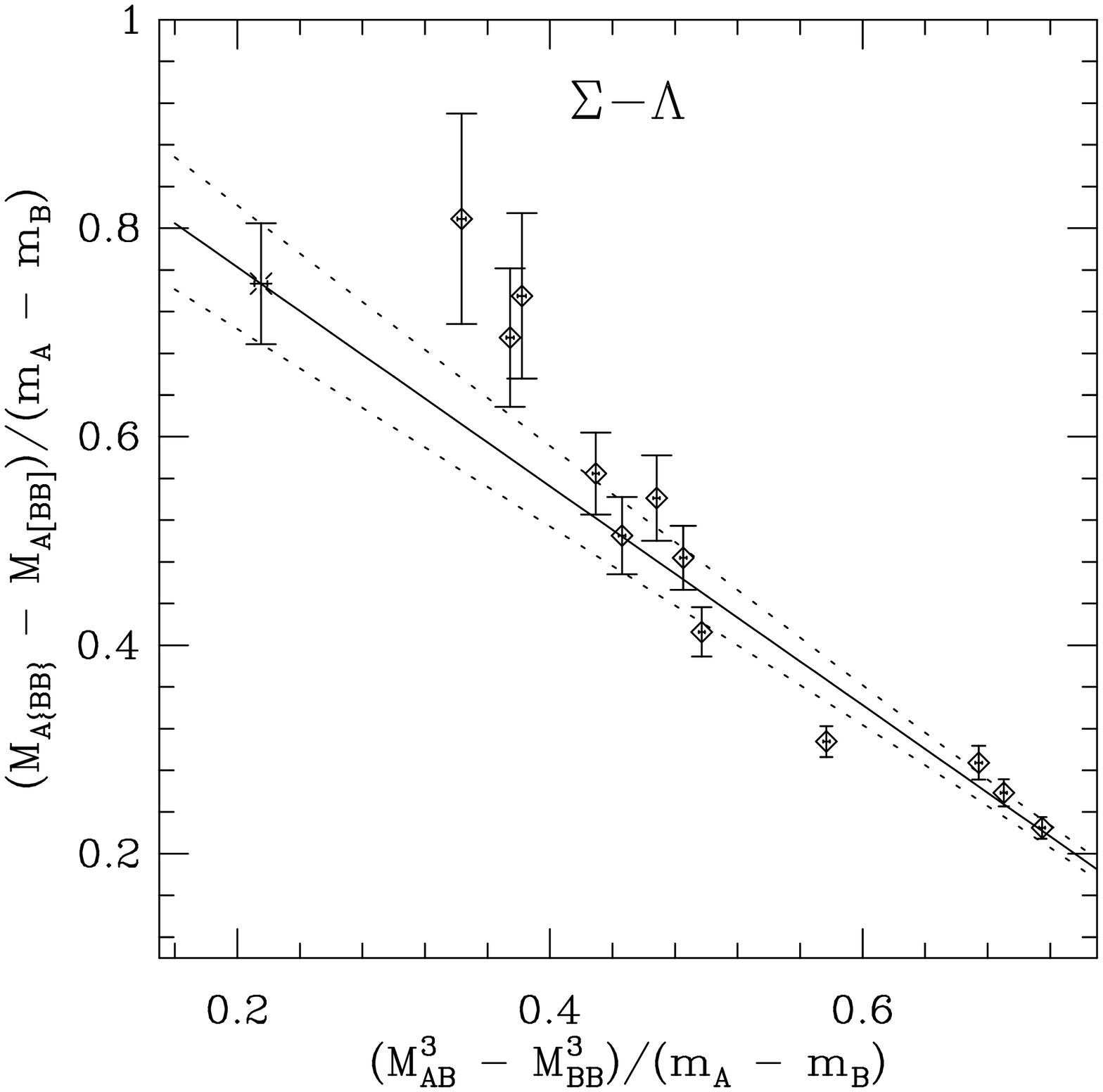}}
{\vtop{\advance\hsize by -2\parindent \noindent 
Test for chiral corrections in $M_\Sigma - M_\Lambda$.
The value extrapolated to the physical point is 
shown by the burst symbol at the extreme left.
In this, and subsequent, figures all masses are in lattice units.}}

We fit our results in two ways. First, we assume that the $d_1$ term
is dominant, and plot our data versus $(M_{AB}^3 - M_{BB}^3)/(m_A - m_B)$.
The outcome is shown in Fig.~\nameuse\fsiglamchiral.
The data should collapse onto a single curve, which should be linear,
and our results are reasonably consistent with this.
What is particularly striking, however, is the size of the slope.
This is a clear sign that terms of higher order than linear in the quark
mass are needed to describe our baryon masses---recall that the linear
term has been divided out in this fit.
If we extrapolate linearly to the physical point ($m_A=m_s$, $m_B=\mbar$)
we find (the fit gives $D=-0.36(3)$, $d_1=-0.19(2)\ \GeV^{-2}$)
\eqn\eDlamsigQ{
M_\Sigma - M_\Lambda =  76(7) \ \MeV \,.
}
It is conventional to quote this result in terms of an effective
$D$ parameter
\eqn\edresult{
D_\eff \equiv {3(M_\Sigma - M_\Lambda) \over 8 (m_s-\mbar)}
\ =\ {-29(3) \MeV \over m_s-\mbar} \ =\ -0.28(3) \,.
}
Here (and in the similar estimates of $D_\eff$ and $F_\eff$ below),
we use the non-perturbative definition of the $\MSbar$ quark mass 
calculated in $\MeV$, but depart from our usual practice by
evaluating it at the scale $2 \GeV$ (instead of $1/a$).
Note that $D_\eff$ differs from the $D$ appearing in
chiral expansions such as Eq. \nameuse\esigmlam, for it absorbs
some of the higher order terms.

\figure\fsiglamdiff{\epsfysize=4in\epsfbox{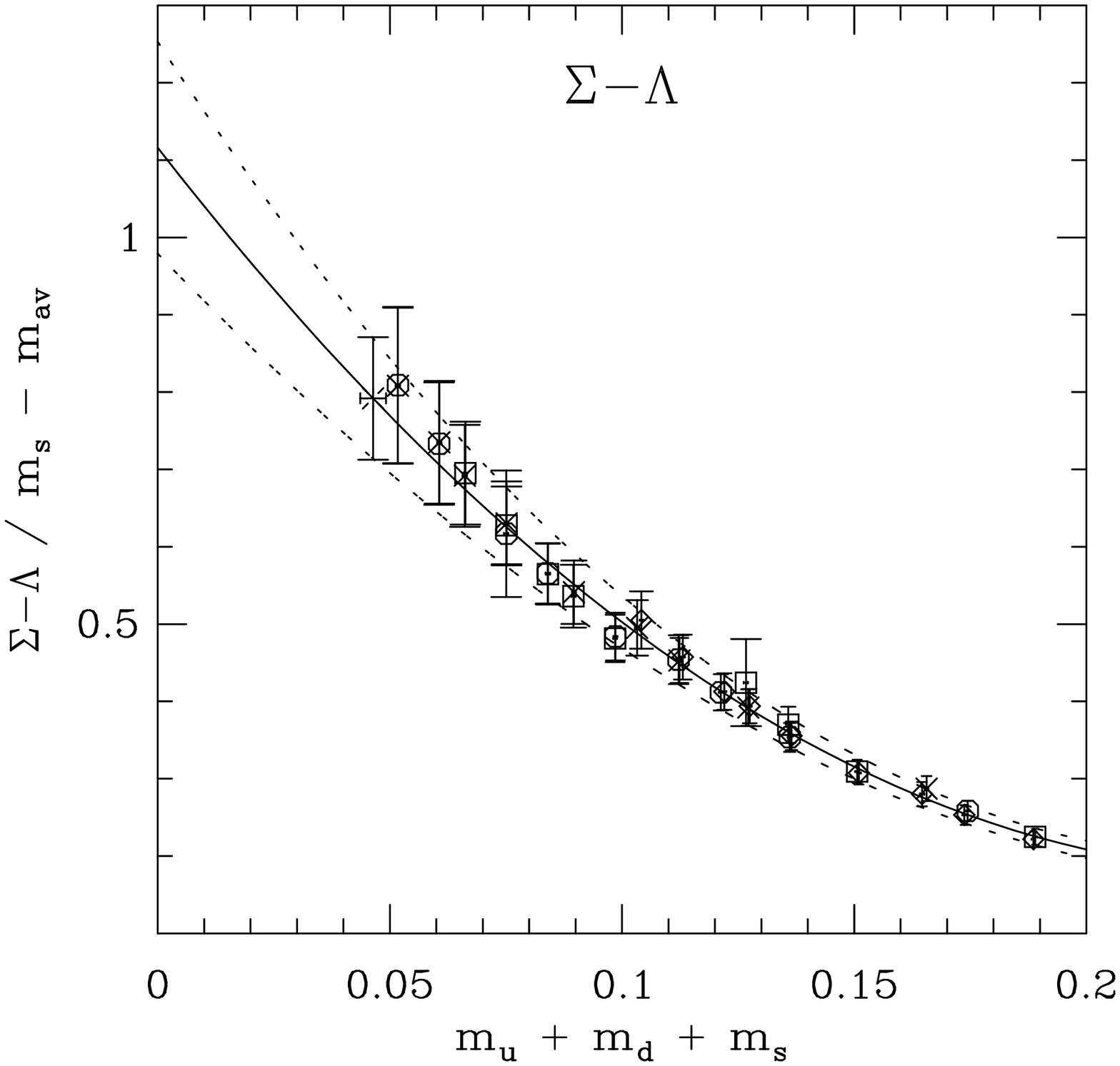}}
{\vtop{\advance\hsize by -2\parindent \noindent 
Quadratic fit to $M_\Sigma - M_\Lambda / (\mbar - m_s) $,
including baryons composed of completely non-degenerate quarks.
The value extrapolated to the physical point is 
shown by the burst symbol at the extreme left.}}

In our second fit we test to see whether our data can be 
represented as well by the analytic terms. We have found, by trial and error,
that Eq.~\esigmlam\ with $d_1=d'_1=0$ and $e_1=e'_1$ does a reasonable job;
the plot is shown in Fig.~\nameuse\fsiglamdiff.
In this plot we have included the results for baryons
composed of completely non-degenerate quarks, based on the discussion 
following Eqs.~\nameuse\emixdeflam\ and \nameuse\emixdefsig.
To emphasize this, we have labeled the
x-axis by $m_u+m_d+m_s$ instead of $m_A+2 m_B$.
The data show definite curvature, so we have extrapolated to the physical
point using a quadratic fit (corresponding to terms up to $m_q^3$ in the
original expression for baryon masses!).
The fit yields
\eqn\eDlamsig{M_\Sigma - M_\Lambda =  80(8) \ \MeV \,,\quad
D_\eff \ =\ {-30(3) \MeV  \over (m_s - \mbar)}  = -0.29(3)  \,.
}
These results are consistent with those from the ``$m_q^{3/2}$'' fit.

\figure\fsignucchiral{\epsfysize=4in\epsfbox{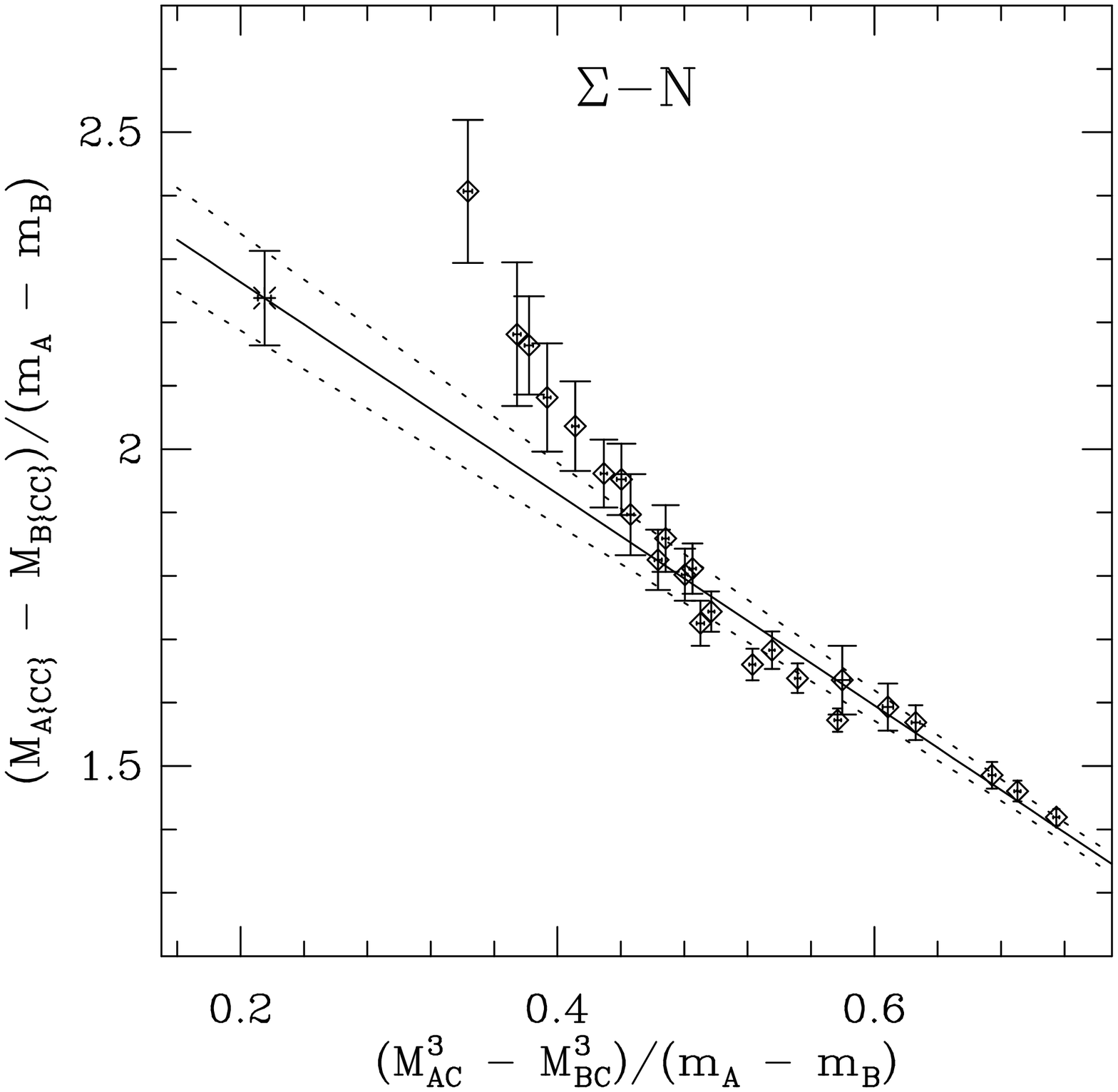}}
{\vtop{\advance\hsize by -2\parindent \noindent 
Test for chiral corrections in $M_\Sigma - M_N$ as described in 
Eq.~\nameuse\esignuc. The physical point is 
shown by the burst symbol at the extreme left.}}

\figure\fsignucdiff{\epsfysize=4in\epsfbox{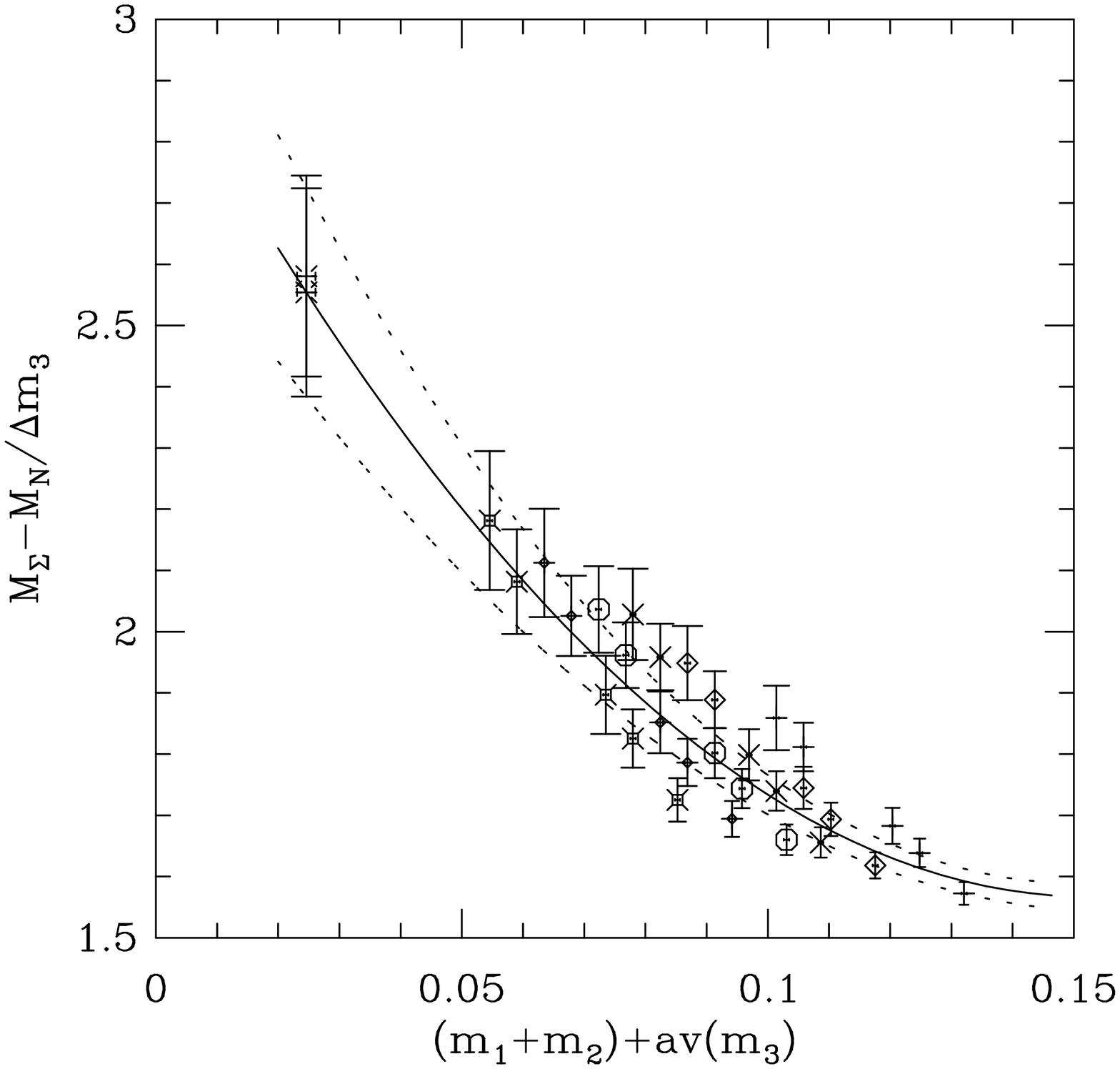}}
{\vtop{\advance\hsize by -2\parindent \noindent 
Quadratic fit to $M_\Sigma - M_N / (m_3^{\Sigma} - m_3^N) $ for the 30
quark combinations. The result of the two ways of extrapolating 
to the physical value described in the text are shown by the burst symbols.}}

Next we consider the ``$\Sigma-N$'' splitting, i.e that between
two $\Sigma$-like states having different strange quark masses.
The chiral expansion is
\eqn\esignuc{\eqalign{
{M_{A\{CC\}} - M_{B\{CC\}} \over m_A - m_B} &= 2(F-D)
+ d_2 { M_{AC}^3 - M_{BC}^3  \over m_A - m_B}
+ d'_2 {M_{AA}^3 - M_{BB}^3  \over m_A - m_B} \cr
&\ + e_2 m_C + e'_2 (m_A + m_B) \,.\cr
}}
Chiral perturbation theory suggests that $|d'_2| < |d_2|$.
Thus we first plot the data assuming $d_2$ is the dominant coefficient
(Fig.~\nameuse\fsignucchiral).
The collapse onto a single curve is reasonable, and a linear fit 
($F-D=1.30(5)$, $d_2=-0.31(3) \ \GeV^{-2}$) gives
\eqn\eFDsignucQ{
M_\Sigma - M_N =  227(14) \ \MeV \,,\quad
F_\eff-D_\eff \ =\ {114(7) \MeV \over (m_s-\mbar)} \ =\ 1.11(7)  \,.
}

We also have investigated various analytic fits, none of
which do a good job of collapsing the data onto a single curve.
Our best attempt, shown in Fig.~\nameuse\fsignucdiff,
assumes $d_2=d'_2=0$ and $e_2= e'_2/4$,
i.e. we plot against the average mass of the quarks in the two baryons.
We again include non-degenerate baryons in this plot.
Since the collapse is not good we use two different extrapolations to the
physical point. The first is a quadratic fit in the average mass as
shown in Fig.~\nameuse\fsignucdiff.
In the second we first extrapolate linearly to $m_B=m_C=\mbar$ 
and then interpolate to $m_A=m_s$.
These two methods give almost identical results,
as shown by the two ``bursts'' in the plot.
The mean of these two points gives
\eqn\eFDsignuc{
M_\Sigma - M_N =  260(20) \ \MeV \,,\quad
F_\eff-D_\eff
\ =\ {130(10)\MeV \over (m_s-\mbar)} \ =\ 1.27(10)   \,.
}
The difference between the two estimates, Eqs.~\eFDsignucQ\ and
\eFDsignuc, is indicative of the fact that neither of the forms
fits all the data well.

\figure\fcasnucchiral{\epsfysize=4in\epsfbox{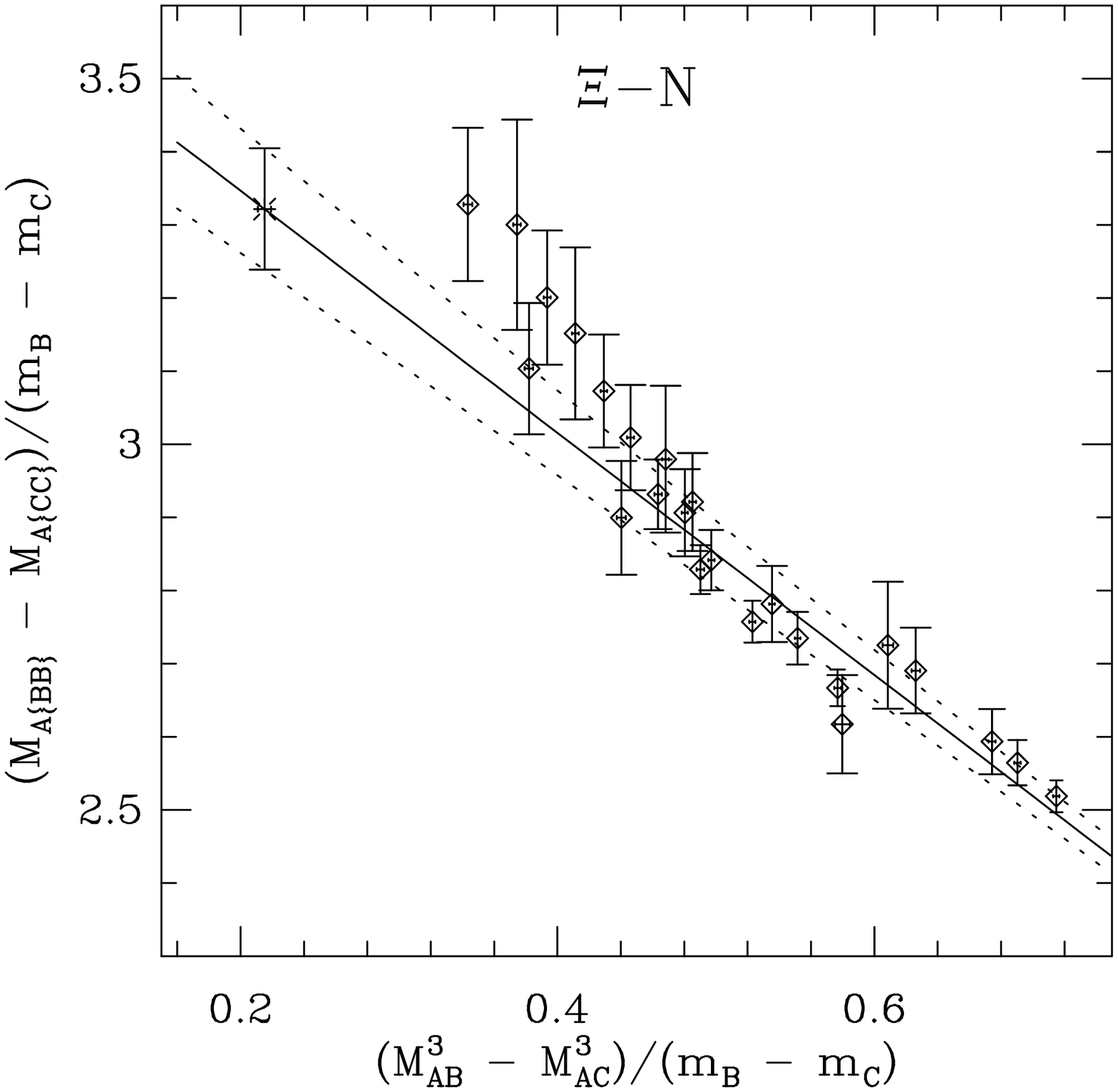}}
{\vtop{\advance\hsize by -2\parindent \noindent 
Test for chiral corrections in $M_\Xi - M_N$ as described in 
Eq.~\nameuse\eximnuc.
The physical point is 
shown by the burst symbol at the extreme left.}}

\figure\fcasnucdiff{\epsfysize=4in\epsfbox{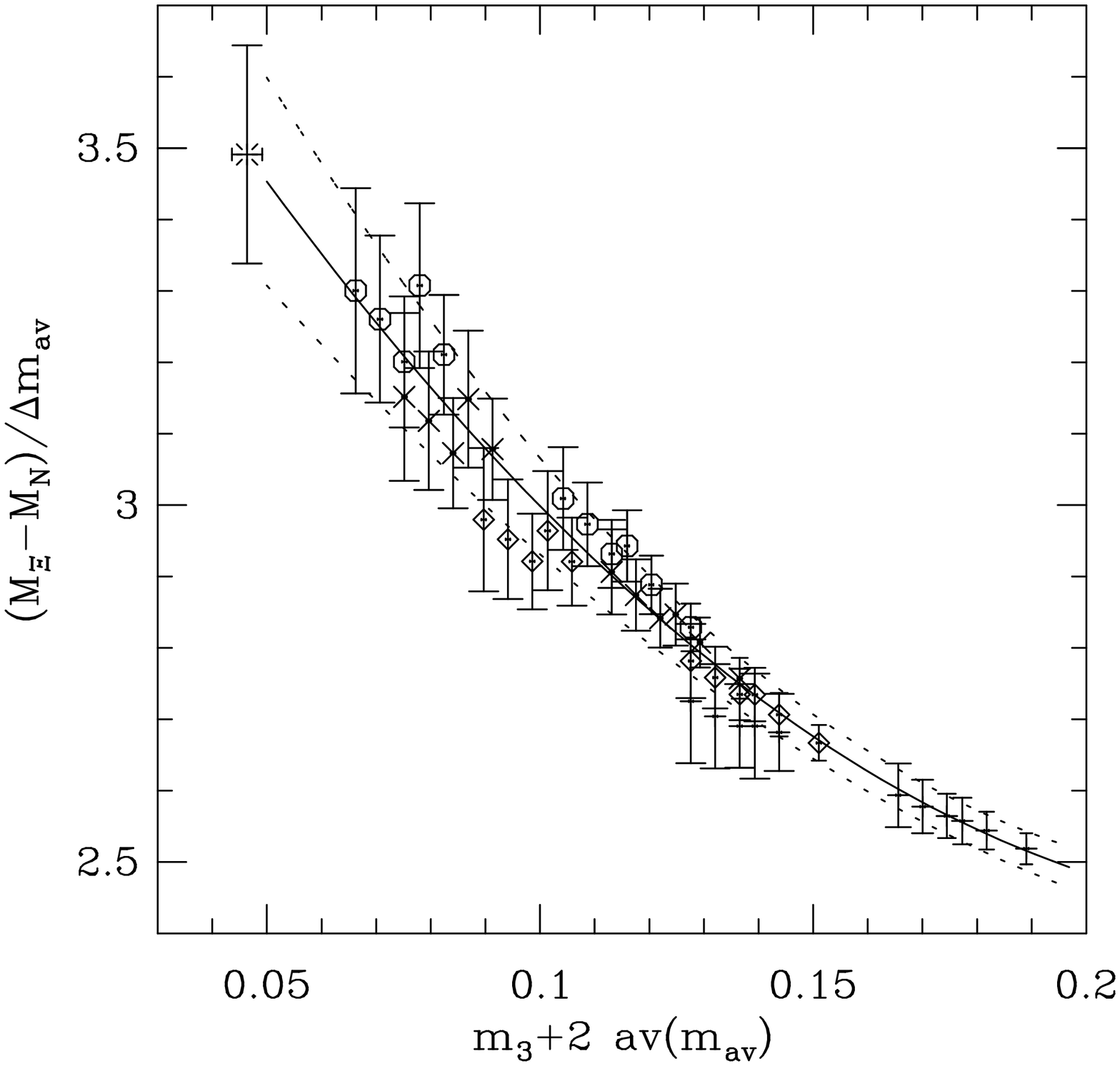}}
{\vtop{\advance\hsize by -2\parindent \noindent Quadratic fit to $M_\Xi
- M_N / (m_{av}^{\Xi} - m_{av}^N) $ for 44 quark combinations. The
result of the quadratic extrapolation to the physical value is shown
by the burst symbol.}}

Thirdly, we consider the difference ``$\Xi-N$'',
in which we study the effect of changing the mass of the two
symmetrized quarks. We expect
\eqn\eximnuc{\eqalign{
{M_{A\{BB\}} - M_{A\{CC\}} \over m_B - m_C} &= 4F
+ d_3 { M_{AB}^3 - M_{AC}^3  \over m_B - m_C}
+ d'_3{ M_{BB}^3 - M_{CC}^3 \over  m_B - m_C} \cr
&\ + e_3 m_A + e'_3 (m_B + m_C) \,. \cr
}}
It turns out that in quenched chiral perturbation theory we expect
$d_3=d_2$ and $e_3=e_2$.
In this case, there is no expectation that $d_3$ and $d'_3$ should be
substantially different in magnitude.
Nevertheless, for simplicity, we first plot the data
assuming $d_3$ is the dominant coefficient---as shown in 
Fig.~\nameuse\fcasnucchiral.
The data collapses reasonably onto a single curve, and a linear fit 
($F=0.92(3)$, $d_3=-0.30(3) \ \GeV^{-2}$) yields
\eqn\eFcasnuc{
M_\Xi - M_N =  338(19) \ \MeV \,,\quad
  F_\eff \ =\ {84(5) \MeV \over (m_s-\mbar)} \ =\  0.82(5)   \,.
}
A quadratic fit to the average quark mass 
(i.e. assuming $e_3=e'_3$) is slightly better,
as shown in  Fig.~\nameuse\fcasnucdiff.  The results are 
nevertheless consistent with those from the first fit:
\eqn\eFcasnuc{
M_\Xi - M_N =  355(21) \ \MeV \,,\quad
F_\eff \ =\ {89(5)\MeV \over (m_s-\mbar)} \ =\  0.87(5)   \,.
}
It is important to note that, once again,
higher order corrections in the chiral expansion are substantial.

To further investigate the mass splittings we have considered
the combination of differences appearing in the Gell-mann--Okubo formula,
$3 M_\Lambda + M_\Sigma - 2 M_N - 2 M_\Xi =0$.
In our notation, this combination is
\eqn\egmo{
\GMO(A,B) = 3 M_{A[BB]} + M_{A\{BB\}} - 2 M_{B\{BB\}} - 2 M_{B\{AA\}} \,.
}
It is constructed to cancel the $O(m_q)$ terms,
\eqn\egmochpt{
\GMO(A,B) = d_4 (M_{AA}^3 - 2M_{AB}^3 + M_{BB}^3)
	   + e_4 (m_A - m_B)^2 + \cdots \,,
}
making it an interesting window on higher order terms.
We have calculated $\GMO(A,B)$ using three methods:
\item{(1)} simply taking the differences of the masses;
\item{(2)} using
\eqn\egmomethodii{\eqalign{
\GMO(A,B) &= 4[M_{A\{BB\}} - M_{B\{BB\}}] - 3 [M_{A\{BB\}} - M_{A[BB]}] \cr
&\quad  - 2[M_{B\{AA\}}-M_{B\{BB\}}] \,, \cr
}}
where the differences within square brackest
are calculated from ratios of correlators; and 
\item{(3)} using another combination of differences (calculated using ratios)
\eqn\egmomethodiii{\eqalign{
\GMO(A,B) &= [M_{A\{BB\}} - M_{B\{BB\}}] + 3[M_{A[BB]} - M_{B\{BB\}}] \cr
&\quad - 2[M_{B\{AA\}}-M_{B\{BB\}}] \,. \cr
}}

\smallskip \noindent
Our results are given in Table~\nameuse\tGMOkubo.
They are consistent with zero for all quark mass combinations,
showing that higher order chiral corrections are not uniformly large.
This is in qualitative agreement with experiment, where the GMO relation
works well: $\GMO=26 \MeV$, or $\GMO=0.011$ in lattice units.
It is difficult to make a quantitative comparison, since we cannot
extrapolate our data to the physical point. We note, however, that
our results for the largest mass splitting, $A/B=S/U_3$, 
are consistent with the experimental value.

\table\tGMOkubo{
\let\ifspace=\iffalse
\def\myskip{\omit&height1.5pt&%
\omit&&%
\omit&&%
\omit&&%
\omit&&%
 &\cr}
\vbox{\hbox{\vbox{
\tabskip=0pt\offinterlineskip
\def\tlr{\noalign{\hrule}}

\halign {\strut#& \vrule\vrule#\tabskip=3pt&
  \hfil$#$\hfil&\vrule#&
  \hfil$#$\hfil&\vrule#&
  \hfil$#$\hfil&\vrule#&
  \hfil$#$\hfil&\vrule#&
  \hfil$#$\hfil&\vrule#\tabskip=0pt\cr\tlr
\myskip
&& A  && B
&& {\rm Method \ 1}
&& {\rm Method \ 2}
&& {\rm Method \ 3}
  &\cr
\myskip\tlr
\omit&height0.5pt&\multispan{ 9                            }&\cr\tlr
%
&& S && U_1
&& 0.000( 01)
&& 0.002( 01)
&& 0.001( 01)
  &\cr\ifspace\myskip&&
&& 
&& 
&& 
  &\cr\fi\myskip\tlr
\myskip
&& S && U_2  
&& 0.003( 03)
&& 0.003( 03)
&& 0.003( 02)
  &\cr\ifspace\myskip&&
&& 
&& 
&& 
  &\cr\fi\myskip\tlr
\myskip
&& S && U_3  
&& 0.008( 08)
&& 0.002( 09)
&& 0.005( 04)
  &\cr\ifspace\myskip&&
&& 
&& 
&& 
  &\cr\fi\myskip\tlr
\myskip
&& U_1 && S
&& 0.002( 02)
&& 
&& 
  &\cr\ifspace\myskip&&
&& 
&& 
&& 
  &\cr\fi\myskip\tlr
\myskip
&& U_1 && U_2
&& 0.001( 01)
&& 0.000( 02)
&& 0.000( 01)
  &\cr\ifspace\myskip&&
&& 
&& 
&& 
  &\cr\fi\myskip\tlr
\myskip
&& U_1 && U_3
&& 0.003( 05)
&& 0.000( 06)
&& 0.002( 02)
  &\cr\ifspace\myskip&&
&& 
&& 
&& 
  &\cr\fi\myskip\tlr
\myskip
&& U_2 && S
&& 0.005( 02)
&& 
&& 
  &\cr\ifspace\myskip&&
&& 
&& 
&& 
  &\cr\fi\myskip\tlr
\myskip
&& U_2 && U_1  
&& 0.001( 01)
&& 0.001( 01)
&& 
  &\cr\ifspace\myskip&&
&& 
&& 
&& 
  &\cr\fi\myskip\tlr
\myskip
&& U_2 && U_3
&& -.001( 04)
&& 
&& 
  &\cr\ifspace\myskip&&
&& 
&& 
&& 
  &\cr\fi\myskip\tlr
\myskip
&& U_3 && S
&& 0.009( 05)
&& 
&& 
  &\cr\ifspace\myskip&&
&& 
&& 
&& 
  &\cr\fi\myskip\tlr
\myskip
&& U_3 && U_1  
&& 0.005( 04)
&& 0.003( 02)
&& 
  &\cr\ifspace\myskip&&
&& 
&& 
&& 
  &\cr\fi\myskip\tlr
\myskip
&& U_3 && U_2
&& 0.003( 04)
&& 
&& 
  &\cr\ifspace\myskip&&
&& 
&& 
&& 
  &\cr\fi\myskip\tlr
\cr}}}}

     }
{\vtop{\advance\hsize by -2\parindent 
\noindent
Tests of the Gell-mann--Okubo mass formula:
results for $\GMO(A,B)$.}}

\table\toctetdiff{
\vbox{\hbox{\indent\vbox{\tabskip=0pt\offinterlineskip
\def\myskip{\omit&height1pt& && && && && &\cr}
\halign {\strut#& \vrule#\tabskip=2pt&
\hfil$#$\hfil&\vrule#&
\hfil$#$&\vrule#&
\hfil$#$&\vrule#&
\hfil$#$&\vrule#&
\hfil$#$&\vrule#\tabskip=0pt\cr\noalign{\hrule}
%
%
\myskip\myskip
&& \hbox{Fit:}
&& m_q^{3/2} \hfil
&& m_q^2 \hfil
&& \hbox{Eq.~\ebarymassquenched}\hfil
&& \hbox{Expt.}\hfil
& \cr
\myskip\myskip
\noalign{\hrule}
%
%
\myskip
&& M_{\Sigma} - M_{N}      && 227(13*) && 260(20*) && 194(12*)  &&  253 &\cr
\myskip			                        
\noalign{\hrule}	                        
\myskip			                        
&& M_{\Xi   } - M_{N}      && 338(19*) && 355(21*) && 306(22*)  &&  375 &\cr
\myskip			                        
\noalign{\hrule}	                        
\myskip			                        
&& M_{\Xi   } - M_{\Sigma} && 107(9*)  && 109(8*)  && 112(14*)  &&  122 &\cr
\myskip			                        
\noalign{\hrule}	                        
\myskip			                        
&& M_{\Sigma} - M_\Lambda  && 76(7*)   && 80(8*)   &&           &&  77 &\cr
\myskip
\noalign{\hrule}
%
%
\crcr}}}}
 
   }
{\vtop{\advance\hsize by -2\parindent 
\noindent 
Estimates of mass splittings in the baryon octet,
using various fits explained in the text.
Experimental results are given for comparison.
All results are in MeV.
}}

A summary of mass-splittings in the octet multiplet is given in
Table~\nameuse\toctetdiff. Several comments are in order.
First, the ``$m_q^{3/2}$'' fits show evidence for curvature which,
if included, would likely make the results agree more closely with
those from the ``$m_q^2$'' fits.
Second, we reiterate that the data is extremely poorly
represented by the first order mass formulae, Eq. \ebarymassquenched---this
would predict that all the curves in Figs. 
\nameuse\fsiglamchiral-\nameuse\fcasnucdiff\ are flat.
If, nevertheless, we fit to the linear terms, we find the results 
listed in the third column.
Finally, we note the good agreement of the results from the $m_q^2$ fit
(which are our most reliable) with the experimental splittings.

To further investigate the applicability of quenched chiral perturbation
theory we have constructed several other quantities.
Although most of these are peculiar to the quenched approximation,
having no counterpart in QCD, they allow us to see how well we
understand the extrapolations that are needed for all quantities,
including those which are physically relevant.
\item{1.}
We begin with the double difference
\eqn\eXsig{\eqalign{
{X_\Sigma(A,B,C) \over m_A - m_B }&= 
        {M_{A\{CC\}} - M_{B\{CC\}} - M_{C\{AA\}} + M_{C\{BB\}} 
	\over m_A-m_B} \cr
&= -2 (F+D) + (d'_2-d'_3) {M_{AA}^3-M_{BB}^3\over m_A-m_B}
	+ (e'_2-e'_3) (m_A+m_B) \,.\cr 
}}
This is interesting for several reasons. First, 
the expected chiral form is simpler than the differences considered above.
Second, for $C=B$ or $C=A$, $X_\Sigma$ reduces to
$M(A\{BB\}) - M(B\{AA\})$, which for $A=s$ and $B=u$ is 
$m_\Xi-m_\Sigma$. Third, $X_\Sigma$ is predicted to be independent
of $m_C$. Strictly speaking, this is true up to quenched artifacts
proportional to $\delta$. Thus the $m_C$ dependence of $X_\Sigma$ is a
window onto such artifacts.
Our data for $X_\Sigma$, plotted in Fig.~\nameuse\fXs, confirms
our expectations. There is no significant dependence on $m_C$---indeed,
there is barely any dependence on $m_A$ or $m_B$ either.
The fit yields $F+D = 0.50(6)$ and $d'_2-d'_3= -0.013(16) \ \GeV^{-2}$ 
assuming $e'_2-e'_3=0$. Clearly we could just as well fit with the $e'$ terms.
If we extrapolate to the physical point (assuming $M_{ss}^2= 2 M_K^2-M_\pi^2$),
we find
\eqn\exisig{
M_\Xi - M_\Sigma =  107(9) \ \MeV \,,\quad
(F_\eff+D_\eff) \ =\ {53(5) \MeV \over (m_s-\mbar)} \ =\  0.52(5)   \,.
}
consistent with the other estimates given in Table~\nameuse\toctetdiff. 
It is noteworthy that the extrapolation required is minimal,
unlike that for most of the mass differences considered above.

\figure\fXs{\epsfysize=4in\epsfbox{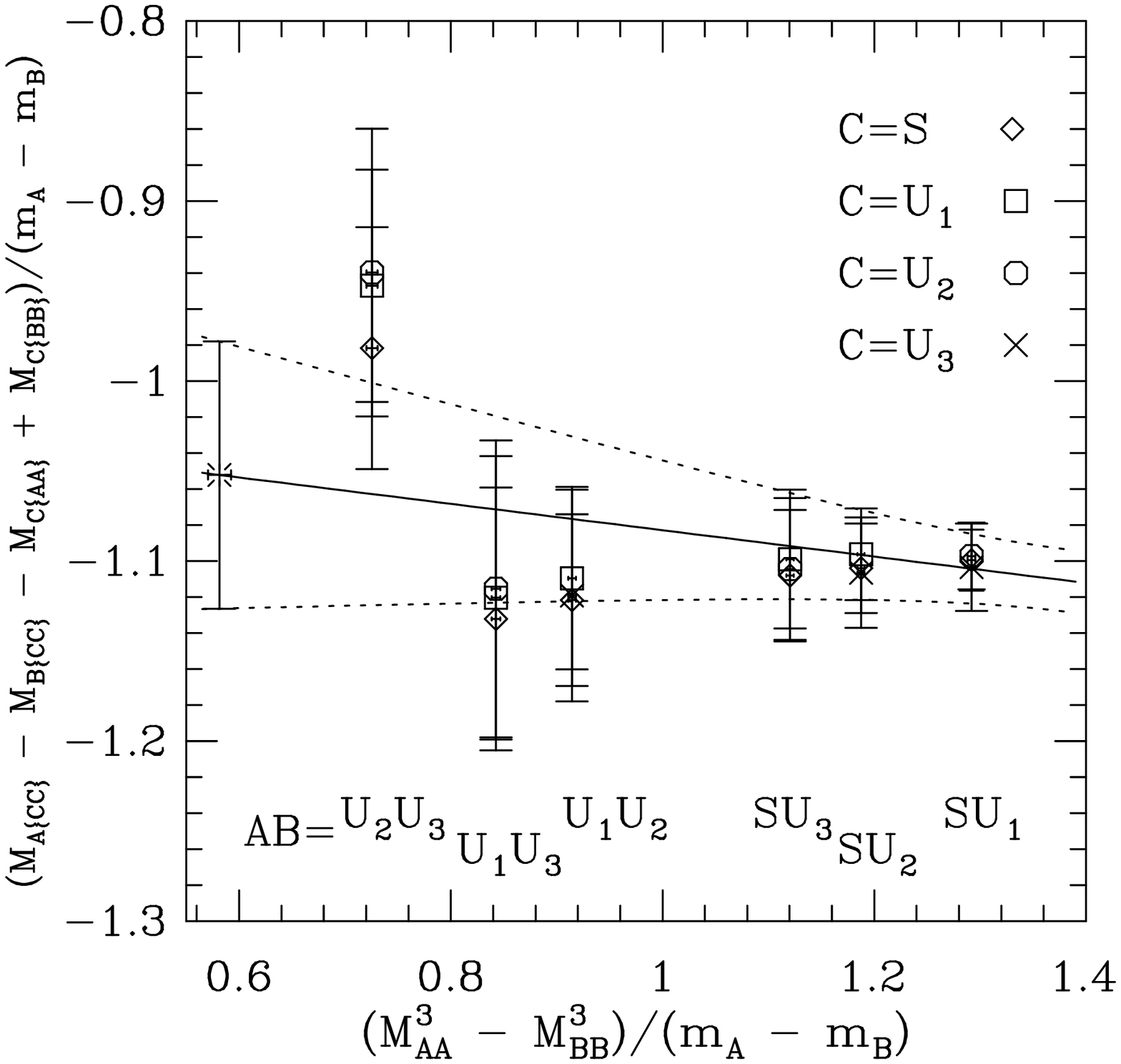}}
{\vtop{\advance\hsize by -2\parindent \noindent 
The quantity $X_\Sigma$ determined from fits to the ratio of correlators. The 
extrapolated value from which $M_\Xi -  M_\Sigma$ is calculated is given by 
the burst symbol at the left.}}

\item{2.}
We next consider the corresponding quantity for the $\Lambda$-like baryons,
\eqn\eXlam{\eqalign{
{X_\Lambda(A,B,C) \over m_A - m_B} &= 
        {M_{A[CC]} - M_{B[CC]} - M_{C[AA]} + M_{C[BB]} \over m_A-m_B} \cr
&= -2F+ 10 D/3 + d_5 {M_{AA}^3-M_{BB}^3\over m_A-m_B}
	+ e_5 (m_A+m_B) \,.\cr
}}
This always involves a quenched particle with no correspondent in
nature, e.g. $M(U[SS])$. We plot $X_\Lambda$ in Fig.~\nameuse\fXl.
Again the expectation of no dependence on $m_C$ is borne out, but this
time there are significant higher order terms.  The fit yields $2F -
10 D/3 = 3.4(3) $ and $d_5=0.26(4) \ \GeV^{-2}$, assuming $e_5=0$.
The data is equally well described using a $e_5$ term with $d_5=0$.

\figure\fXl{\epsfysize=4in\epsfbox{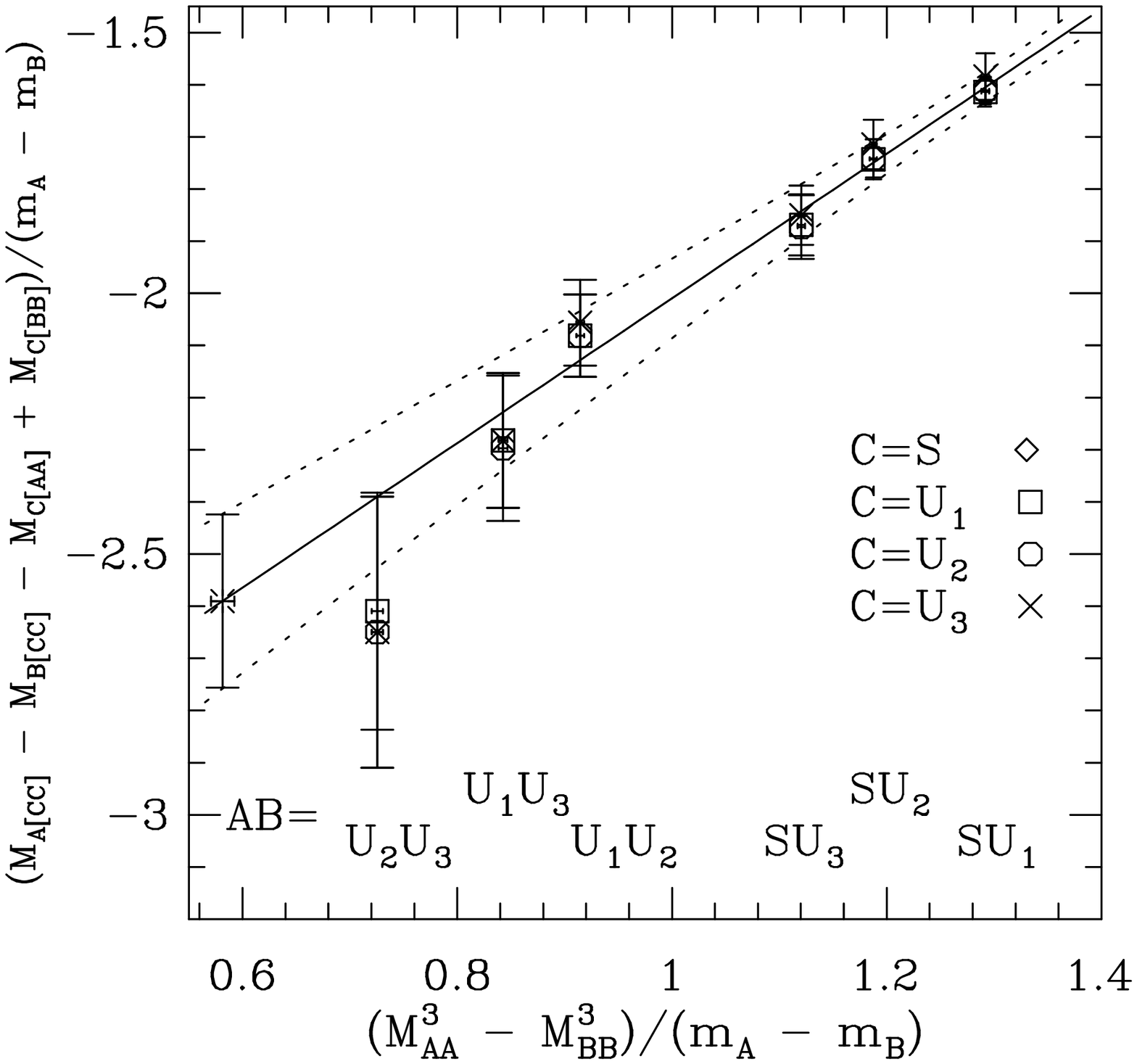}}
{\vtop{\advance\hsize by -2\parindent \noindent 
The quantity $X_\Lambda$ determined from the masses of the individual
states. The point for physical quark masses is the shown
with the burst symbol at the left.
}}

\item{3.}
In the quenched approximation, we can form two additional double
differences, analogous to $\GMO(A,B)$, in which the $O(m_q)$ terms
cancel
\eqn\eYs{\eqalign{
Y_\Sigma(A,B,C,D) &= 
        M_{A\{CC\}} - M_{B\{CC\}} - M_{A\{DD\}} + M_{B\{DD\}} \,,\cr
Y_\Lambda(A,B,C,D) &= 
        M_{A[CC]} - M_{B[CC]} - M_{A[DD]} + M_{B[DD]} \,.\cr
}}
$Y_\Sigma$ picks out the $d'_2$ and $e'_2$ terms in Eqs.~\esignuc\ and \eXsig.
The predicted form is
\eqn\eYsig{
{Y_\Sigma(A,B,C,D) \over M_{AC}^3 - M_{BC}^3 - M_{AD}^3 + M_{BD}^3}
= d_2 
 + e_2 {(m_A-m_B)(m_C-m_D)\over M_{AC}^3 - M_{BC}^3 - M_{AD}^3 + M_{BD}^3}
 \,.
}
We test this in Fig.~\nameuse\fYs.
What is most noteworthy is that, unlike the GMO relation, there are
significant higher order corrections.
We find a reasonable fit if $d_2\approx -5.4(1.4) \ \GeV^{-2}$ 
and $e_2\approx 44(11) \ \GeV^{-1}$.
The large errors in the fit parameters are due to the fact that
the two variables $M_{AC}^3 - M_{BC}^3 - M_{AD}^3 + M_{BD}^3$
and $(m_A-m_B)(m_C-m_D)$ are nearly proportional for our set of quark masses.
There is thus a significant cancellation between the two terms in the fit.
Indeed, we can obtain as good a fit setting $e_2=0$,
and replacing it with a higher order term proportional to
$m_A+m_B+m_C+m_D$.

\figure\fYs{\epsfysize=4in\epsfbox{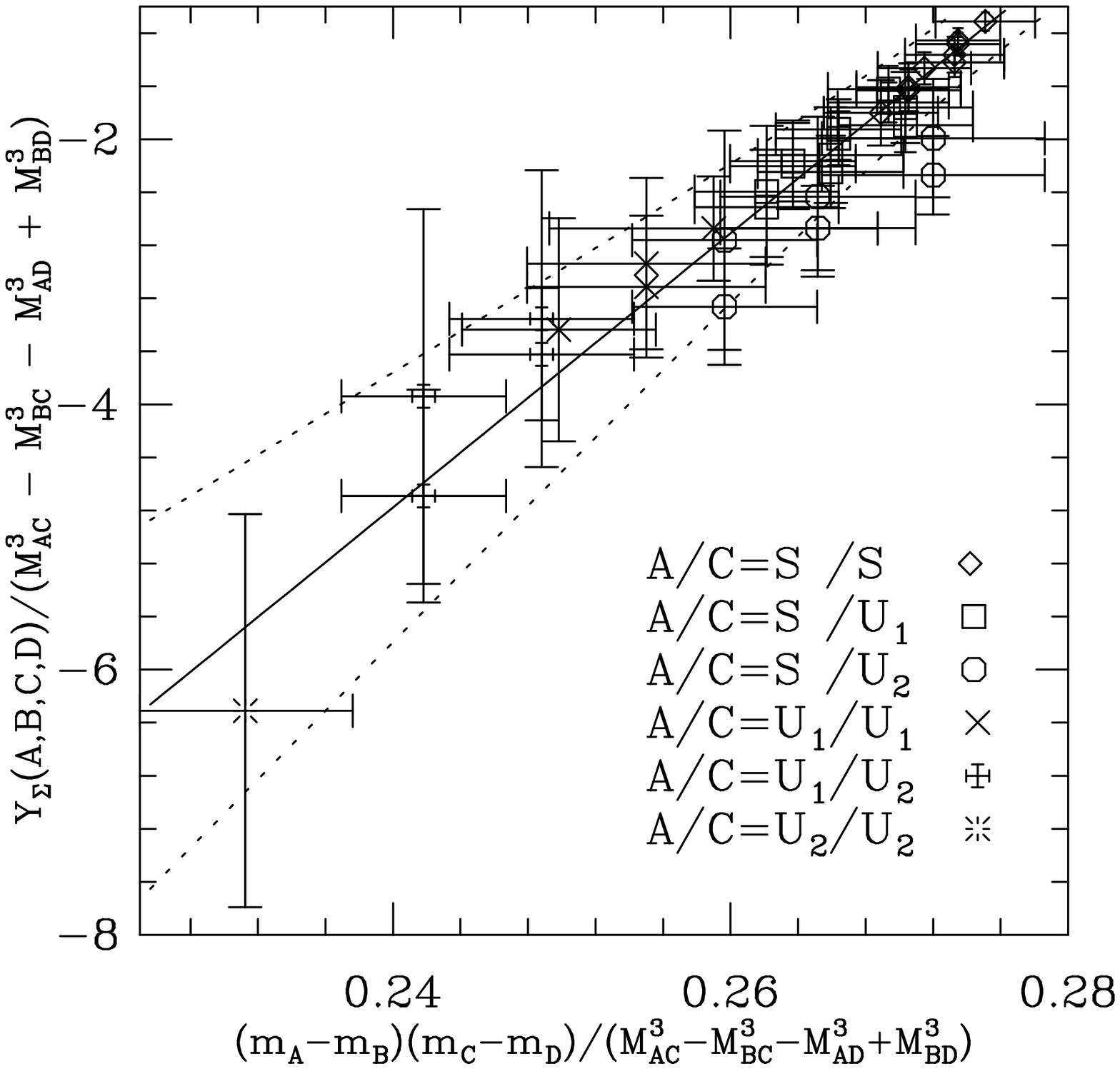}}
{\vtop{\advance\hsize by -2\parindent \noindent 
Testing the mass dependence of $Y_\Sigma$
(itself determined from fits to the ratio of correlators).
}}

\item{4.}
Finally, we consider $Y_\Lambda$, which is expected
to have the same functional form as $Y_\Sigma$.
The data, shown in Fig.~\nameuse\fYl, is again consistent with
the expected form, although the errors are larger since $Y_\Lambda$ 
is calculated from masses instead of mass differences. 
The fit gives $-4.3(2.4) \ \GeV^{-2}$ for the intercept 
and $35(20) \ \GeV^{-1}$ for the slope.

\figure\fYl{\epsfysize=4in\epsfbox{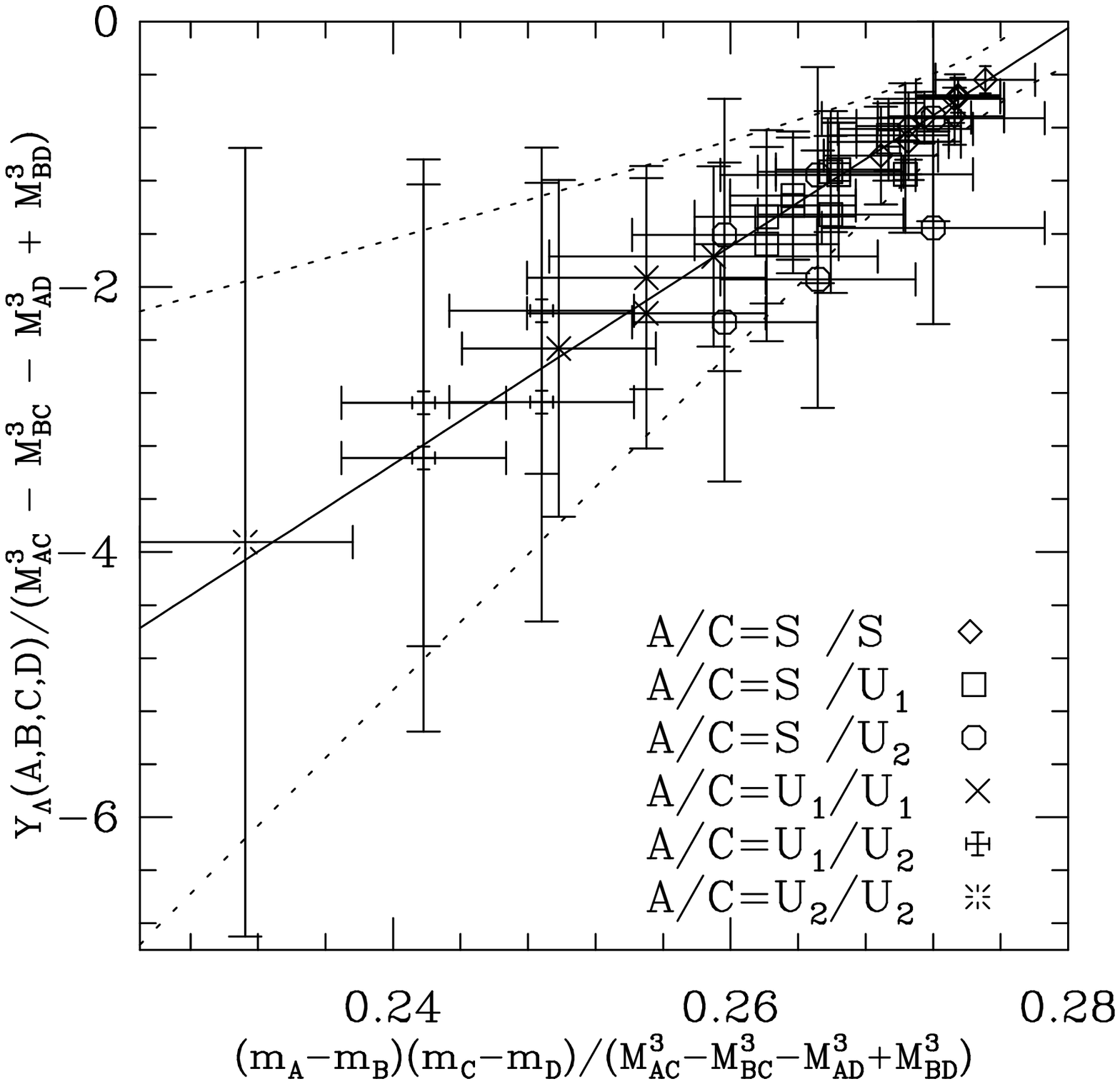}}
{\vtop{\advance\hsize by -2\parindent \noindent 
The quantity $Y_\Lambda$ determined from the masses of the individual states.
}}

\figure\fnucvsmnp{\epsfysize=4in\epsfbox{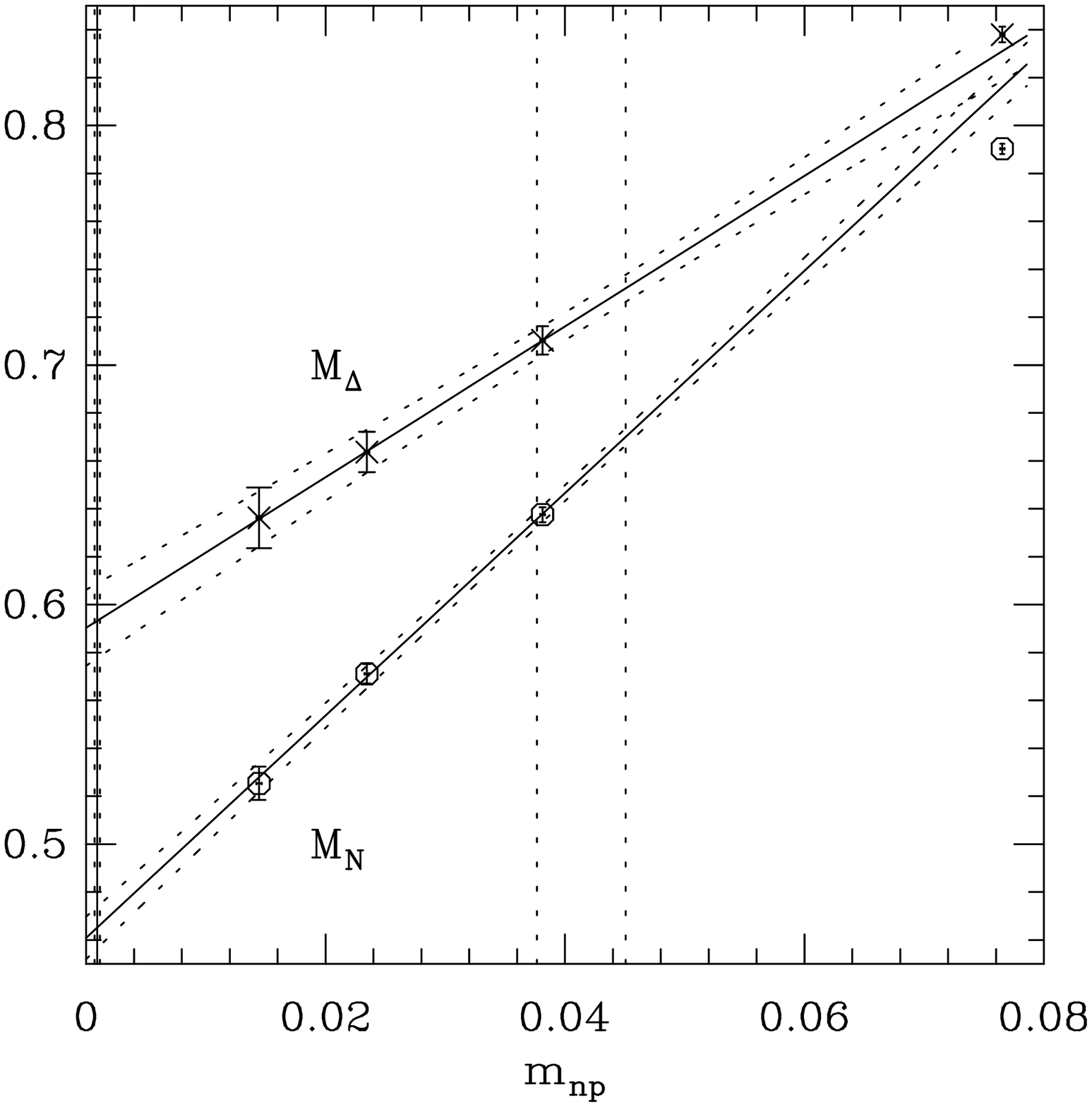}}
{\vtop{\advance\hsize by -2\parindent \noindent 
Linear fit to $M_N$ and 
$M_\Delta$ for the three degenerate quark combinations. The fourth 
degenerate point $SSS$ is also shown.}}

\medskip
In summary, we have demonstrated that terms beyond linear order
in the quark mass are necessary to fit the octet baryon mass splittings.
The lever arm provided by our four ``light'' quark masses is not, however,
strong enough to disentangle the $m^{3/2}$ and $m^2$ contributions.
Indeed, it is important to note that, if we did not have results for the
baryons containing the $S$ quark, our evidence for higher order terms would
have been much weaker, and our data would have been reasonably well
fit by an $O(m)$ term alone. This would, however, have led to 
underestimates of all the mass differences, particularly $m_\Sigma-m_N$.

\subsec{Nucleon mass}

We now turn to the overall mass scale of the spin-1/2 baryons, which
we set using the nucleon mass.  As for the mass splittings, we expect to
need terms of higher order than linear in the chiral expansion to describe
our data, so we analyze the data with and without them.
We first fit to the masses of the baryons containing three degenerate quarks.
The data are shown in Fig.~\nameuse\fnucvsmnp\ 
(along with those for the spin-3/2 baryons).
We begin with a linear fit to the lightest three baryons, yielding
the fit shown in the Figure. We find
$$
M_{N} = (0.461(9) + 4.7(2)\mbar) = 1084(27) \MeV \,.
$$
Next, we fit all four masses including an $m_q^{3/2}$ term, 
yielding $M_{N} = 1070(35) \MeV$.  
Finally, we fit all four masses including an
$m_q^2$ term, which gives $M_{N} = 1072(31) \MeV$.  
In the last two cases we have included the same form for the higher order 
corrections in $M_\rho$ for the determination of the scale and $\mbar$.  
Thus the inclusion of curvature systematically reduces $M_N/M_\rho$ 
by about $0.5\sigma$, although the precise functional form of the
higher order terms is not resolved. 
For our best estimate we take the mean of the
last two estimates, $i.e.$ $M_{N} = 1071(35) \MeV$, corresponding to 
$a^{-1}(M_N) = 2062(56)\ \MeV$.

The analysis of the previous paragraph
is based on our full sample of lattices.
Most of our results for mass differences came, by contrast,
from our sub-sample of 110 lattices.
Thus it is interesting to repeat the extraction of $m_N$ on this
sub-sample. We find that the results are slightly lower, though
consistent within errors.
For example, the linear ``3 point fit'' to the lightest degenerate baryons
yields $M_{N} =(0.446(9) + 5.04(21)\mbar) = 1051(26) \MeV$,
about $1\sigma$ below the corresponding result from the full sample.
Using the sub-sample, however, we can attempt a global fit including
baryons with non-degenerate quarks. We have done this using
the $\Sigma$-like baryons composed of $U_i$ quarks, 
fitting them to the linear chiral form of Eq.~\ebarymassquenched,
We include all eighteen $U_A \{U_B U_C\}$ correlators,
using the prescription of Eq.~\enondegenfix.
The fit, shown in Fig.~\nameuse\fnucfit, gives
$$
M_{N} = 0.452(9*) + 1.91(8*)m_A + 3.01(16*)(m_B+m_C)/2 = 1064(26*) \MeV \,.
$$
This is in excellent agreement with the 3 point fit. 
The fact that the fit is reasonable underscores the point made above
that the need for higher order terms is not apparent using the $U_i$ quarks
alone. A similar statement holds for the $\Lambda$-like states.

\figure\fnucfit{\epsfysize=4in\epsfbox{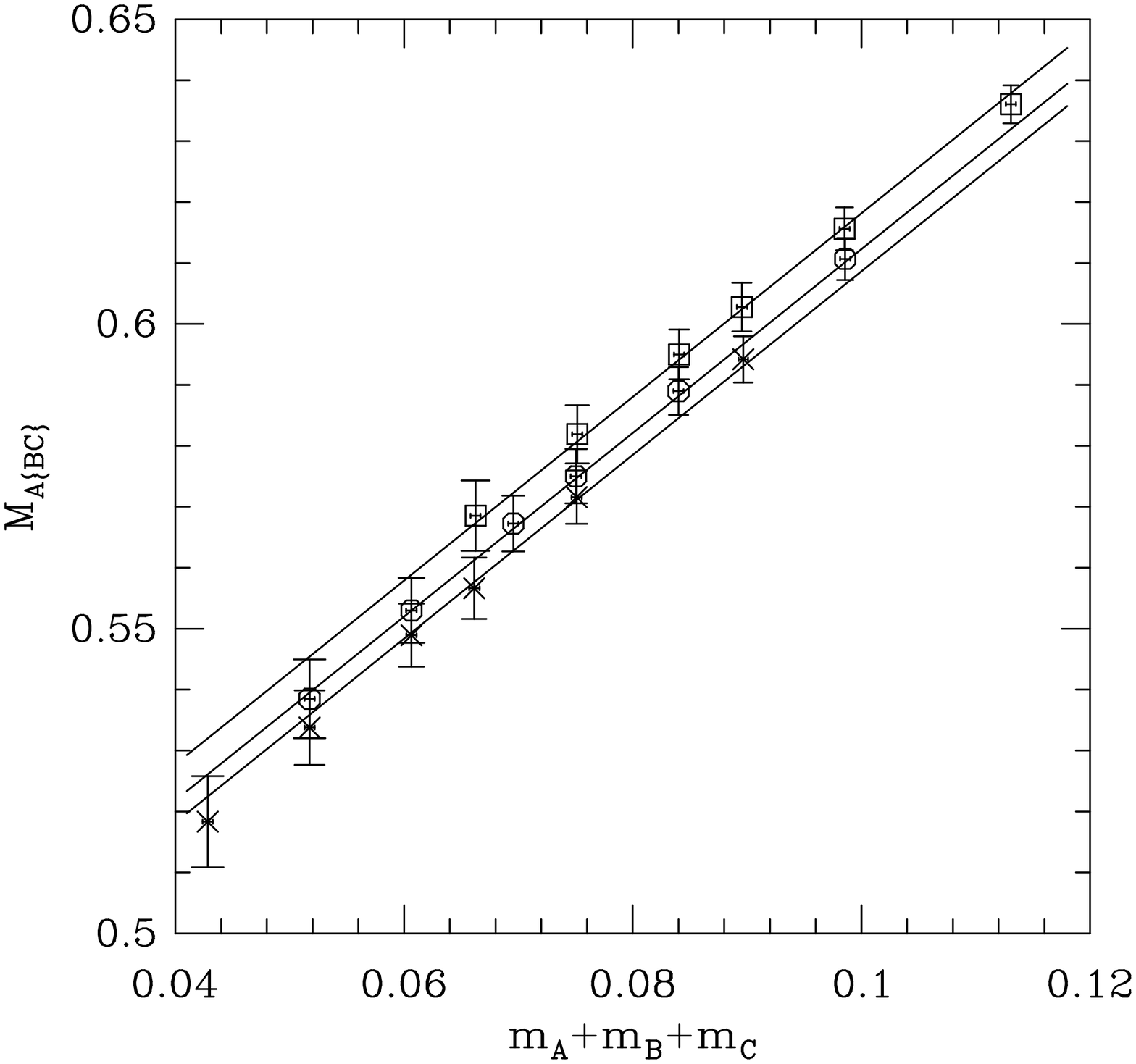}}
{\vtop{\advance\hsize by -2\parindent \noindent 
Global fit to nucleon data using Eq.~\nameuse\ebarymassquenched. The 
six cases for $m_A = U_1$ are shown by square symbols,  $m_A = U_2$ by 
octagons, and  $m_A = U_3$ by crosses.}}

\subsec{Spin-3/2 baryon mass splittings}

The analysis of the masses of the spin-3/2 baryons is more
straightforward.  As noted above, the correlators are labeled
$\{ABC\}$, and are completely symmetric between the three flavors.  We
can form 20 states with our four masses, and none of these mix with
each other.

\figure\fdecupvsmnp{\epsfysize=4in\epsfbox{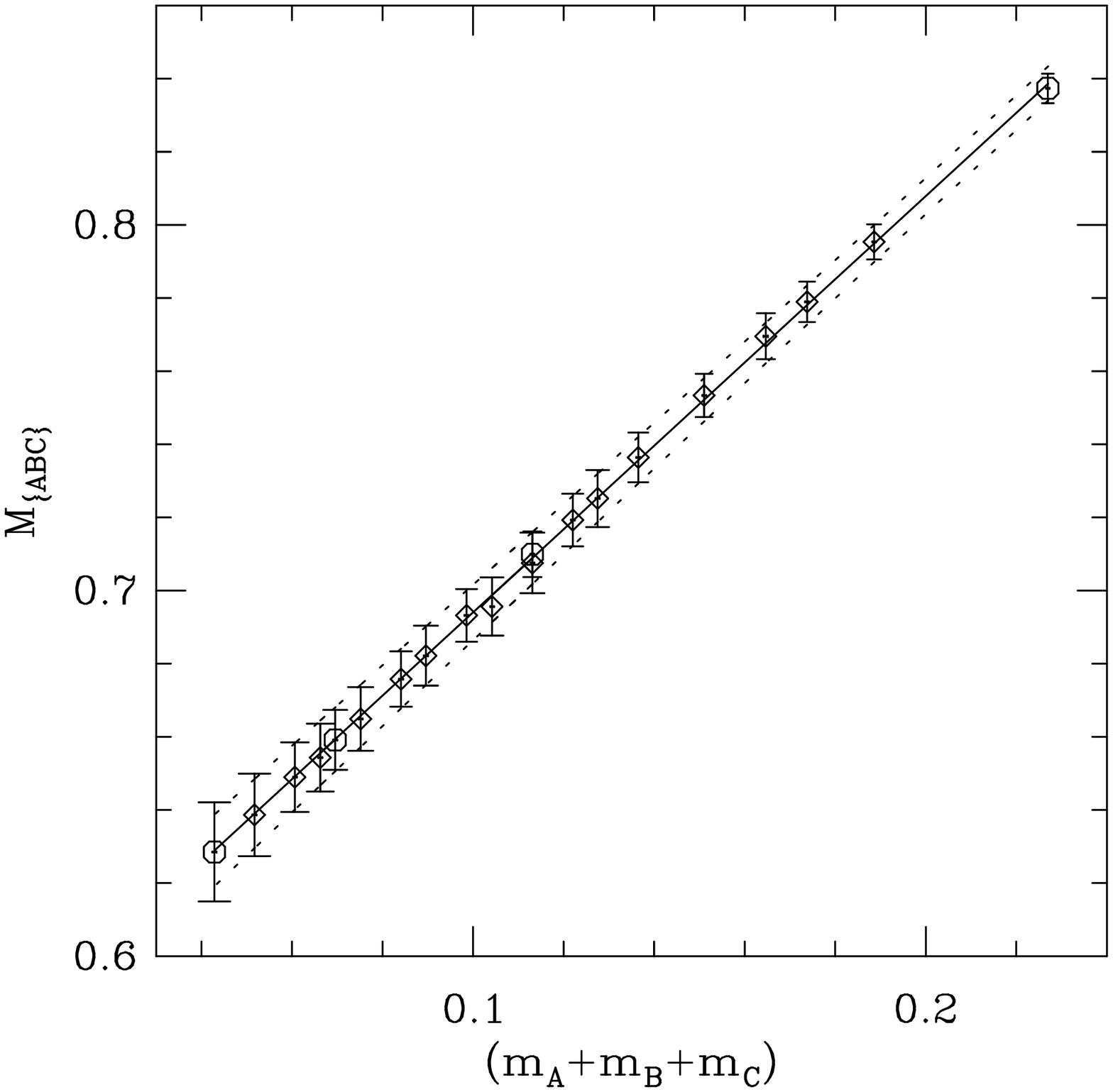}}
{\vtop{\advance\hsize by -2\parindent \noindent 
Linear fit to $M_{decuplet}$ for the 20 quark combinations. The four 
degenerate cases are shown with an octagon symbol.}}

As for the spin-1/2 baryons, we first concentrate on mass splittings.
The chiral expansion for the spin-3/2 masses has been worked out
for the case of 3 non-degenerate quarks \rSRSchiral, and is
(excluding quenched artifacts proportional to $\delta$)
\eqn\eDECmass{\eqalign{
M\decu ABC &= M_0 + \Delta M + c^\Delta (m_A + m_B + m_C) \cr
           &{}+ d_1^\Delta (M_{AB}^3+M_{BC}^3+M_{CA}^3) 
	+ d_2^\Delta (M_{AA}^3+M_{BB}^3+M_{CC}^3) \cr
           &{}+ e_1^\Delta (m_A m_B + m_B m_C + m_C m_A) 
	+ e_2^\Delta (m_A^2 + m_B^2 + m_C^2) \,.
}}
$\Delta M$ is the decuplet-octet splitting in the chiral limit.
In contrast to the spin-1/2 baryons, the terms of higher order than
linear in the quark mass are small. 
This is shown in Fig.~\nameuse\fdecupvsmnp,
in which we give the result of a linear fit to all 20 masses.
We have also done linear fits to the 10 baryons composed of $U_i$ quarks,
and to the 3 lightest baryons composed of degenerate quarks.
We refer to these three fits 
as ``20-point'', ``10-point'' and ``3-point'' fits, respectively.
The results are consistent with one another:
\eqn\edecupletfit{\eqalign{
M_{decuplet} &= 0.590(16\hphantom{*}) + 3.18(34\hphantom{*}) m_q
\hbox{\rm \hskip 1in 3 point fit (full sample)} \,,\cr
M_{decuplet} &= 0.578(16*) + 3.49(33*) m_q
\hbox{\rm \hskip 1in 3 point fit} \,,\cr
M_{decuplet} &= 0.578(16*) + 3.50(34*) (m_A + m_B + m_C)/3
\hbox{\rm \hskip 1in 10 point fit} \,,\cr
M_{decuplet} &= 0.580(12*) + 3.42(14*) (m_A + m_B + m_C)/3
\hbox{\rm \hskip 1in 20 point fit} \,,\cr
}}
although there is a small systematic difference between the
results from the full sample and the sub-sample.

The higher order terms, though small, are nevertheless present.
We have investigated them using fits to ratios of correlators.
We first consider $M_\Omega-M_\Delta$, which is obtained by
fitting our data to
\eqn\eDEComegadelta{\eqalign{
{M\decu AAA - M\decu BBB \over m_A - m_B} &= 
	3 c^\Delta + 3(d^\Delta_1+d^\Delta_2) {M_{AA}^3 - M_{BB}^3 \over m_A-m_B} \cr
        &{} + 3(e^\Delta_1+e^\Delta_2) (m_A+m_B) \,.\cr
}}
As before, we can fit our data keeping either the $d^\Delta_1+d^\Delta_2$ or 
the $e^\Delta_1+e^\Delta_2$ term. The former fit is shown in 
Fig.~\nameuse\fomegadelta,
and has parameters $c^\Delta=0.94(12)$ and $d^\Delta_1+d^\Delta_2=0.023(16)\ \GeV^{-2}$.
The slope is small and marginally significant.  Extrapolating to
the physical point (using $M_{ss}^2 = 2 M_K^2-M_\pi^2$) yields
$M_\Omega-M_\Delta=308(24) \MeV$.

We have also done the extrapolation using a more traditional method:
we calculate $M(\{AAA\})-M(\{BBB\})$ for $B=U_i$ and $A=S$ or $U_1$,
extrapolate linearly to $m_B=\mbar$, and then interpolate linearly to
$m_A=m_s$. We refer to this as the ``3 point, ratio'' method. The result is
$M_\Omega-M_\Delta=316(18) \MeV$ 
and is consistent with our $m_q^{3/2}$ estimate.
The result from our subsample, $310(19*) \MeV$, is consistent.

We have repeated this analysis for the other two physically
interesting mass splittings, and collect the results in 
Table~\nameuse\tdecdiff.
We only quote results from the ``3 point, ratio'' extrapolations,
since, with such mild deviations from linearity,
any reasonable extrapolation method using our ratio data gives 
almost the same result,
We also include the results of extrapolating using the parameters from
the 20-point linear fit.
The small difference between the two sets of results
is due to the higher order terms. Both sets of results are
significantly smaller than the experimental splittings, and
the inclusion of the higher order terms makes the disagreement worse.
These features are in marked contrast to those we saw in the spin-1/2 baryons.

\figure\fomegadelta{\epsfysize=4in\epsfbox{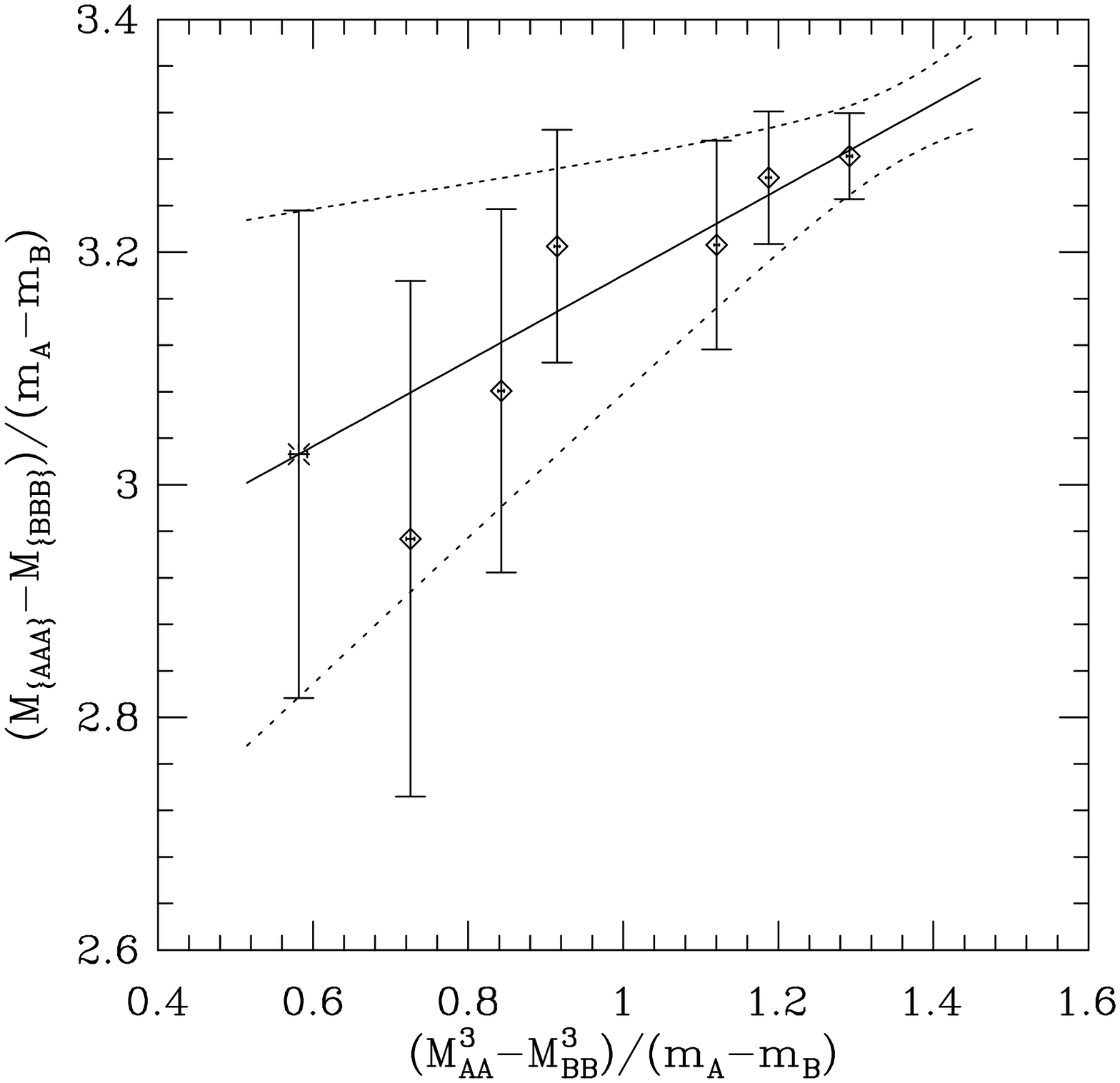}}
{\vtop{\advance\hsize by -2\parindent \noindent 
The $M_\Omega-M_\Delta$ mass-splitting.
The physical point is shown by the burst symbol at the left.
}}

\table\tdecdiff{
%
\vbox{\hbox{\indent\vbox{\tabskip=0pt\offinterlineskip
\def\myskip{\omit&height1pt& && && && &\cr}
\halign {\strut#& \vrule#\tabskip=2pt&
\hfil$#$\hfil&\vrule#&
$#$\hfil&\vrule#&
$#$\hfil&\vrule#&
$#$\hfil&\vrule#\tabskip=0pt\cr\noalign{\hrule}
%
%
\myskip\myskip
&& \hbox{}
&& \hfil\hbox{Ratio}
&& \hfil\hbox{Eq.~\edecupletfit}
&& \hfil\hbox{}
& \cr
\myskip\myskip
&& \hbox{State}
&& \hfil\hbox{3 point }
&& \hfil\hbox{20 point}
&& \hfil\hbox{Expt.}
& \cr
\myskip\myskip
\noalign{\hrule}
%
%
\myskip
\myskip
&& M_{\Sigma^*} - M_\Delta && 107(6*)  &&  116(7*)  &&  151 &\cr
\myskip
\noalign{\hrule}
\myskip
&& M_{\Xi^*   } - M_\Delta && 215(12*) &&  232(14*) &&  299 &\cr
\myskip
\noalign{\hrule}
\myskip
&& M_{\Omega  } - M_\Delta && 316(18)  &&  347(21*) &&  436 &\cr
\myskip
\noalign{\hrule}
%
%
\crcr}}}}

}
{\vtop{\advance\hsize by -2\parindent 
\noindent 
Estimates of mass splittings in the baryon decuplet. 
The different methods are explained in the text.
All results are in MeV.
}}

We can study the higher order terms in more detail by looking at
violations of the ``equal-spacing rule''
\eqn\eeqsp{
M\decu AAA - M\decu AAB\ =\ 
M\decu AAB - M\decu ABB\ =\
M\decu ABB - M\decu BBB
\,.
}
This rule holds when keeping up to the linear term in the chiral expansion.
Experimentally, the three splittings are $M_\Omega-M_{\Xi^*}=137.5 \MeV$,
$M_{\Xi^*}-M_{\Sigma^*}=148.1 \MeV$ and $M_{\Sigma^*}-M_{\Delta}= 151 \MeV$.
The violations of the rule are thus small, and it is interesting
to see how well the quenched approximation reproduces the magnitude
and pattern of these violations.
We consider two double differences
\eqn\eDECgmo{\eqalign{
{\rm ES1}(A,B) &= (M\decu AAA - M\decu AAB) - (M\decu ABB - M\decu BBB) \,,\cr
{\rm ES2}(A,B) &= (M\decu AAA - M\decu BBB) - 3(M\decu AAB - M\decu ABB) \,.\cr
}}
The first becomes $(M_\Omega-M_{\Xi^*})-(M_{\Sigma^*}-M_{\Delta})=-13 \MeV$
when $m_A=m_s$ and $m_B=\mbar$.
The expectation from quenched chiral perturbation theory is that
\eqn\eESi{
{\rm ES1}(A,B) = 2 d_1^\Delta (M_{AA}^3 + M_{BB}^3 - 2 M_{AB}^3) + 2
e_1^\Delta (m_A-m_B)^2 \,, } i.e. the form of the higher order terms
is the same as that appearing in the GMO relation (Eq.~\egmochpt).
They are shown in Fig.~\nameuse\fdecGMO, 
and are consistent with the expected chiral
form ($d^\Delta_1 = 0.46(43)\ \GeV^{-2}$, $e^\Delta_1=-8(7)\ \GeV^{-1}$),
with large errors..
Extrapolating to the physical point, we find ${\rm ES1} = -4(7) \MeV$,
marginally inconsistent with the experimental value.

\figure\fdecGMO{\epsfysize=4in\epsfbox{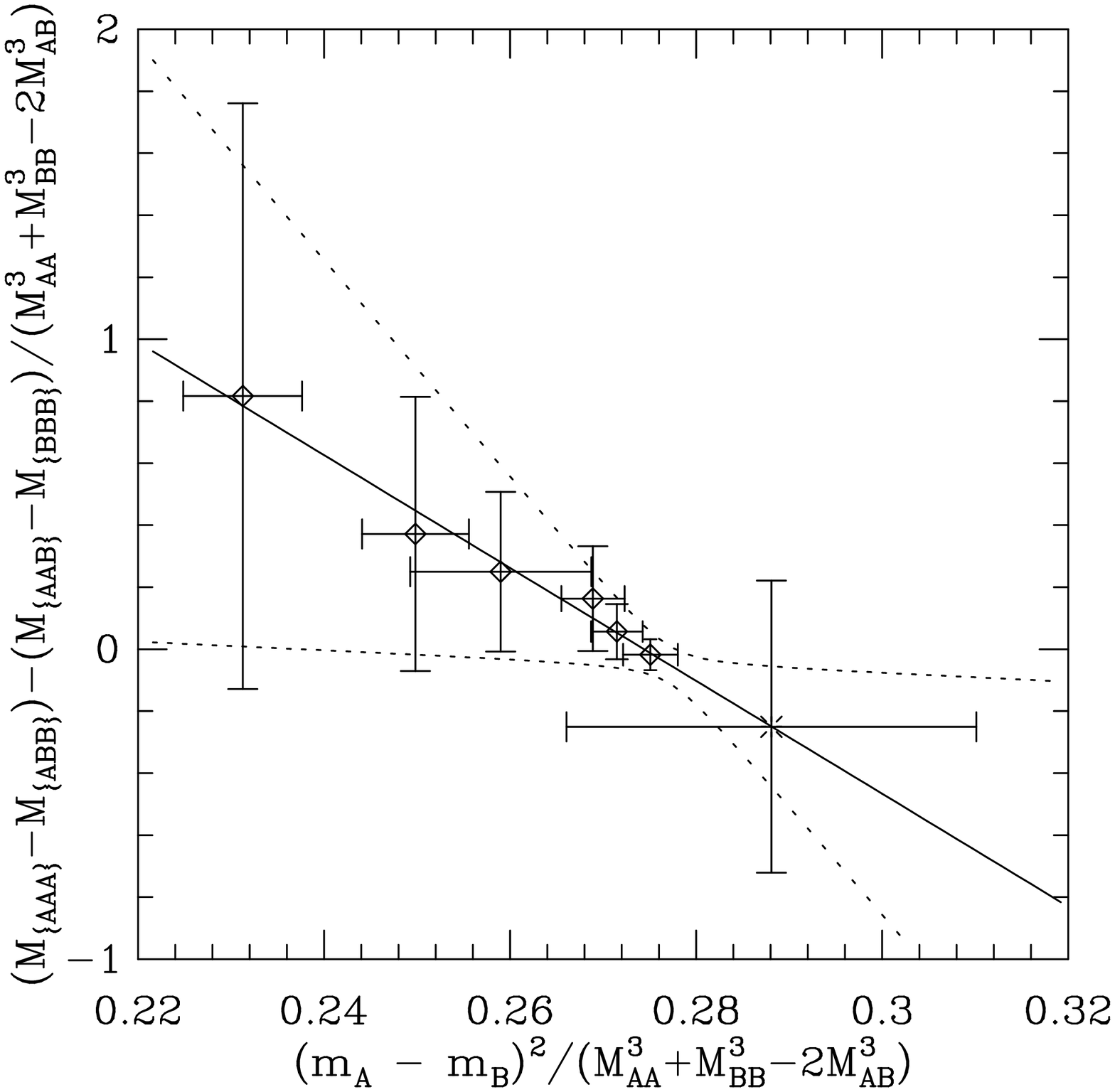}}
{\vtop{\advance\hsize by -2\parindent \noindent 
Test of violations of the equal spacing rule, Eq.~\eESi.
The physical point is indicated by the burst symbol, here at the right.
}}

The second double difference, ${\rm ES2}(A,B)$, is predicted to vanish
by the general form of Eq. \eDECmass. The same is true in chiral perturbation
theory in full QCD \ref\rjenkins{E. Jenkins, 
 \spiresjournal{Nuc.+Phys.}{B368}{190} {\NPB{368} (1992) 190}.}.
Thus it is a window onto the quenched artifacts proportional to $\delta$,
whose form is predicted to be \rSRSchiral
\eqn\eESii{\eqalign{
{\rm ES2}(A,B) &= \delta f^\Delta {\rm chiral}(A,B) \,,\cr
{\rm chiral}(A,B) &= (m_A-m_B)
\left(\ln(M_{AA}^2/M_{BB}^2) + 2 - 
{2 M_{AA}^2 \ln(M_{AA}^2/M_{BB}^2) \over M_{AA}^2 - M_{BB}^2 }\right)
\,.
}}
As explained in section Sec.~\nameuse\secquench,
we have omitted such artifacts from previous expressions, 
but we include them here so as to have a form to fit with.
Note that the function ${\rm chiral}(A,B)$ is antisymmetric under
$A\leftrightarrow B$, as it must be
in order to match the antisymmetry of the l.h.s..
Note also that it diverges logarithmically when $m_A\to 0$---an example of
the sickness of the quenched approximation in the chiral limit.

Our results, shown in Fig.~\nameuse\fdecequalspace, are statistically different
from zero. We show a linear fit of ${\rm ES2}(A,B)/{\rm chiral}(A,B)$
versus $m_A+m_B$, which yields an intercept of $\delta f^\Delta = 0.29(19)$
and a slope of $-0.9(0.7) \GeV^{-1}$. Interpolating to the physical point,
we find $ES2 = -29(18) \MeV$.
This has the same sign as the experimental value
$$
M_\Omega -M_\Delta - 3(M_{\Xi^*} - M_{\Sigma^*}) = -8 \MeV \,,
$$
but its magnitude is much larger.
It is important to realize, however, that our values for
${\rm ES2}$ range from $-1$ to $-10 \MeV$.
The large value after extrapolation is due to the divergence of
${\rm chiral(A,B)}$ as $\mbar\to0$.
We have attempted analytic extrapolations, using a linear combination
of $(m_A-m_B)^3$ and $(m_A-m_B)(m_A+m_B)^2$, but this ansatz does not
fit our data.

\figure\fdecequalspace{\epsfysize=4in\epsfbox{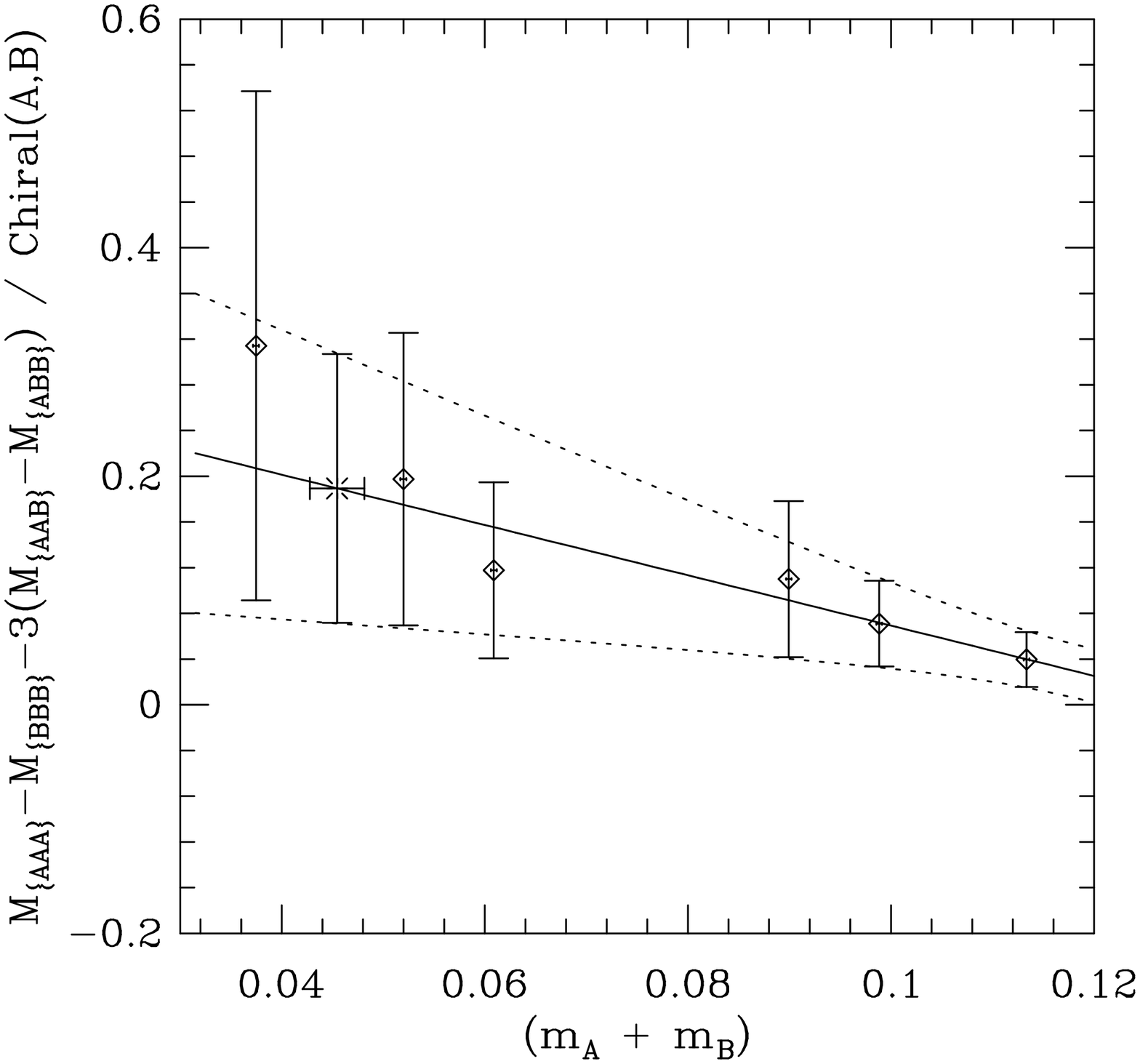}}
{\vtop{\advance\hsize by -2\parindent \noindent %
Test of violations of the equal spacing rule, Eq.~\eESii.
The physical point is indicated by the burst symbol (second from left).
}}

The most important conclusion to be drawn from this analysis is that
the decuplet splittings are smaller than the experimental values.
This discrepancy is only made worse if we use other definitions of the
strange quark mass, e.g. $m_s(M_K)$.  
We have also found some evidence for artifacts due to quenching in
the baryon spectrum. Given the size of our errors, however, this
result is far from definitive.

\subsec{$M_\Delta$ and $M_\Delta-M_N$}

We close this section with our results for the overall scale of the
spin-3/2 baryons, and its relation to that of the spin-1/2 baryons.
Our preferred value for $M_\Delta$ comes from the 3-point linear
fit to the full sample (shown in Fig. \nameuse\fnucvsmnp),
yielding $M_\Delta =1382(36)$ MeV.
Fits to the sub-sample, the parameters of which are given in 
Eq.~\edecupletfit, yield values that are consistent with one another,
but lie slightly below that for the full sample.
For example, the 20-point fit gives $M_\Delta = 1362(36*) \MeV$.
We have also tried 4-point fits to the degenerate states, including
$m^{3/2}$ (or $m^2$) corrections in both $\Delta$ and $\rho$.
These increase the estimates of $M_\Delta$ to $1410(48)$ ($1395(43)$), 
the bulk of the increase coming from the change in scale due to 
the curvature in $M_\rho$.

We have calculated $M_\Delta - M_N$ from the ratios of
correlators using the full data sample.  Using linear extrapolation
from the three light mass points we find $318(30) \MeV$.
Including an $m_q^{3/2}$ or $m_q^2$ term in both $M_\Delta - M_N$ and 
$M_\rho$ fits we get $365(44)$ and $347(39)$ respectively.
These values are slightly higher than the experimental value $293\ \MeV$. 
A summary of our
baryon mass results (without extrapolation to $a = 0$) is shown in
Fig. \nameuse\fbaryexpt\ along with the experimental data. 

\figure\fbaryexpt{\epsfysize=4in\epsfbox{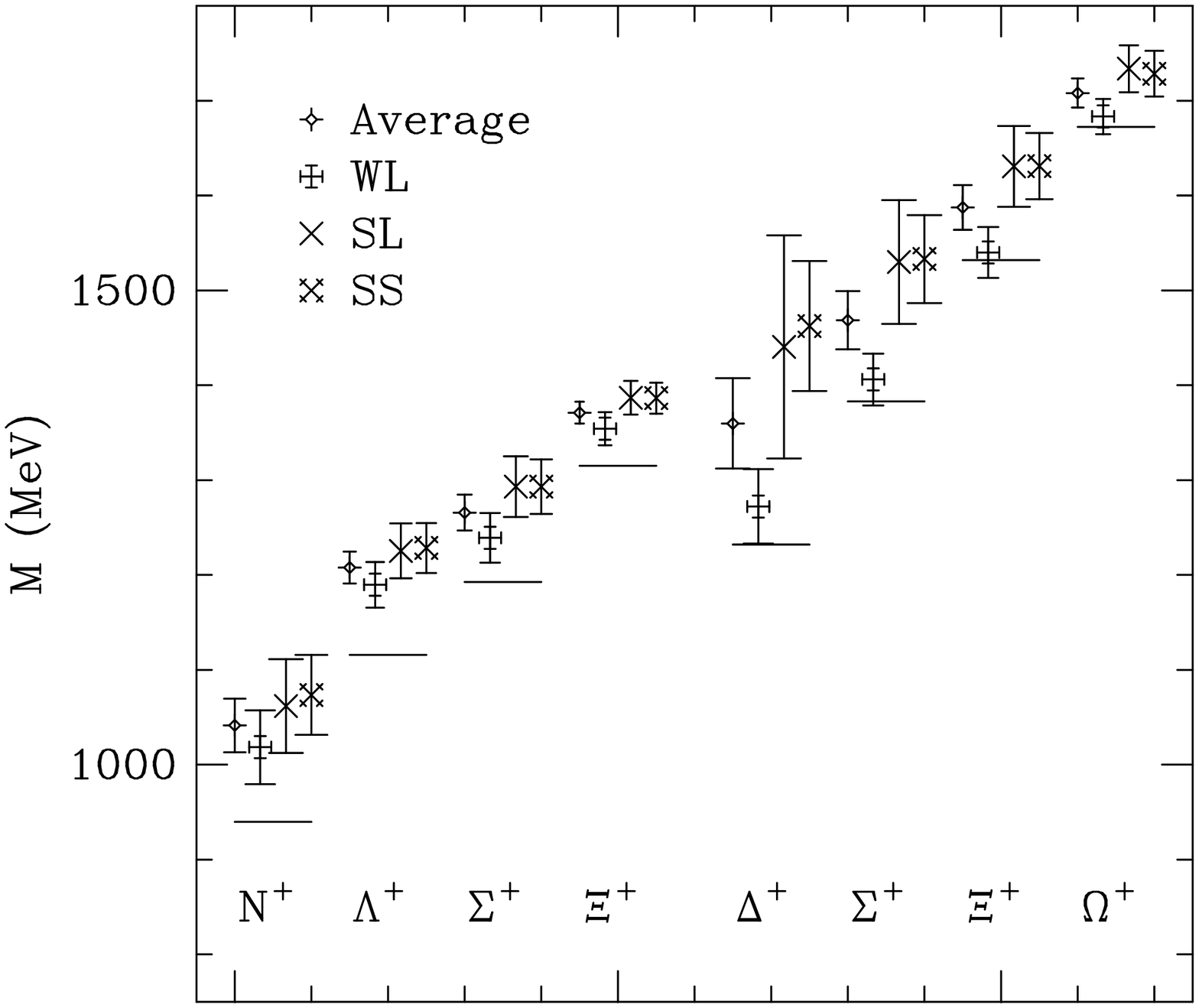}}
{\vtop{\advance\hsize by -2\parindent \noindent A comparison of the
baryon masses obtained in this calculation with the experimental data 
indicated by the horizontal lines.
The scale $a$ is set by $M_\rho$ and the data for all states are from
the sub-sample of 110 lattices.
}}

We have also made fits to the negative parity baryon states.  The
signal in the correlators is much poorer, falling below the noise by
$t=12$. For this reason we only present the summary shown in
Fig. \nameuse\fbaryexptM.

\figure\fbaryexptM{\epsfysize=4in\epsfbox{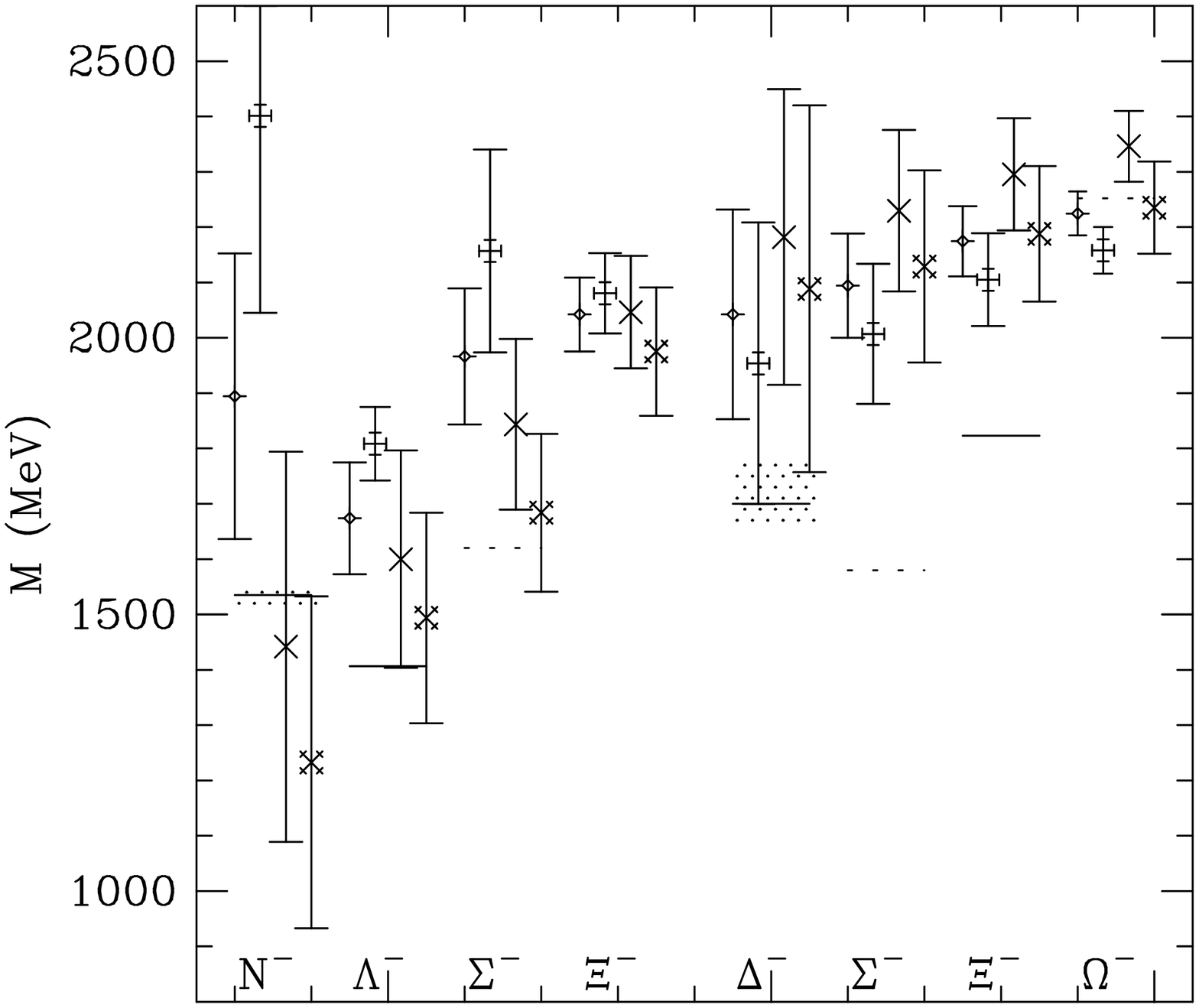}}
{\vtop{\advance\hsize by -2\parindent \noindent %
A comparison of the negative parity 
baryon masses obtained in this calculation with the experimental data 
indicated by the horizontal lines. Shaded bands show the experimental 
uncertainty.
The scale $a$ is set by $M_\rho$ and the data for all states are from
the sub-sample of 110 lattices.
}}

\newsec{Infinite volume continuum results}

There exist three other high statistics calculations of the spectrum with 
Wilson fermions at $\beta = 6.0$ on lattices of size $24^3$
\ref\APEwhm{APE Collaboration, \spiresjournal{Phys.+Lett.}{B258}{195}
{\PLB{258} (1991) 195}.}\ 
\ref\QCDPAXwhm{\spiresjournal{Nucl.+Phys.+Proc.+Suppl.}{26}{281}
{QCDPAX Collaboration, \tsukuba, 281},
and private communications.}, and \jlqcd. 
Their results are given in Table~\nameuse\tcompareold\ along with our
best estimates.  
The data indicate that there are no significant
differences between the $24^3$ and $32^3$ lattices. Thus we do not
apply any finite size corrections to our data. 

\table\tcompareold{
\vbox{\hbox{\indent\vbox{\tabskip=0pt\offinterlineskip
\def\myskip{\omit&height2pt& && && && && && && &\cr}
\def\lata{24^3 \times 32}
\def\latb{24^3 \times 54}
\def\latc{32^3 \times 64}
\def\latd{24^3 \times 64}
\halign {\strut#& \vrule#\tabskip=2pt&
\hfil$#$\hfil&\vrule#&
$#$\hfil&\vrule#&
\hfil$#\quad$&\vrule#&
$#$\hfil&\vrule#&
$#$\hfil&\vrule#&
$#$\hfil&\vrule#&
$#$\hfil&\vrule#\tabskip=0pt\cr\noalign{\hrule}
%
%
&& \kappa
&& \hfil\hbox{Size}
&& \omit\hfil\hbox{Statistics}\hfil
&& \hfil m_\pi
&& \hfil m_\rho
&& \hfil m_N
&& \hfil m_\Delta
& \cr
\myskip
\noalign{\hrule}
%
%
\myskip\myskip
&& 0.155   && \lata  && 78   && 0.298(2)  && 0.428(4)  &&  0.647(6)  && 0.745(14) &\cr
&& 0.155   && \latb  && 200  && 0.296(1)  && 0.417(5)  &&  0.645(6)  && 0.728(7)  &\cr
&& 0.155   && \latc  && 170  && 0.296(1)  && 0.422(2)  &&  0.638(3)  && 0.710(6)  &\cr
&& 0.155   && \latd  && 1000 && 0.2964(4) && 0.423(2)  &&  0.642(3)  &&           &\cr
\myskip
\noalign{\hrule}
\myskip
&& 0.1558  && \lata  && 78  && 0.234(3) && 0.397(5)  &&  0.574(8)  && 0.686(25) &\cr
&& 0.1558  && \latc  && 170 && 0.234(1) && 0.387(3)  &&  0.571(4)  && 0.664(8)  &\cr
\myskip
\noalign{\hrule}
\myskip
&& 0.1563  && \lata  && 78  && 0.184(3) && 0.378(10) &&  0.522(14) && 0.636(45) &\cr
&& 0.1563  && \latb  && 200 && 0.184(2) && 0.343(14) &&  0.530(16) && 0.629(19) &\cr
&& 0.1563  && \latc  && 170 && 0.185(1) && 0.361(5)  &&  0.525(7)  && 0.636(13) &\cr
\myskip
\noalign{\hrule}
%
%
\crcr}}}}
 
     }
{\vtop{\advance\hsize by -2\parindent 
\noindent 
Comparison of hadron masses obtained using Wilson fermions at $\beta=6.0$. 
Results from $32^3\times 64$ lattices are from the present work,
while $24^3 \times 32$ data is from Ref.~\APEwhm, 
$24^3 \times 54$ data is from Ref.~\QCDPAXwhm, and 
$24^3 \times 64$ data is from Ref.~\jlqcd.}}

Our best estimates at $\beta=6.0$ for various mass ratios of interest are 
\eqn\emassrat{\eqalign{
{M_N      \over M_\rho} = 1.412(35) \qquad \hbox{\rm Expt: 1.22}
\,, \cr
{M_\Delta \over M_\rho} = 1.800(47) \qquad \hbox{\rm Expt: 1.60}
\,, \cr
{M_\Delta \over M_N}    = 1.275(36) \qquad \hbox{\rm Expt: 1.31}
\,. \cr
}}
These are obtained by extrapolating individual masses linearly to
$\mbar$, and then taking ratios within the jack-knife procedure.  All
extrapolations are done using only the $U_i$ quarks.  To check for
extrapolation errors we have also calculated the ratios for each quark
mass and then linearly extrapolated these to $\mbar$.  This
yields $M_N / M_\rho = 1.42(3)$ $M_\Delta / M_\rho = 1.79(4)$, and 
$M_\Delta / M_N = 1.25(3)$ consistent with the first method.

Our results for $M_N/M_\rho$ and $M_\Delta/M_\rho$ are larger than the
experimental values. It has been stressed by the GF11 collaboration,
however, that these ratios have a significant dependence on $a$. In an
attempt to check this we combine our results with those from the GF11
Collaboration \weinhm\ at $\beta=5.7,\ 5.93$, and $6.17$. The two
calculations have similar statistics, while the physical volume of our
lattices is larger.  The GF11 collaboration have made two separate
extrapolations to the continuum limit using different gaussian smeared sinks.
In their notation, ``012'' is a combination of results from three
sinks with smearing radii $0,\ 1, 2$, while ``4'' refers to the use
of a single larger smearing radius of size $4$.  In the final analysis they
prefer to use the ``012'' result as it has smaller statistical errors.
We update these two linear fits (using the same 3 data points used in
\weinhm\ and ours at $\beta=6.0$), and the results are shown in
Fig.~\nameuse\fwein. After
including our point, the extrapolated values for the ``012'' data
(dotted lines) change from $M_N / M_\rho = 1.28(7)$ to $1.30(6)$ and
from $M_\Delta / M_\rho = 1.61(8)$ to $1.62(7)$.  The $\chi^2/_{dof}$
for the new fits are $2.1$ and $0.85$ respectively.  The analogous
numbers for the sink ``4'' data (solid line) are $ 1.33(9) \to
1.38(7)$ and $ 1.68(10) \to 1.73(10)$ with $\chi^2/_{dof}$ for the new
fits equal to $1.2$ and $0.86$ respectively.

As is evident from Fig.~\nameuse\fwein, the main difference between
the two fits comes from the difference in the ``012'' and ``4'' data
at $\beta=5.7$.  On the basis of $\chi^2/_{dof}$, we find that
combining our results with the sink ``4'' GF11 data is preferred, in
which case there is very little $a$ dependence.  If we neglect the
point at strongest coupling, $\beta=5.7$, then the remaining three
points again show no clear $a$ dependence for both $M_N / M_\rho$ and
$M_\Delta / M_\rho$, and give very similar values for the fit
parameters.

The ambiguity in the extrapolation makes it clear that data at more
values of $\beta$ are needed in order to reliably extrapolate to the
continuum limit.  
Our preferred estimates
are $M_N / M_\rho = 1.38(7)$ and $M_\Delta / M_\rho = 1.73(10)$ from
fits shown by a solid line in Fig.~\nameuse\fwein. This suggests
that the quenched approximation is good to only $\sim 10-15\%$.

\figure\fwein{\epsfysize=4in\epsfbox{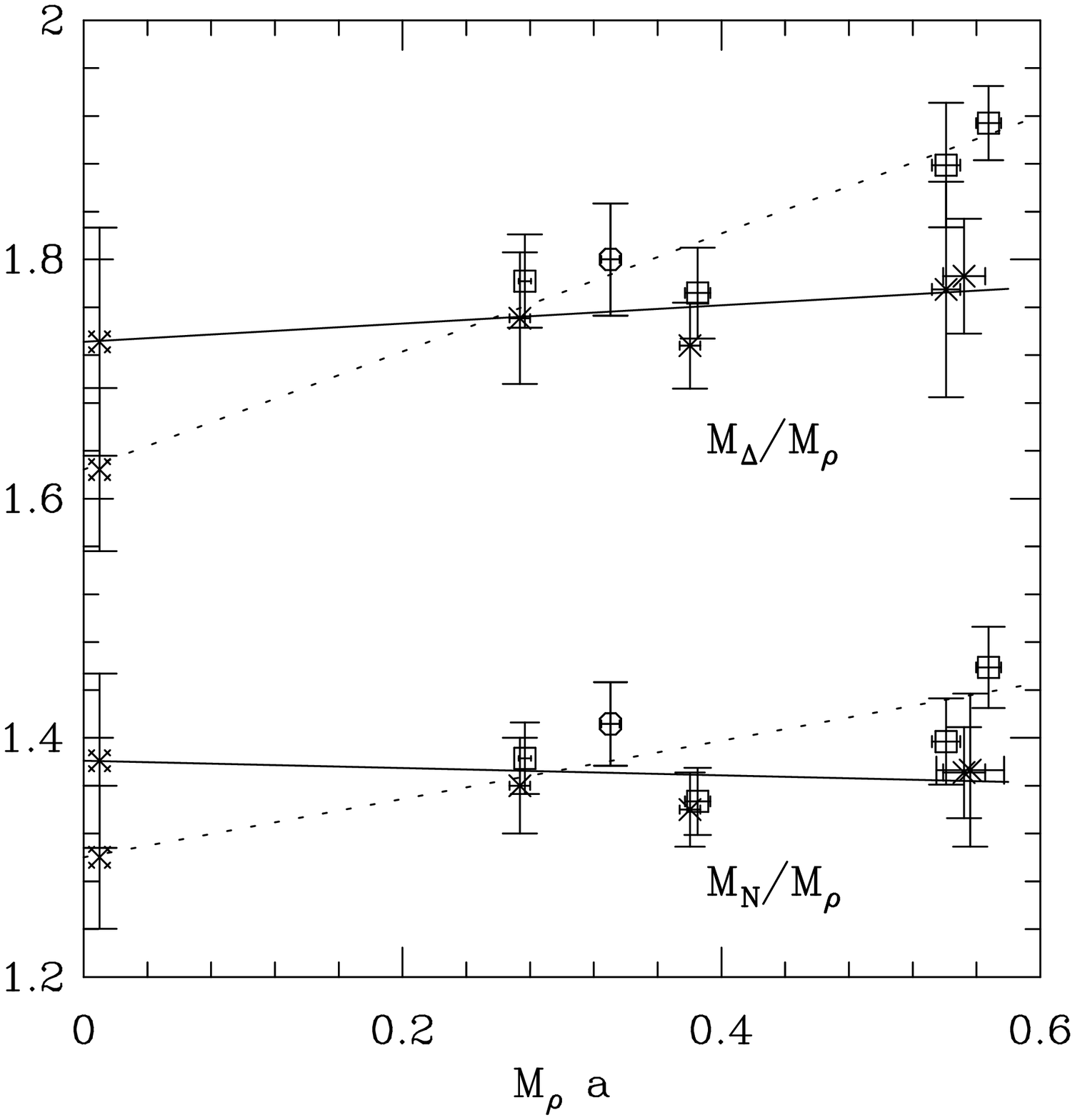}}{\vtop{%
\advance\hsize by -2\parindent \noindent Linear extrapolation 
to the continuum limit of the ratios $M_N / M_\rho $ and $M_\Delta /
M_\rho $. Our data is shown with symbol octagon, and the rest of the
points are from the GF11 Collaboration \weinhm.  The ``012'' sink
points are labeled by squares and the sink ``4'' points by crosses.
The extrapolated values are shown by symbol fancy cross and are
shifted from $M_\rho a = 0$ for clarity. At $\beta=5.7$ ($M_\rho a 
\approx 0.56$) we have shown GF11 collaboration's data for both $16^3$
and $24^3$ lattices even though they used only the $16^3$ data in the
fit and shifted their continuum result by the difference to account
for finite size effects.}}


\newsec{Conclusions}

We have presented a detailed analysis of the hadron spectrum in
quenched QCD at $\beta=6.0$ with Wilson fermions, focusing on states
composed of light quarks. Our small statistical errors, and our
use of several moderately light quarks, have allowed
us to improve the extrapolations to the chiral limit.
This is particularly true of the mass splittings amongst the octet baryons.
Here we find substantial contributions from
terms of higher order than linear in the quark mass.
Motivated by quenched chiral perturbation theory, we have
found good variables with which to extrapolate to the physical quark
masses. Our results show that the splittings are larger than previously
thought, and are comparable with their experimental values.
These results emphasize the importance of calculating masses of
baryons composed of several combinations of non-degenerate quarks.

The extrapolations required for the $\pi$, $\rho$, $N$ and $\Delta$
are less sensitive to higher order terms.
There is a clear curvature in the $\rho$ and nucleon channels, 
and it can be accommodated either by including a term of $O(m_q^{3/2})$ 
(which would result from chiral loops) 
or a term of $O(m_q^2)$. 
The effect on the extrapolated values for
the $M_\rho$ and $M_N$ is, however, small---roughly comparable to the
statistical errors.
Higher order terms are also small for the decuplet baryons.

It is not surprising that higher order terms are needed when 
considering quarks with masses ranging up to and beyond that of the
physical strange quark. In previous calculations of 
baryons composed of degenerate quarks, these terms were small and 
often neglected. 
What is striking is how this is not true for many of the mass 
differences involving baryons composed of non-degenerate quarks.

One caveat concerning the results for mass splittings is 
the fact that there are substantial systematic errors in the
extraction of $m_s$.
Different methods lead to results differing by up to $\sim 20\%$.
This is presumably an error due to quenching, although some part of it
could be due to discretization.
Our favored choice for $m_s$ is that determined by matching to the ratio
$M_\phi/M_\rho$. 
This gives the largest estimate for $m_s$.

How does the spectrum at $\beta=6$ compare with experiment?
We find that the ratio $M_N/m_\rho = 1.41(4)$ is too high, while
$M_\Delta / M_N = 1.27(4)$ is consistent with experiments.
Using the larger estimate of $m_s$, the octet baryons splittings
agree with experiment, while those in the decuplet are too small by 30\%.

Of course it is quite possible that there is substantial variation
in some of these ratios as we extrapolate to the continuum limit.
Combining our data with that of the GF11 collaboration, however,
we conclude that there remains considerable uncertainty in this extrapolation.
Our preferred extrapolation gives $M_N / M_\rho = 1.38(7)$ and
$M_\Delta / M_\rho = 1.73(10)$,
but the systematic errors exceed those from statistics.
Thus, in our view, it remains an open question how well the quenched
approximation represents full QCD when extrapolated to the continuum limit.
The errors could well be as large as $\sim 10-15\%$.

\newsec{Acknowledgments}

These calculations have been done on the CM5 at LANL as part of the
DOE HPCC Grand Challenge program, and at NCSA under a Metacenter
allocation.  We thank Jeff Mandula, Larry Smarr, Andy White and the
entire staff at the two centers for their tremendous support
throughout this project.

\listrefsnomod
\end